\documentclass[acmsmall]{acmart}

\AtBeginDocument{%
  \providecommand\BibTeX{{%
    \normalfont B\kern-0.5em{\scshape i\kern-0.25em b}\kern-0.8em\TeX}}}

\setcopyright{acmlicensed}
\acmJournal{CSUR}
\acmYear{2021} \acmVolume{1} \acmNumber{1} \acmArticle{1} \acmMonth{1} \acmPrice{15.00}\acmDOI{10.1145/3446662}

\acmJournal{CSUR}
\acmVolume{37}
\acmNumber{4}
\acmArticle{111}
\acmMonth{8}

\usepackage{amsmath,amsfonts}
\usepackage{tabularx}
\usepackage{float}
\usepackage{graphicx}
\usepackage{booktabs}
\usepackage[font=small,skip=0pt,aboveskip=0pt,belowskip=0pt]{caption}
\usepackage{subfigure}
\usepackage[mathscr]{eucal}
\usepackage{color, soul}
\usepackage{multirow}
\usepackage{array}
\usepackage{hyperref}
\usepackage{chngcntr}
\usepackage{caption}
\newcommand{\tnote}[1]{$^{\text{#1}}$}
\begin{document}

\title{User Response Prediction in Online Advertising}

 \author{Zhabiz Gharibshah}
 \authornotemark[1]
 \email{zgharibshah2017@fau.edu}
 \affiliation{%
   \institution{Dept. of Computer \& Elect. Eng. and Computer Science, Florida Atlantic University}
   \streetaddress{777 Glades Road}
   \city{Boca Raton}
   \state{FL}
   \postcode{33431}
 }

 \author{Xingquan Zhu}
 \authornotemark[1]
 \email{xzhu3@fau.edu}
 \affiliation{%
   \institution{Dept. of Computer \& Elect. Eng. and Computer Science, Florida Atlantic University}
   \streetaddress{777 Glades Road}
   \city{Boca Raton}
   \state{FL}
   \postcode{33431}
 }
\begin{abstract}
\textbf{Abstract}-Online advertising, as the vast market, has gained significant attentions in various platforms ranging from search engines, third-party websites, social media, and mobile apps. The prosperity of online campaigns is a challenge in online marketing and is usually evaluated by user response through different metrics, such as clicks on advertisement (ad) creatives, subscriptions to products, purchases of items, or explicit user feedback through online surveys.
%It typically gets started by clicking on ad creative to the last action like buying or subscribing a product.
%Following a performance dependent approach, advertisers and publishers are charged and paid if users respond positively to ads. Therefore, predicting user responses in different forms, such as click-through rate or conversion rate, plays an essential role in digital advertising. 
Recent years have witnessed a significant increase in the number of studies using computational approaches, including machine learning methods, for user response prediction. However, existing literature mainly focuses on algorithmic-driven designs to solve specific challenges, and no comprehensive review exists to answer many important questions. What are the parties involved in the online digital advertising eco-systems? What type of data are available for user response prediction? How to predict user response in a reliable and/or transparent way? In this survey, we provide a comprehensive review of user response prediction in online advertising and related recommender applications. Our essential goal is to provide a thorough understanding of online advertising platforms, stakeholders, data availability, and typical ways of user response prediction. We propose a taxonomy to categorize state-of-the-art user response prediction methods, primarily focus on the current progress of machine learning methods used in different online platforms. In addition, we also review applications of user response prediction, %the model evaluation tools, 
benchmark datasets, and open-source codes in the field.
\end{abstract}



% \keywords{datasets, neural networks, gaze detection, text tagging}

\maketitle
% \counterwithin{table}{section}
% \counterwithin{figure}{section}
\section{Introduction}
Online advertising~\cite{zhu:2017:fraud}, as a multi-billion dollars business, provides a common marketing experience when people are accessing to online services using electronic devices, such as desktop computers, tablets, smartphones \textit{etc}. 
% \ttnote{a} is b
% Using the internet as a means of advertising, there are various ways to show and place advertisements to the eyes of viewers. Search engines and social network platforms are the common ways to do so. In search engines, it is treated to place ads alongside the search
% results while in social media marketing and display advertising, some spots in web pages are dedicated to display ads within the original content.
Using Internet as a means of advertising, different stakeholders act in the background to provide and deliver advertisements to users through numerous platforms, such as search engines, news sites, social networks, where dedicated spots of areas are used to display advertisement (ad) along with search results, posts, or page content. 

%that are treated to place ads alongside the search results or social media marketing and contextual advertising where some spots in web pages are dedicated to display ads within the contents of web sites. 

Similar to the traditional media, such as printing magazines and newspapers where specific spaces are assigned to be sold for ads, a portion of online services and websites are filled with clickable components to display marketing messages. Under such circumstances, the ads to be displayed to audience (\textit{i.e.}, users) are either pre-sold (\textit{i.e.}, negotiated) by sellers (publishers) to buyers (advertisers) or they are dynamically selected through a real-time bidding (or auction)~\cite{SurveyYuan2014,DisplayWang2016}. In online advertising, advertisers are bidding an ad opportunity, but only the winner has the chance to serve their ads to users (so only the winner needs to pay to the publisher for the purchase of the auctioned ad opportunity). During the whole process, the effectiveness of the online advertising is typically evaluated through signals made by users towards the displayed ads. These signals are typically considered as user responses starting with a click on ads in web-pages or a tap on screen in mobile apps. Once displayed ads are clicked by users, the payment/revenue is generated between advertisers and publishers. As a result, for both advertisers and publishers, it is crucial to design a user response based pricing model.

%Contrary to conventional digital advertising platforms where the content of advertisement in publisher websites are previously selected from pre-paid contents, online advertising systems run under a complicated ecosystem to select ads mainly using a real-time bidding (or auction) based mechanism. In online advertising, multiple advertisers are bidding an ad opportunity, and only the winner(s) has the chance to deliver their ads to users. During the whole process,  are  They evaluate the relevancy of ads from signal made by users. These signals are typically considered as user responses starting with a click on ads in web-pages or a tap on screen in mobile apps. It leads to develop a user response based pricing model that advertisers pay to online platforms when displayed ads are clicked by users.

%determined by the probability users represent positive responses. These responses simply start with signal sent by users.  with the hope to grasp positive user responses. This develops a pricing model a commercial response based model called cost-per-click pricing to pay publisher when ads are clicked by users. 
%Clicks on the advertisements are awarded to the published by a model called Cost-per-click(CPC). This model defines the price of each add based on the location of the ad on the webpage and importance of the webpage and etc. 
%The current online advertising systems follow a commercial response based model called cost-per-click pricing to pay publisher when ads are clicked by users. 
Predicting a click, as the first measurable user response, is an important step for many digital advertising and recommendation systems to capture the user propensity to following up actions, such as purchasing a product or subscribing a service. Based on this observed feedback, these systems are tailored for user preferences to decide about the order that ads should be served to them.
 %  connect It is found beneficial to use a modern online advertising in different aspects.
%From a simple static banner advertising~\cite{sherman:2001:banner} to user response based models, online advertising and recommendation systems have been rapidly evolved to benefit both users and businesses. Compared to conventional approaches, using the web as the means of advertising make the way possible for business owners to find prospective customers at the massive scale across the globe. In industry, various recommender systems are extensively developed in e-commerce platforms to predict users interest and guide them with list of products among huge possible options. %they might get interested into. , leads to one of great advantages of these systems to have a capability 
%One of the great advantages of these systems is that they can use historical user data captured from web cookies to target a particular group of people potentially interesting in the products or services, with minimum delay or in real-time.  
In the era of search engines and social websites, companies like Google introduce paid search advertising~\cite{Rutz:2011:generic} via user intents recognized through the query keywords. In social media marketing, platforms like Facebook provide advertisers with user demographic information from user-generated content for viral marketing~\cite{Chu:2011:viral}. In conventional advertising in TV or printed newspapers, monitoring the effectiveness of ads is difficult. 
%Contrary to conventional advertising in TV and printed newspapers in which business owners could not monitor the effectiveness of ads and should follow negotiations for long term contracts, 
However, online advertising leverage performance metrics for targeting ad audience, so stakeholders can immediately obtain ad advertising feedback, through clicks, conversions, and other types of user response, to adjust their budget, price for bidding etc. ~\cite{DisplayWang2016}.  
%. in terms of number of time users visit and follow the ads 
%in addition to the
%As a result, they provide tremendous flexibility in the content type and budget to establish ad campaigns for users in determined time.
 
The essential goal of different types of advertising systems, either traditional media based or modern online advertising based, %types %mentioned above 
is to find the best matching between audience (users) and ads, given contextual features in each platform. From computational perspective, this is equivalent to finding a way to accurately predict positive or negative user responses to an ad, given observed user data. It is shown that the accurate prediction of user response metrics can directly determine the revenue for both publishers and advertisers~\cite{SimpleChapelle2015,EstimatingAgarwal2010}. The variation of the problem is defined by the availability of context in different platforms. The context in search engines are query generated by users. In display advertising, the context is considered as websites visited by users, and in-app advertising the context is the specific logical stage in mobile apps for marketing. 

For years, industry and academia have developed numerous approaches to use holistic data to predict positive response of users where the positive response is typically defined in the form of the estimation of click-through rate on ads or user interactions for purchasing a product, \textit{i.e.} a conversion. Such approaches vary from data hierarchy~\cite{EstimatingAgarwal2010,EstimatingAgarwal2007,PredictingOentaryo2014,EstimatingLee2012}, clustering~\cite{ImprovingHillard2010,PredictingRegelson2006,PersonalisingReps2014,ClickWang2011}, collaborative filtering~\cite{ResponseMenon2011,CollabLiu2014}, classification~\cite{FMRendle2010,ScalableDalessandro2014,LearningXu2019,HigherOrderBlondel2016,FIBINETHuang2019}, to graph and network based analysis~\cite{RippleNetWang2018,MKRWang2019}.
%So far 

As data are becoming rapidly available, machine learning based approaches have been used in nearly all domains to solve different types of challenges for knowledge discovery~\cite{zhu:2007:knowledge}. For online advertising, this is especially true. Since the very beginning, industry has been actively seeking effective and efficient computational methods to tackle the data volumes and real-time decision challenges. Many approaches, such as deep learning and factorization machine based methods, demonstrate a great potential to accurately estimate user responses~\cite{FFMJuan2017,DeepFMGuo2017}, but the data intensive nature and the real-time requirement have made the accurate user response prediction for online advertising extremely challenging. Here, we briefly summarize the major challenges as the following fourfold: % of user still exist that can be summarized as follows.
%[PBODL paper]

\begin{itemize}
\item\textbf{Scalability:} In real-world advertising eco-system, the number of visited web-pages is extremely large. Combining with factors like the number of unique visiting users and the amount of ads, it results in a giant dataset for analysing. In many studies \cite{DeepChen2016} machine learning has been applied to predict user response and boost the personalizing of digital advertising. %However, confronting this data, training process is time-consuming which needs a lot of computing resources. %It cannot usually be used for online scenarios. Hence, 
It is important to design solutions for %user response prediction algorithms to estimate the click-through rate or conversion rate values via studying user interaction logs gathered from 
large scale advertising data~\cite{SimpleChapelle2015,PracticePi2019,BillionWang2018,ImageGe2018}.

\item\textbf{Response rarity:} %Although in the beginning years of online advertising in 1990s online ads got a lot of attention from users that generated high rate of click-through rate. But 
Statistics shows that the rate of click and conversion of all types of ads is not more than 2 percent over all displaying ads. 
%The latest statistics reported in  \cite{InternetAFLi2004} shows that the rate of click and conversion of all types of ads is not more than 2 percent over all displaying ads. %It means that considering the advertising events in the form of a network including two types of vertices of users and ads and the links between them as clicks, only very few numbers of link are observed. It is indicated that the distribution of click over every showing of ad follows power law pattern\cite{EstimatingAgarwal2007}. %Some web-pages are more popular which account for ads many more times. However, the overall rate is a few percent. 
Therefore, finding a way to overcome class imbalance issue and mitigate the adverse effects on prediction results is a challenge for the prediction algorithms.

\item\textbf{Data sparsity:} This issue in online advertising and recommender systems stems from two factors. First, majority of input data consist of categorical features which need to use binary representation, resulting in high dimensional vectors with very few non-zero values. Additionally, interactions between users and items follow the power law distribution, meaning majority users are interacting with a small number of items and products.% compared to all items and their related ads. 

\item\textbf{Cold start:} This is the common challenge for new new ads, products, and services, because no historical user information available is available to be used for estimation. %Therefore, there are frequent stages in system that the majority of available data are accounted to small number of ads and users while remaining ad items incorporate the small portion of samples. 
%Therefore, it brings about challenges for machine learning and data mining methods to deal with these items which are not in users behavior history.  
\end{itemize}

Indeed, many solutions have been proposed, but primarily focus on new methods for user response prediction. Several works propose to study current business model and technologies evolved from traditional media buying~\cite{SurveyYuan2014,InDepthChen2016}, or review display advertising literature and new directions~\cite{OnlineChoi2019}.
%and number of reviews are available in literature on online advertising none of which consider the existing machine learning methods for user response prediction. The focus of these studies are focused on current business model and technologies evolved from traditional media buying. \cite{SurveyYuan2014,InDepthChen2016,OnlineChoi2019}. 
In~\cite{SurveyYuan2014}, the authors go over the business model of real-time bidding by introducing keys actors in the market for ad delivery. From economic perspective, \cite{OnlineChoi2019} outlines the eco-system of display ad market and non-guaranteed selling channels provided to buy and sell ads in real-time. It reviews the disciplines regarding the ad pricing decision made by different actors like advertisers and publishers and other intermediary nodes. The study \cite{InDepthChen2016} reviews the technologies provided for online and mobile advertising, including pricing models implemented between advertisers and publishers, inherent networking schemes by addressing the user privacy and malicious ad related activities. %It also reviews studies covering the human preferences factor and power consumption constraints in hand-held devices in connection to mobile advertising. 

Unfortunately, all existing works, including the literature review, do not provide a complete overview about types of user response and underlying technical solutions in online advertising. Answers to many key questions remain unclear for both industry and academia, especially for someone who just steps into the online advertising field. What are the main advertising platforms? What type of user response can be modeled/predicted using computational approaches? What are the features and the source of features useful for use response prediction? How to utilize features for use response prediction? What are the main types of technical solutions for user response prediction? Are there any benchmark and online resources (datasets/software) available for evaluation purposes?

%\subsection{Contribution}
In this paper, we provide a comprehensive literature review of the latest computational methods for user response prediction in online advertising, with a focus on machine learning based approaches. To the best of knowledge this is the first survey study which is focusing on computational approaches for user response prediction. Our review covers different types of user response prediction tasks  %a focus on
ranging from click-through rate prediction to user post-click experience evaluation. Our survey includes the description of the online advertising eco-system, platforms, data sources, and early studies for user response prediction. We also consider the most recent work in this context which propose more advanced algorithmic designs and feature extraction methods. %The contribution of survey can be summarized as follows:

\section{Advertising Eco-System \& User Response}
In this section, we introduce key components and important concepts of online advertising eco-system. %We first review business model and online platforms which provide users with personalized advertisement. The architecture includes the ad network and several key components in dealing with ad requests. After that, we will briefly summarize types of user response, including user clicks, user engagement, and video completion. 
For ease of understanding, we summarize key concepts and their descriptions in Table~\ref{tab:notations}.% and notations  In this model, the criteria to present best matched advertisements for users is designed from analyzing different type of user responses.

\subsection{Online Advertising Eco-System}
%Traditionally, digital advertising is known by %followed typical inventory-centric pre-paid advertising in which 
%prepaid ads being displayed to the online audience without measuring their feedback. 
Online digital adverting heavily relies on real-time bidding (auction)~\cite{SurveyYuan2014} for advertisers to make decisions to display ads in online portals. 
%To this end, it follows the idea of stock exchanges to have an infrastructure that ads can be bought and sold on demand in real-time. 
%As you can see, in the ecosystem of online advertising, 
In this architecture, an ad exchange network connects sellers (publishers) and buyers (advertisers), so they can negotiate to respond to ad requests in real time. 
In order to participate in the ad bidding, publishers and advertisers connect to the ad exchange network through SSP (Supply-Side Platform)s and DSP (Demand-Side Platform)s, respectively, to cast auctions (for SSP) and manage bids (for DSP), therefore ads are eventually delivered to different media platforms, \textit{e.g.} a  third-party website, search engine result page, or the web-page of social networks. %From advertisers perspective, ad network take on the responsibility to help for selecting media and ad placement. 

\begin{figure*}[h]
\begin{small}
  \centering
  \includegraphics[width=0.65\linewidth]{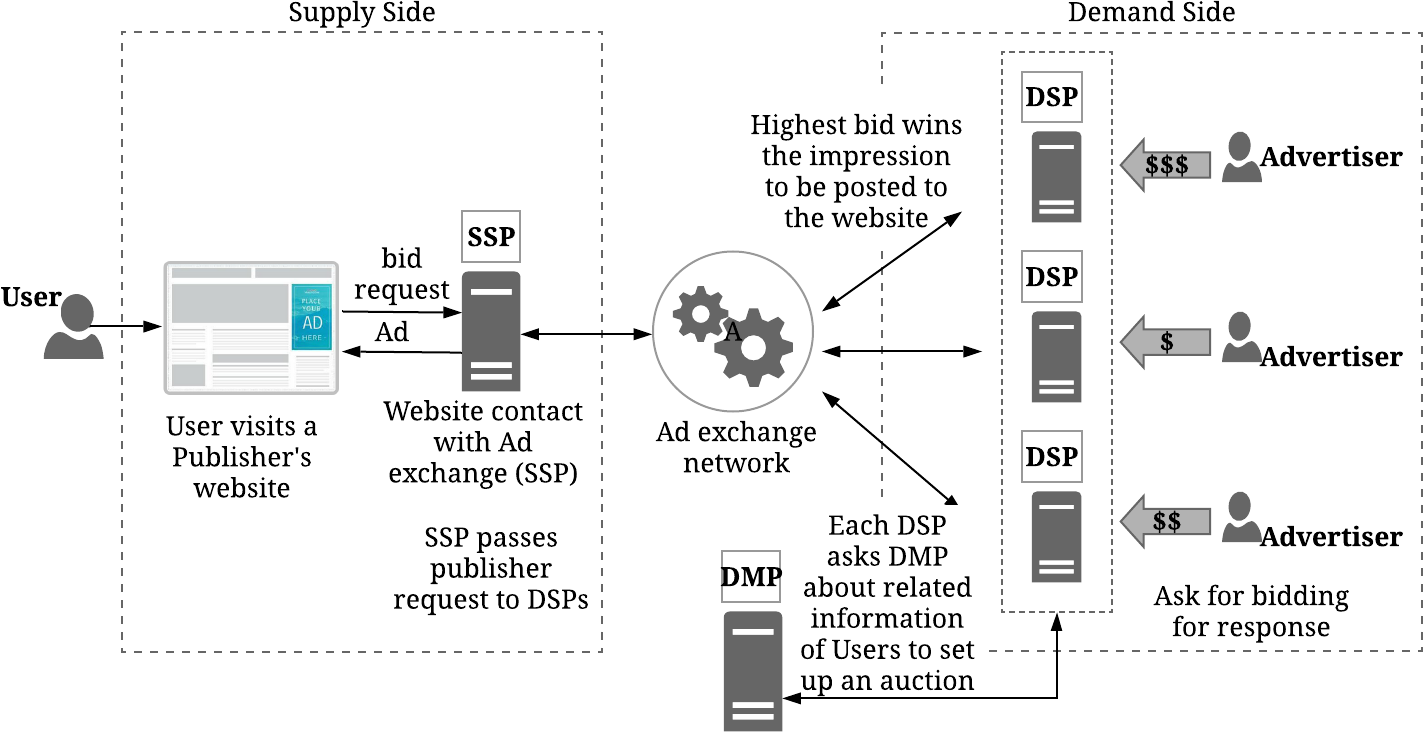}
  \caption{Advertising Eco-system. 
  %\textcolor{red}{please revise this figure, and also include content description} The conceptual view of online advertising.
  From left to right, a process is triggered when users start to interact with online services through either visiting a web-page, searching an item, or checking the social media in publisher website. In the case that the web-page has web placement available, the publisher sends ad request through SSP node to the Ad Exchange. The bid request related to requested ads are forwarded from Ad Exchange to DSP nodes which represent advertisers. After getting relevant user information, such as user profile and their previous interaction, through DMPs, the auction is set up to gather bid among DSPs. The bidder with the highest bid wins and its ad script is forwarded to the publisher to be embedded in the page requested by users.}
  \label{fig:ecosystem}
%   \Description{The 1907 Franklin Model D roadster.}
\end{small}
\end{figure*}

In Figure \ref{fig:ecosystem}, we illustrate an online advertising eco-system. The workflow starts with an event when a user, \textit{i.e.} an audience, launches an URL request from a publisher's web page. The ad request for ad placements is sent to SSP to trigger an ad auction call (\textit{i.e.} an opportunity). If the requested web-page contains available ad placement, the ad call will be submitted to Ad Exchange Network, leading to negotiations with advertisers through DSP based on bidding mechanism. %the Ad call creates an opportunity for advertisers to propose the bid prices. 
The winning advertiser will insert  %for the publisher to find an advertiser 
ad script in the user requested page, so the ad is eventually delivered to the user. In the case that the displayed advertisement matches to the user preference, user response in the form of click or further user engagements, like purchasing or subscription, is generated.

The revenue of advertisers and publishers, in the online advertising, is based on the user response such as clicks or conversions. Therefore, serving users with ads best matching to their preference is of interests to both advertisers and publishers. Under such circumstances, using context to find users' preference plays an essential role for user response prediction. %understanding achieved from %information gathered about contexts and end-users' feedback. 
The information from publisher websites is usually obtained from crawling the web-pages to summarize the context. It is then complimented by online analysis of cookie data and browsing history made by users. Such information allows system to identify user interest and response regarding ad impression. %With the goal to develop this capability,
%Advancing this capability, 
%researchers designed different data analytics and advanced learning methods in the literature.
% \raggedbottom
\subsection{User Response Types} \label{sec:user_res_types}
In web applications like web search engines, display advertising, recommendation systems, or e-commerce platforms, a user response to advertisements starts with a simple click on the ad or a touch on the screen in mobile app. This action is considered as an implicit positive user response which will direct users to a landing web-page. % or external advertiser landing web-sites. %The notion of implicit users feedback here refers to that users have expressed initial interests with regard to items but it does not mean they really like selected items. However, when
If ad content matches to user preferences, it encourages users to follow up promoted messages by generating the next clicks which can end up with desired activity such as a purchase. %In social media, user activities are expanded to show diverse positive and negative responses. For example, users are given a dismiss option to express their tendency to not seeing a commercial anymore. Similarly, in Twitter, users have an access to take further types of positive actions like sharing a post or tweet or replying back to promoted post~\cite{TwitterLi2015}. 
In online advertising the initial click or final purchase actions over displayed advertisements are considered as the critical measures to evaluate the performance of user response predictive models. Online advertising systems are generally integrated with recommendation systems in e-commerce platforms to provide users with ranked items based on explicit user's rating and implicit feedback. These feedback can be measured using different metrics to show the performance of the advertising systems. In the following subsections, we define prevalent metrics in this domain. % In following sections, we cover common user response types that are used as the metric to evaluate predicting models for online advertising and recommender systems.     

\subsubsection{Click-Through Rate} 
Click-through rate (CTR) value is of the most important metrics to evaluate the quality of ads and the performance of campaign ads. %CTR is simply defined as the proportion of clicks made by users after being serving a number of ads. %
Two elements to calculate the click-through rate values are clicks and impressions. The click-through rate is typically defined as the number of click events over impressions or the percentage of served advertisement ending up with user click events. 
\vspace{-0.1cm}
\begin{equation}
\small
\label{eq:ctr}
\begin{aligned}
CTR=\dfrac{\#~ of~ Clicks}{\#~ of~ Impressions}
\end{aligned}
\end{equation}
The number of impressions are perceived as the number of times an ad or a promoted product is served to the users' device which is engaged to an active online platform, where the publisher can be the website of search engine, a social media, or a third-party website. %The displaying of ad in the publisher website is generally treated to be seen by users, although they may not specifically examined. 
A click event is an indicator of user engagement, which can be a mouse click on ad creatives on a desktop system or touching them on mobile devices. The definition of click event is extended in different applications like the number of downloads ~\cite{AutoGroupLiu2020}, or in the social media context as positive and negative actions like reply, commenting, sharing, dismiss, etc. ~\cite{TwitterLi2015}. Common issue that frequently exists regarding this metric is data class imbalance problem where the number of clicks compared to the number of impressions is very few. Some studies ~\cite{DoZheng2010,DidKumar2018} suggest that relying on this metric to evaluate the performance of e-commerce search results can be noisy and generate misleading outcomes. %However, CTR is still a good indicator for many display advertising and sponsored search advertising systems. 
% \textcolor{red}{(1) define what is a click, what is an impression, (2) show CTR prediction = number of clicks divided by number of impressions. }\textcolor{blue}{Done!!}

\subsubsection{Conversion Rate}
%In online advertising, %click-through rate (CTR) value has been known as a major metric to measure user response to ads. Working based on cost-per-click paradigm, the value of CTRs are used to determine the revenue and payment of advertisers.
In order to evaluate user experience and activities after the click, metrics are introduced to evaluate ad campaigns following cost-per-conversion business model. The desired actions for advertisers like purchases, subscription of service, registrations, and installation of a software, are considered examples of conversion events. 

%In summary, 
Conversion rate is simply defined as the proportion of users who visited ad creative in online portal and chose to take any above mentioned actions after opening the landing website.
\vspace{-0.07cm}
\begin{equation}
\small
\label{eq:cvr}
\begin{aligned}
CVR=\dfrac{\#~of~Conversions}{\#~of~Impressions}
\end{aligned}
\end{equation}

A conversion is generally considered as a user response following a temporal order of events starting with the page visit, ad display, click, to the conversion. In the case that the sequence of conversion, click, and visit of ads are all available, the prediction of conversion rate is defined as a post-click conversion prediction. The essential goal of the problem is to estimate the probability of a conversion event given clicks and the context~\cite{EntireMa2018,ConversionWen2019}.

\subsubsection{User Engagement}
Recommender systems have been commonly used in different application platforms range from social networks and news feed services to e-commerce portal and entertainment data stream services. The common problem in these system is the overload of information that users are confronted with the high volume of items being overwhelming to browse. The priority of these systems is to attract more users by replying their requests with a relevant list of items matched with their preferences. So, the common objective is to recommend a small set of items which includes promoted ones to get immediate implicit user feedback~(e.g. CTR) while keeps users activating. User engagement objectives have been studies differently in prior research. There are some studies which model active users by following churn-rate and dwell time analysis. Recent studies have modeled user engagement using multi-objective optimizations. So two recommending and online advertising are optimized together to satisfy user experience in the long-term~\cite{Jointly2020,Maximizing2020}.

%User engagement is defined as how much user is interested in specific ad, and how likely he/she will be click on the ad.
With the advent of smartphones and the increase in their popularity among users, there is a surge of interest in developing softwares operating on this platform. As a result, a new online advertising, called in-app advertising, emerges where specific spots on screen before completing a transition in the app are designed for commercial ads. In this context, some studies proposed to provide personalized ads~\cite{PrivilegedXu2019,DidKumar2018}
which are evaluated by studying different users activity patterns to model users' engagement.
%which are evaluated by user response metrics, such as users' engagement. 
%including click-through rate and conversion rate values. 
Because smartphone platforms are personalized with respect to individual users, user response can be extended to the user engagement concept with the general questions to learn the factors which can (1) retain user being active to use an online service, like streaming providers and (2) also help gain revenue through directing people to take a desirable action with regard to ads. Therefore, several researches~\cite{CharacterizingLiu2019,MeasuringRutz2019} investigate features leading to user engagement with regard to mobile apps, and a recent work \cite{ModellingBarbaro2019} proposes to study factors which are resulting in users being disengaged from mobile apps through hierarchical clustering models.
% \paragraph{\underline{Video Completion Rate:}}

In video streaming platformss like YouTube, %Videos in online video streaming services like YouTube are playing an important role in the media platform to provide users with their needs for education, amusement, and entertaining proposes. 
video ads have become a modern effective way of conveying commercial messages via telling a story to users. In this context, video completion rate value is a metric designed to evaluate the effectiveness of video advertisements and user engagements%Online video content is the predominant traffic data in the internet. 
 %Taking advantage of entertaining factor, video ads are one of the most popular form of advertising have been seen in many platforms. 
%The common approach to display a video ad is roll-out advertising in which ads are played at the beginning, in the middle, or at the end of the video contents. The metric to evaluate the effectiveness of video ads and adjust advertising strategies is completion rate, which 
\footnote{It is defined as the percentage that videos are watched to the end. % or in other words, the proportion of video completion number over whole number of time video ads starting to play. 
The more time the video ad is watched by users, the higher the chance it may influence users to take follow-up actions.}. As it is shown in \cite{UnderstandingKrishnan2013} the content, position and length of video ad along with the length and the provider of host videos in addition to user connection information (geography and connection devices) are key factors to impact video completion rates and evaluating the effectiveness of video ads.

% \subsubsection{Recommended Point-Of-Interests}
% This is another metric used in recommendation systems which is generally formulated as a ranking prediction task. Rather than predicting one value to represent immediate positive user feedback like the probability of click(CTR) or conversion(CVR) to a desired activity like purchasing or subscription, the goal for recommender systems is providing a personalized list of point-of-interest(POI) objects. Here POI can be considered as items in different contexts which may have a specific tag, category and location and content. By applying a designed ranking mechanism, a top-K recommended list of POIs with higher chance to get further activities are returned. This can be obtained by calculating the predicted rank score to provide the ordered list. 

In the context of e-commerce system, %ranking models are designed for recommendation applications. The 
use engagement is generally evaluated by ranking metrics used in information retrieval systems. %The evaluation of ranking models in e-commerce systems can be done using different commonly-used metrics introduced in Information retrieval and recommendation systems study domains. 
The performance of ranking in the produced ordered lists is established by considering a samples of users who have positive interaction with items. Generally, there is a chance that items preferred by users are missing in the list. %The metric $HR@K$ determines the proportion of pair of user and products gathered from ground-truth which have a hit in the top-K list. 
Mean average precision at rank K ($MAP@K$), mean average recall at rank K ($MAR@K$) and Normalized Discounted Cumulative Gain at rank K, are the frequent metrics which give more details about the ranking performance. The $MAP@K$ assesses how much system can incorporate relevant items in the list. The second one $MAR@K$ checks how well model can create a list from all available items being relevant to user preferences. Due to the fact that the relevancy of items to user preferences are not the same, $NDCG@K$ consider the performance to put the more relevant items before the others in the recommended list $S$. It gives more significance to hit rates happened at higher ranks of the recommended list. According to Eq.~(\ref{eq:ndcg_at_k}), for each item in the recommended list, $rel_{s,r}=1$ represents a case that the item $s$ ranked at $r$ is matched with the ground-truth otherwise it would be $rel_{s,r}=0$. %It evaluates that model can put the sample matched with the ground-truth in the higher rank in the recommended list. 
A log factor is used to assign a penalty with regard to position of items in the list.
% \begin{equation}
% \scriptsize
% \label{eq:map_at_k}
% \begin{aligned}
% HR@K=\dfrac{\#~of~recommended~items~being~relevant}{\#~of~recommended~items}
% \end{aligned}
% \end{equation}
\begin{equation}
\scriptsize
\centering
\label{eq:map_at_k}
\begin{aligned}
MAP@K=\dfrac{\#~of~recommended~items~being~relevant}{\#~of~recommended~items} &&
MAR@K=\dfrac{\#~of~recommended~items~being~relevant}{\#~of~relevant~items}
\end{aligned}
\end{equation}
% \begin{equation}
% \scriptsize
% \label{eq:mar_at_k}
% \begin{aligned}
% MAR@K=\dfrac{\#~of~recommended~items~being~relevant}{\#~of~relevant~items}
% \end{aligned}
% \end{equation}
\begin{equation}
\scriptsize
\label{eq:ndcg_at_k}
\begin{aligned}
NDCG@K=\dfrac{\sum\nolimits_{s \in S} \sum\limits_{r=1}^K\dfrac{rel_{s,r}}{\log_2(r+1)}}{\#~of~recommended~items}
\end{aligned}
\end{equation}

\subsubsection{Explicit user feedback}
%The user response types introduced above are gathered from the data which incorporate implicit user feedback such as a click event on items, the view event or watching a video referring to user preferences in recommender and online advertising systems. % to check out item specification which have been studied to develop recommendation systems in different online platforms. 
%This data which is abundant on online services are generally used to develop model to provide relevant items for users. User preferences here are inferred in an indirect way through user activities and behaviour. 

In contrast to implicit user feedback, explicit rating score information allows users to express their interests or opinions through online methods like surveys. Compared to implicit user feedback, this information are scarce since they require users to provide additional input with regard to items via surveys and online forms. In addition, they may come with bias in user's opinions. Implicit feedback are frequently analyzed through models for classification tasks where explicit user responses are adopted %This data is generally analyzed to develop recommendation systems in form of a 
for regression tasks such as rating prediction so that user rating score with regard to new items are estimated by the system. 

Table \ref{tab:user_resopnse_types} summarizes characteristics and challenges of different user response types.

\begin{table}[H]
\scriptsize 
\setlength{\tabcolsep}{0.5em}
\centering
\caption{The characteristics of different user response types.}
\label{tab:user_resopnse_types}
\begin{tabular}{|@{\hskip3pt}p{.14\linewidth}@{\hskip1pt}|@{\hskip2pt}p{.07\linewidth}@{\hskip5pt}|@{\hskip2pt}p{.06\linewidth}|p{.035\linewidth}@{\hskip9pt}|p{.035\linewidth}@{\hskip9pt}|>{\raggedright}p{.11\linewidth}|@{\hskip2pt}p{.4\linewidth}|} 
\hline
\multirow{2}{*}{Metric} & \multirow{2}{*}{Abundance} & \multirow{2}{*}{Accuracy} & \multicolumn{2}{c|}{User feedback} & \multirow{2}{*}{\shortstack[l]{Illustration of\\ user preferences}} & \multirow{2}{*}{Description}  \\ 
\cline{4-5}
&&&Implicit&Explicit&&\\ 
\hline
Click-Through Rate             & High      & Low      & \checkmark                      &                        & Positive                         & Often not the final goal                                                   \\ 
\hline
Conversion Rate                & Low       & High     & \checkmark                      &                        & Positive                         & Needs a domain specific definition                                         \\ 
\hline
User Engagement                & High      & High     & \checkmark                      &                        & Positive                         &Assumes a direct trend between retention \& engagement\par
Takes short-term or sequential user behavior and intents \\ 
\hline
% Recommended Point-Of-Interests & High      & High     & \checkmark                      &                        & Positive                         & Takes short-term or sequential user behavior and intents                   \\ 
% \hline
User rating scores                & Low       & High     &                        & \checkmark                      & Pos. and Neg.            & Sparse data                                                                \\
\hline
\end{tabular}
\end{table}

% https://arxiv.org/pdf/1906.00355.pdf

% https://content.iospress.com/articles/data-science/ds190027

% https://www.sciencedirect.com/science/article/pii/S0167811619300023

\section{Features for User Response Prediction}
User response prediction plays an essential role for online advertising and recommender systems~\cite{EstimatingLee2012}, where the prediction is typically defined as the probability of users making a positive response on promoted item in a marketplace, ad, or news article in  online platforms~\cite{ResponseMenon2011,NeuralShioji2017,DisplayWang2016}. The performance-based advertising is the paradigm mainly followed in online advertising systems, where the predicted probability is not only used as an indicator to present user preferences, it is also involved in bidding strategies to determine the revenue of advertiser and publishers~\cite{PredictingRichardson2007}. 

\begin{figure*}[!h]
  \centering
  \includegraphics[width=.85\linewidth]{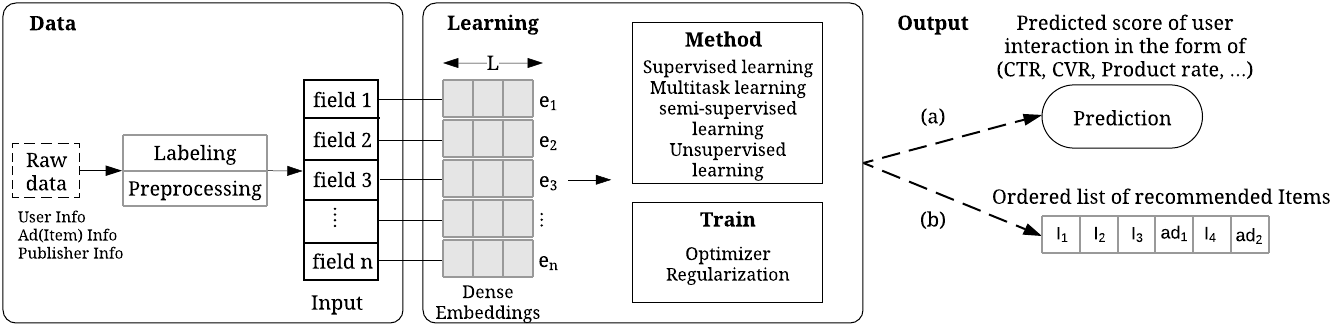}
  \caption{The schema of user response prediction workflow. Embedding layer is the common paradigm to deal with high dimensional binary representation in user response prediction. They can either be set by pre-defined values or be trained as internal parameters in end-to-end models like deep learning methods. The output can be considered as two types of user responses a) a scalar value of predicted score for an interaction between given user $u_i$ and item $I_j$ b) a ranked list of regular and promoted items ordered by predicted user response scores.}
  \label{fig:UserResPredict}
%   \Description{The 1907 Franklin Model D roadster.}
\end{figure*}

Figure \ref{fig:UserResPredict} shows the workflow of typical user response prediction models consisting of two main stages. The first stage is related to data collected from different data sources (Users, Advertisers and Publishers) in online advertising systems. After the pre-processing and labeling steps, data samples are described with series of features (fields) along with label (class) values which are normally specified as binary user response value such as 1 for click, conversion, purchasing, etc. and 0 otherwise. For recommendation systems, the output in Figure \ref{fig:UserResPredict} is an ordered list of promoted/recommended products. For the prediction task, it will output probability of users making an interaction (\textit{e.g.} a click) on items in the list. Like typical machine learning problems, the input data should be described through feature vectors to capture the class correlation, meaning that features need to be discriminative for the prediction task. Therefore, during the second (learning) phase, features are extracted using different approaches, such as (1) using data fields to represent users, pages, etc. and create sparse features; or (2) using embedding based approaches to create dense features. %Therefore the relevant feature extraction step(Embedding step) is necessary before designing any predicting models. 

%In Figure~\ref{fig:UserResPredict}, the extracted features are usually represented as dense vectors. These vectors are then fed into the proper machine learning model. 
%So  the second step includes embedding layer followed by a proper learning model. 
%In the next subsection, we start with an overview of data sources, followed by commonly used type of features in user response prediction. Technical details about learning frameworks for user response prediction, including embedding component and learning models %in Figure \ref{fig:UserResPredict}
%will be discussed in the next section (User Response Prediction Framework).
% \vspace{-0.5cm}
% \vspace{-0.5cm}

\subsection{Type of Features}
%Nowadays web applications like recommender systems and online advertising are usually working on user response information measured as click-through rate values.
%The performance of recommendation systems and online advertising is directly evaluated by user responses. 
In order to accurately predict user response, it is important to train models using discriminative features. %It can show the importance of valuable features to create a predictive model to estimate the probability of a user feedback to a given ad or commercial items. 
%In literature, different methods ranging from  logistic regression based methods and factorization machines based ones to advanced deep learning based models, are proposed to model feature interaction to design offline and online user response prediction systems. 
In the following subsection, we will discuss features studied by various methods.
\subsubsection{Multi-field categorical features} \label{sec:mulcat}
\begin{figure}[h]
  \centering
  \includegraphics[width=.45\linewidth]{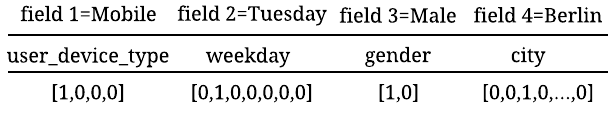}
  \caption{The characteristics of multi-field categorical features as the input to user response prediction models. The binary representation of multi-categorical features is created using one-hot-encoding}
  \label{fig:MulField}
%   \Description{The 1907 Franklin Model D roadster.}
\end{figure}

The typical input data fed into online advertising systems are generally formed as multi-field categorical values. Contrary to continuous features which are generally found when dealing with images or audios, the input data contains an array of categorical fields including Gender, City, Age, Id, $\cdots$ and device type and ad category, $\cdots$, to describe users and ads or the other related objects in the system. An event representing the user interaction with online advertising includes features from different actors like users, publishers, advertisers and the context in online advertising systems. %such as advwetiser, publisher and According to online advertising ecosystem.
%An event describes the case when an ad is shown to user in a publisher web-page and receives the possible user click responses. 
A representative list of categorical features corresponding to user profile and behavior, advertisement and publisher's web-page is provided in Table \ref{tab:cate_features}. The one-hot encoding is the conventional approach to deal with this type of data \cite{PracticalHe2014}. As shown in Figure \ref{fig:MulField}, each field is shown as a binary vector. The dimension of vector is determined by the number of unique values which are taken in the field in which one entry is set as one while the remaining as zero. In this example, fields like gender has the length of 2 and the length of weekday is 7. The simple way to represent features is the concatenation of these vectors which typically creates a high dimensional sparse binary vector. In the mathematical way, considering the input data with $n$ feature fields and $x_i$ is the hot-encoded vector of the field feature $i$ with dimension of $K_i$ where $\sum_{j=1}^{n}|x_i|=k$. In the case $k=1$, we have one-hot-encoded vectors while $k>1$ refers to multi-hot-encoding \cite{DeepGharibshah2020,DINZhou2018,StructuredNiu2018} that feature field is represented by more than one value entries. To handle the high dimensionality issue, the common approach for many classification based methods is employing the embedding step to generate condensed embedding vectors. These vectors can be concatenated like $x=[x_{1},...,x_{i},...,x_{n}]$ 
% $x=[x_{embed_1},...,x_{embed_i},...,x_{embed_n}]$
to create input layer of different user response predicting models.
%Stacking the condensed embedding vector as the combination of all feature field like $x=[x_{embed_1},...,x_{embed_i},...,x_{embed_n}]$ feed to classifier model for modeling. \cite{ProductBasedNNQu2016, FFMJuan2017}
\vspace{-.1cm}
\vspace{-.3cm}
\begin{table}[h]
\scriptsize 	
\begin{tabular}{|l|p{.75\linewidth
}|}%{\linewidth}{|l|L|}
 \hline
 Object&Features\\
% \multicolumn{2}{|l|}{\centering{features}} \\
%  \hline
%  Country Name     or Area Name& ISO ALPHA 2 Code &ISO ALPHA 3 Code&ISO numeric Code\\
 \hline
 User & Id, location(Area, city, country), IP, Network Spec,  Browser cookie, Gender, Age , Date\\
 Advertiser & Ad id, Ad group Id, Campaign Id, Ad Category, Bid, Ad Size, Creative, Creative Type, Advertiser Network\\
 Publisher& Publisher(Id), Site, Section, Ad Placement, Content Category, Publisher Network, Device type, Page Referrer\\
 Context & Serve time, Response Time\\\hline
%  Algeria    &DZ 012\\
%  American Samoa&   AS 016\\
%  Andorra& AD 020\\
%  Angola& AO 024\\

\end{tabular}
\caption{Representative categorical features corresponding to the main online advertising objects}
\label{tab:cate_features}
\end{table}
\raggedbottom

\subsubsection{Textual features} 
In search advertising, ads are displayed in the search result pages, incorporating textual data such as headline, relevant keywords and the body to highlight the details of promoted products. %This type of ads are generally served to users who searching a subject. 
Many research proposes to treat click-through rate prediction task as the similarity learning between users' query keywords and keywords of ads using their proposed text based similarity. %The textual features are extracted from two source of data in many proposed models. 
For example, keywords in the title and body of advertisements \cite{UsingBaqapuri2014,ClickEffendi2016} and keywords in user queries are considered as two sources of data to extract textual features in many designed models~\cite{DeepEdizel2017,DKNWang2018,DeeplyGligorijevic2018}. Relying only on ad textual content and user query at character- and word-level, a deep CTR prediction model~\cite{DeepEdizel2017} collects data from textual letters of query along with the title, the description, and the ad URL. They are organized to feed into system as a one-hot-encoded matrix. 

%authors in \cite{DeepEdizel2017} conducted a research to present a deep CTR prediction model. Input samples are collected from the textual letters of query along with the title, the description and the URL of Ad. They are organized to feed into system in the form of a one-hot-encoded matrix.
% \subsubsection{}
% \subsubsection{Banner Features}

% \subsection{Suppler Side Features}
% supplier is referred to the publishers, banners, sites.

% Mobile app advertising (or in-app advertising) is a type of suppler side features

% \subsection{Demand Side Features}
% Demand is refer to DSP, including advertiser.

% \subsection{Third Party Features}
% \subsubsection{Ad Exchange Features}
% Ad exchange is referring to the ad exchange, bidding volumes, price, etc. 
% \subsubsection{DMP Features}
% This refer to third party data management platform features, including user featuers, traffic featureas etc. Users are referring to ad viewers. Such as user segmentation, keynotes, cookies
\subsubsection{Visual features} 
E-commerce platforms, which are available through web portals and mobile apps, are hosts of many categorical ads and items. Each item is generally described by texts and images, acting as visual features to attract users' attention. Categorical features are generally used for model user behavior history. However, the data sparsity issue in categorical data encourages to consider intrinsic visual information in images for the development of further methods~\cite{ImageGe2018,DeepChen2016}. There is also an increasing interest to develop video ads for digital streaming platforms in which user responses generally happen by clicking on the image section~\cite{DeepShi2020}. Very recently, user facial information along with user behavior history is proposed to use for modeling user purchase interest. Analyzing this type data can provide an estimation of some user profile information such as gender, age and ethnicity which in turn draw inferences about user background and status and their manner for purchasing\cite{FaceLiu2020}.   %Deep learning based models generally employ convolutional neural networks for feature extraction and generate embedding vectors from visual features%, and feed features to the deep neural networks 
% can be used to learn response prediction models.% They incorporate information to be conveyed to users.
 %Although many studies consider to utilize categorical features to form the user behavior history for analyzing the user preferences, the data sparsity issue in categorical data encourages to consider intrinsic visual information in images for the development of further methods~\cite{ImageGe2018,DeepChen2016}. Deep learning based models generally employ convolutional neural networks for feature extraction and generate embedding vectors%, and feed features to the deep neural networks 
%can be used to learn response prediction models.% get processed in deeper layers.  

\subsection{Organization of Features}
Earlier studies to analyze user responses in online advertising mainly use one type of multi-field, visual or textual features for model designing, mainly because of their transparency and easy to interpret. Advanced models are later studied to extract complex features for better prediction accuracy. In the following section, we will go over a couple of models which take advantage of two important layout of features such as sequential and hybrid features to improve the performance of predictive models.
\subsubsection{Temporal and sequential (user behavior) features}
%In section \ref{sec:mulcat}, different methods are proposed to consider feature conjunctions between input raw features which are generally categorical and have some challenges like data sparsity and high dimensionality. These methods take multi-field categorical features for one sample of ad and and user profile to estimate the potential click-though rate value. 
%The format of input features for user response prediction is not limited to multi-field categorical features including a series of categorical discrete ids. 
Users activities are commonly recorded as data logs available in many online data provider services. Considering the sequence of user actions \textit{w.r.t.} different types of ads are valuable features for analyzing user response prediction~\cite{DIENZhou2019,ExploitingQui2020,JointXu2020}.
The majority of proposed methods %in order to predict user feedback 
are categorized into recurrent neural network based and network based models (detailed in Sections \ref{sec:rnn} and \ref{sec:network_based}). Some studies in literature showed that the history of previous visited pages, clicked ads \cite{PrivilegedXu2019}, not-clicked ads \cite{DeepOuyang2019} sorted by time in system can be leveraged to model sequential dependency between input features. Several studies have shown the importance of these features to enhance the performance of various user response prediction tasks~\cite{DINZhou2018,DSINFeng2019,DIENZhou2019,TimeGligorijevic2019}.    %Related to two main actors in online advertising user and ad(item), there are two categories of historical features to represent user interest and ad temporal/spatial dependency. 

%To capture user interest to target ad campaign or item, there are studies in which the input features are organized for deep learning modeling. 
In \cite{DINZhou2018} user behavior features are represented as a list of visiting events of ads, each of which are described by categorical features about goods, shop, and page categories of past user-defined time points. Each time point is described using multi-hot-encoding.

Sessions are also used to represent user's behavior history~\cite{DSINFeng2019}. Separated by occurring time, user activities are considered homogeneous in very short time slots within sessions but different with regard to other sessions. User interaction patterns over time are evolving in short-term and long-term trends. A common scenario is to deal with virtually short-term sequences of user behaviors when user profile (\textit{e.g.} ID) of active users is not accessible via log-in to system. So, the task of predicting user responses is defined to extract relevant patterns based on limited actions of anonymous users. In this case, the complete user interaction histories are also organized into sessions~\cite{ExploitingQui2020,PreferenceZhang2020}. The long-term user interaction can also be studied to create user profile behavior. It can not only provide indication of user intent change over time which can be used to improve the predication user responses, the popularity pattern regarding products can also be identified to remind users about a product according their previous interactions. 
%Having access to the trails of major user activities before and after ad click event which is gathered from different data sources, a study \cite{TimeGligorijevic2019} consider a sequence of activity events and corresponding passed time-slot to investigate the potential user conversion intents. 
%There is a study in which
In ~\cite{TimeGligorijevic2019} authors consider a sequence of user activity events before and after the ad click and corresponding passed time-slot to investigate the potential user conversion intents. %~\cite{TimeGligorijevic2019}.
They analyzed the effect of elapsed time as a feature for conversion rate prediction and using targeting and retargeting\footnote{A cookie based advertising which tracks users clicking or visiting ad in a website who have not taken further actions against promoted products. Using this paradigm advertising systems remind users their previous interest about a promoted product.} paradigm for different users in online advertising systems.% They included the sequence of elapsed time between subsequent events as the training features to justify the difference of using targeting and retargeting\footnote{This is a cookie based advertising which tracks prospective users clicking or visiting a advertiser website who have not taken further actions against promoted products. As statistics suggest, there is a delay between user visit and their conversion to purchasing. So advertising systems remind users their previous interest about a promoted product.} advertising paradigms for various users. Re-targeting approach generally lends itself for users with shorter elapsed time slots in the recent past and close to conversion event whereas targeting deals with each user as a new potential customer regardless of their recent past activities. 

\subsubsection{Hybrid Features}
Combining different types of features are also studied to enhance user response prediction. % methods as hybrid features. %As a result, hybrid approaches emerge to consider different feature extractor blocks in the model. 
Some studies use textual features along with multi-field categorical features to improve the performance of recommendation systems in e-commerce platforms \cite{StructuredNiu2018} and the sponsored search marketing \cite{DeepCovington2016}. Some research consider the compounds of categorical features and image data~\cite{ImageGe2018,DeepChen2016} and the combination of categorical features with video data \cite{DeepShi2020} to improve predictions. The combination of different features in modeling lead to various compound embedding layer for input data %. It entails using an aggregation module in the design 
to generate a condensed feature representation, with pooling being employed to reduce parameters and cope with over-fitting. Max and sum pooling are also studied as the aggregation mechanism in some studies~\cite{DeepCovington2016,ImageGe2018}. The concatenation of feature embedding vectors is a straightforward approach commonly used in many studies~\cite{ProductBasedNNQu2016,DeepCrossingShan2016,FNNZhang2016,FeatureGenLiu2019}. Recently, an adaptive approach to combine most relevant features from different feature types is employed based on attentive mechanism~\cite{ImageGe2018,DINZhou2018}.

\section{User Response Prediction Frameworks}
%Using different features as input, user response prediction intends to accurately estimate/predict user response \textit{w.r.t} ads. 
For years, user response prediction, in online advertising, has been continuously evolving. Early approaches usually reply on hand crafted features to dissect data into different segments, where each segment contains users with similar response. Therefore, the click-through rate values or conversion rate values estimated on each segment can be used to estimate future (new) users' CTR values. Following similar approaches, clustering or collaborative filter based approaches are also proposed to recommend ads to users. In the context of recommendation systems, the ordered list of items including promoted products are proposed to users by predicting how likely the list contains items matching to user preferences. The evaluation of these systems are examined with different ranking and regression metrics~(Detailed in Section \ref{sec:user_res_types}). Typical types of recommender systems have analyzed past user interactions to detect a connection between users and products either through studying users with same tastes or similar items visited by different users. Recently, machine learning, especially deep learning, based approaches, are becoming increasingly popular for user response prediction, mainly because these approaches can simultaneously accommodate a large number of features, and learn to create new features, for accurate user response prediction. 

In Table \ref{tab:taxonomy}, we propose a taxonomy of user response prediction, which includes hierarchical based methods, collaborative filtering based approaches, supervised, semi-supervised and unsupervised learning studies. In the case that labeled data are available, supervised learning algorithms leverage the label in the definition of the loss function in their learning procedure. Unsupervised learning rely on unlabeled data for the loss function optimization. Semi-supervised models are between supervised and unsupervised models where their objective functions are optimized considering both data with and without labels. %With regard to user response prediction models for recommendation systems and online advertising, 
The supervised methods can be further categorized into basic predictive models and ensembles ones whereas semi-supervised and unsupervised category consists of network-based and clustering based methods. The last category in the taxonomy includes stream-based methods. Following subsections will discuss and review representative methods in each category. 
\begin{table*}[t]
\centering
\begin{scriptsize}
% \scriptsize
% \caption {Input field feature type}
% \cline{2-3}
% \begin{tabular}{|l|l|l|l|}
% \hline

\begin{tabular}{|>{\raggedright}p{0.15\textwidth}>{\raggedright}p{0.13\textwidth}p{0.24\textwidth}|p{0.38\textwidth}|}
% \toprule[1pt]
\hline
% \toprule[1pt]
\multicolumn{3}{|l|}{\textbf{Category}}              & 
\textbf{Publication}               \\ 
\hline
\multicolumn{3}{|l|}{Data Hierarchy Based} & \cite{EstimatingAgarwal2010,EstimatingAgarwal2007,PredictingOentaryo2014,EstimatingLee2012,TemporalKota2011,OnlineChao2019}\\
\hline
\multirow{3}{*}{\parbox{1.8cm}{Collaborative-Filtering Based}} &\multicolumn{2}{l|}{Hierarchy Based} & \cite{ResponseMenon2011,CollabLiu2014}\\
\cline{2-4}
&\multicolumn{2}{l|}{Network Based} & \cite{FactoricationKoren2008,RippleNetWang2018,MKRWang2019,NGCFWang2019,LightGCNHe2020}\\
\cline{2-4}
&\multicolumn{2}{l|}{Hybrid Methods} & \cite{NeuralCFHe2017,NTFWu2019,AttentiveCFHe2017}\\
\hline
\multirow{9}{*}{Supervised Learning} &{\multirow{6}{*}{\parbox{1.8cm}{Predictive Models}}} & Logistic Regression & \cite{EstimatingLee2012,EffectsKumar2016,EvaluatingDalessandro2012,PracticalHe2014,SimpleChapelle2015,CostVasile2017,UserRen2016,ZhangBidAware2016,EnsembleAryafar2017,ZhuSEM2017}\\
\cline{3-4}
& & Factorization Machines& \cite{FMRendle2010,PredictingOentaryo2014,SparseFMPan2016,FFMJuan2017,FwFMPan2018,RobustFMPunjabi2018,PBODLLiu2017,FIBINETHuang2019,Field-AwareZhang2019,FLENChen2019,PredictingOentaryo2014,FNNZhang2016,ProductMultiQu2018}\\
\cline{3-4}
& & Deep Learning Methods & \cite{DeepCovington2016,DeepCrossingShan2016,DINZhou2018,DeepOuyang2019,AutoIntSong2019,PrivilegedXu2019,ImageGe2018,LearningYang2019,ResembeddingZhou2019,DIENZhou2019,DSINFeng2019,PracticePi2019,OnlineChao2019,PredictionBigon2019,FeatureGenLiu2019,ConvolutionalLiu2015,DeepEdizel2017,DeepChen2016,DeeplyGligorijevic2018,StructuredNiu2018,OnlineChao2019,NeuralCFHe2017}\\
\cline{3-4}
& & Hybrid Methods & \cite{DeepFMGuo2017,ProductBasedNNQu2016,WideAndDeepCheng2016}\\
\cline{2-4}
&{\multirow{4}{*}{Ensemble Methods}} & Cascading & \cite{FNNZhang2016}\\
\cline{3-4}
& & Stacking& \cite{EnsembleAryafar2017}\\
\cline{3-4}
& & Boosting& \cite{IterativeLevine2020}\\
\cline{3-4}
& & Mixed & \cite{MultiWen2019}\\
\hline
\multirow{7}{*}{\parbox{2cm}{Un-supervised \&\\ Semi-supervised Learning}} &{\multirow{6}{*}{\parbox{1.8cm}{Network Based}}} & Network Embedding: \textit{(Node Embedding, GNN Based methods, User Intention Network Modeling)} & \cite{BillionWang2018,FiGNNLi2019,GINLi2019,ExploitingQui2020,JointXu2020,GraphYing2018}\\
\cline{3-4}
& & Knowledge Network Based: \textit{(Node Embedding, Meta-path Based methods, GNN Based methods)}& \cite{DKNWang2018,MetapathFan2019,Knowledge-AwareWang2019,End-to-EndQu2019,NGCFWang2019,LightGCNHe2020,HetGANWang2019,HetGNNZhang2019}\\
% \cline{3-4}
% & & Hybrid Methods & \cite{DeepFMGuo2017,ProductBasedNNQu2016,WideAndDeepCheng2016}\\
% \cline{3-4}
% & & Deep Learning Methods & \cite{DeepCovington2016,DeepCrossingShan2016,DINZhou2018,DeepOuyang2019,AutoIntSong2019,PrivilegedXu2019,ImageGe2018,LearningYang2019,ResembeddingZhou2019,DIENZhou2019,DSINFeng2019,PracticePi2019,OnlineChao2019,PredictionBigon2019,FeatureGenLiu2019,ConvolutionalLiu2015,DeepEdizel2017,DeepChen2016,DeeplyGligorijevic2018,StructuredNiu2018,OnlineChao2019}\\
\cline{2-4}
&Clustering Based &  & \cite{ImprovingHillard2010,PredictingRegelson2006}\\
\hline
\multicolumn{3}{|l|}{Stream-Based Data} & \cite{Real-timeDiaz2012,TwitterLi2015,AddressingKtena2019,PredictingZhou2016,PromotingLalmas2015,ImprovingBarbieri2016,GAGQui2020}\\

\hline
\end{tabular}
\end{scriptsize}
\caption{A taxonomy of user response prediction in online advertising, along with representative publications}
\label{tab:taxonomy}
\end{table*}
% \vspace{-0.5cm}
% \begin{table*}[t]
% \begin{small}
% % \scriptsize
% % \caption {Input field feature type}
% % \cline{2-3}
% % \begin{tabular}{|l|l|l|l|}
% % \hline

% \begin{tabular}{lll|l}
% \toprule[1pt]
% % \hline
% % \toprule[1pt]
% \multicolumn{3}{l|}{Context Category}              & 
% Publication                \\ 
% \hline
% \multicolumn{3}{l|}{Display Advertising} & []\\
% \hline
% \multicolumn{3}{l|}{Mobile(in-app) Advertising} & []\\
% \hline
% \multicolumn{3}{l|}{Sponsored Search Advertising} & []\\
% \hline
% \bottomrule[1pt]

% % \hline
% \end{tabular}
% \end{small}
% \caption{Taxonomy of user response prediction in different contexts of advertising along with the representative publications}
% \end{table*}

\subsection{Data Hierarchy Based Approaches}
%In this subsection, we will provide an overview of hierarchical based algorithms for user response prediction. 
Using unstructured input features, data sparsity and cold start are common issues in online advertising and recommender systems. Data hierarchy based methods refer to approaches that organize data %interaction among input data are followed after organizing them 
in a hierarchical format~\cite{liu:2017:ULTR}. The motivation is to build a tree structured hierarchy, using some selected features, such that each leaf nodes represents a user groups sharing similar response. This hierarchy provides valuable information to show correlation between user responses at different level of granularity, which alleviates the adverse effect of limited historical information about users.

\subsubsection{Data Hierarchy}
\begin{figure}[h]

% \centering
\subfigure[Advertiser hierarchy]{%
  \includegraphics[width=4.7cm]{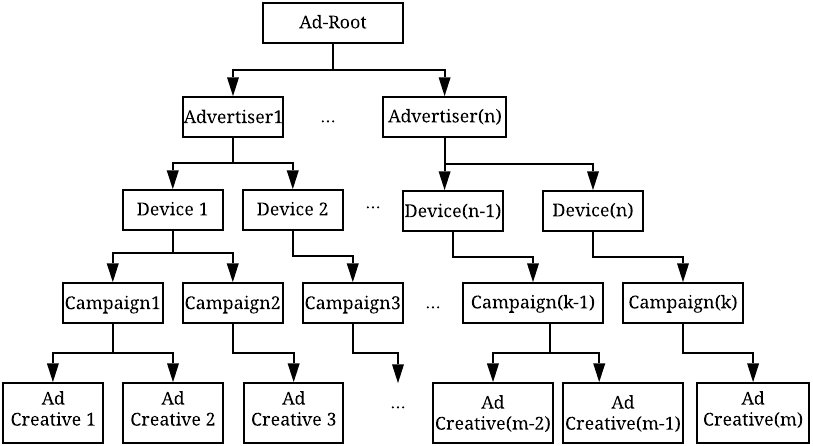}%
%   \label{fig:evaluation:revenue}%
}\qquad
\subfigure[Publisher hierarchy]{%
  \includegraphics[width=3.8cm]{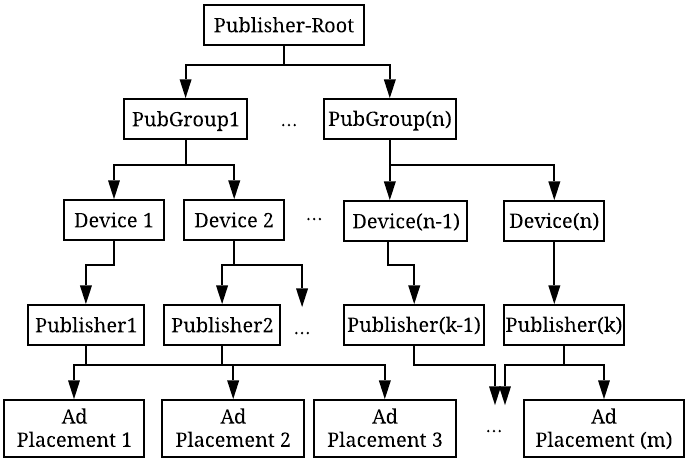}%
%   \label{fig:evaluation:avgPrice}%
}
% \qquad
% \subfigure[Unfolded single layer LSTM network]{%
%   \includegraphics[width=3.6cm]{figs/lstm_layer_new3.png}%
% %   \label{fig:evaluation:avgPrice}%
% }
\caption{%Hierarchical structure of Advertiser and Publisher data: %\textcolor{red}{for figure (a) the hierarchy: Ad-root, Publisher, Device, Campaign, Creative. For figure (b) the hiearchy: Publisher-root, advertiser, device, campaign, placement.} 
The sample taxonomy representing a hierarchical structure for Advertiser and Publisher data where a) demonstrates grouping ad creatives through multi-level joint points when they are of same campaigns and are designed for the same devices by an individual advertiser b) indicates the publisher hierarchy from ad placement in web-pages, the running devices and grouping of publishers}
\label{fig:3_hierarchies}
\end{figure}

As the first attempt to cope with data sparsity and limited historical data in online advertising, hierarchical structures of publisher web pages, ads and end-users are commonly used to address correlation between input features~\cite{EstimatingAgarwal2010,EstimatingAgarwal2007,PredictingOentaryo2014,EstimatingLee2012,liu:2017:ULTR}. In this case, users, web-pages and ads are grouped based on different factors, such as demographic or geographic information about users, domain and content of web pages and the context and campaign of ads. An example of the data hierarchy is shown in Figure \ref{fig:3_hierarchies}. From advertiser perspective, the hierarchy can be created by classifying ads based on  campaign, content type, and advertisers. For publishers, web-pages can be grouped using simply URL path or the content category. Users can also be organized as hierarchical data using third party information like user geographic, ad and web-page visit history \textit{etc.} Studies show that data hierarchy for ads, pages, and users provides useful knowledge to handle data rarity in click-through prediction~\cite{EstimatingLee2012}. Partitioning input space using tree structure represents similarity between connected nodes with respect to user responses in local areas~\cite{PredictingOentaryo2014,EstimatingAgarwal2007}. In industry, these data hierarchies are created and maintained by domain experts. %is that information associated to nodes are close to each other like
% \vspace{-.25cm}
\subsubsection{Representative Hierarchy Based CTR Prediction Frameworks}
Input features of online advertising systems consist of various sparse categorical features, which contribute to generate rare user responses such as clicks or conversions. %These two challenges make it difficult to create models to get trained properly to predict user responses. 
To address these issues, many methods propose to create a hierarchical structure from input features to estimate user response from previous similar available samples~\cite{EstimatingLee2012,EstimatingAgarwal2007}. %They are attributed by features describing similar targeted users, advertisements and web-pages. 

\paragraph{Problem Definition}%Using hierarchical information in input data, 
For ads being served to users multiple times, the baseline problem to predict user response is defined as: given a pair of web-page $j$ and ad $k$, the probability of response, like a mouse click, is calculated through the probability formula
$P_{jk}=Pr(Click|Impression;j,k)$. This probability, \textit{a.k.a} Click-through rate value (CTR), can be computed via binomial maximum likelihood estimation (MLE) where $V_{jk}$ indicates the number of times ad $k$ is displayed on the web page $j$, and $C_{jk}$ is the indicator of click number, respectively~\cite{ClickWang2011,PredictingRichardson2007}.
For the case $V_{jk}=0$ or equals to a small value, the value of MLE estimate for CTR values becomes unreliable. Therefore, in literature, different methods are proposed to exploit hierarchical information for smoothing out the MLE predictions. 

For instance, authors in \cite{EstimatingAgarwal2007} proposed to utilize two hierarchical structures between input features related to web-pages and ads to improve the prediction of user click responses in online advertising. %They considered a probability following binomial distribution to model click response given a web-page $p_j$ and advertiser $a_k$ as $p(Click=1| p_j, a_k)$. 
In this study, they tackled a form of sparsity issue in the input data with a few number of available clicks and impressions. %This issue was originated from two reasons. The first went back to regular updates and modifications in input data that change the content of web-pages and advertisements. Besides, a technical issue and computational costs restricted to use crawling methods to extract consistent features from web-pages. Therefore, in the pre-processing step, they gathered data from those web-pages which have user visits and were able to be crawled in a feasible threshold time. Then they proposed a method to project the results over the entire dataset.   %in addition to nature of online advertising to have scarce clicks and the implementation issue of page crawling methods to have enough impressions in the hierarchy. 
In order to reduce the variance made by the sparse clicks and/or impressions, a sampling approach is used to alleviate the rarity issue via negative sampling of majority class, \textit{i.e.} web-pages without a click response. To control the effect of the bias made by sampling, a two-step method is used to predict the click-through rate. In the first step, a maximum entropy model is optimized based on an iterative proportional fitting method to estimate the actual number of impressions at all defined levels in the hierarchical structure. A tree-shaped Markov model is then used to predict the click-through rate value in the whole levels of the hierarchy using correlations between sibling nodes.  %Considering the correlation captured in different levels of the hierarchy between different nodes to impute extra impression based on maximum entropy approach, 

Further, a log-linear model (LMMH)~\cite{EstimatingAgarwal2010} is introduced to improve user response prediction by exploiting correlations between sibling nodes at different data hierarchy levels. To increase the scalability of model to large dataset, a spike and slab variable selection method is proposed to control number of parameters in the regression model. 
This method deals with rare response rates by pooling data along a directed acyclic graph (DAG) obtained through a cross-product of multiple hierarchies.
%%%%%%%%%%%%%%%%%%%%%%%%%%%%%%%%%%%%%%%%%%%%%%%%%
%They relaxed the dependency between user $u_i$ information and (advertiser $a_k$, publisher $p_j$) pairs to calculate the click-through rate values in the following:
% \vspace{-.05cm}
% \begin{equation}
% \small
% % \label{obj_function}
% \begin{aligned}
% p(Click=1|u_i, p_j, a_k) \propto p(Click=1|u_i) p(Y=1| p_j, a_k)
% \end{aligned}
% \end{equation}

% They proposed to estimate $p(Click=1|u_i)$ by regression methods on user features including demographic, geographic, and temporal (frequency) information. The probability $p(Click=1| p_j, a_k)$ was predicted by using a probabilistic inference on pre-defined strict hierarchies of publisher and advertiser data based on the assumption that the estimated rates for cells in hierarchical are correlated with their close siblings.
%%%%%%%%%%%%%%%%%%%%%%%%%%%%%%%%%%%%%%%%%%%%%%%
Another study \cite{TemporalKota2011} advances the LMMH method using higher order feature interactions by fitting local LMMH models to relatively homogeneous subsets of the data. Given a relatively homogeneous partitioning of the feature space, several local LMMH models are fitted to data subsets on different nodes of a decision tree. %To ensure scalable computation, these subsets are induced through a decision tree. 
To address over-fitting issue in the model, models are coupled with a temporal smoothing procedure designed based on a fast Kalman filter style algorithm.% which can prevent system from over-fitting.

% Moreover, a study \cite{EstimatingLee2012} investigated conversion rate estimation which relies on utilizing past performance observations  along user, publisher and advertiser data hierarchies. The conversion event is modelled at different select levels of the cross-product of user, publisher and advertiser data hierarchies with separate binomial distributions and estimate the distribution parameters individually. Rather
% than using eq \ref{eq:argmaxbernoli} to try to pick the best one among a couple of estimators, These individual estimators are combined using logistic regression to accurately identify conversion events. 

Last but not least, a study \cite{EstimatingLee2012} investigates the data hierarchy for three objects of users, advertisers and publishers to deal with the data sparsity % along the dimensions
%and the mass data 
and class imbalance problem for conversion prediction. Taking conversion event as Bernoulli random variable with two possible values of conversion and no conversion, a binomial distribution is used to model the conversion given a triple of user $u_i$ and ad ${a_k}$ and web-page $p_j$. To address the data sparsity, they propose to capture the correlation in the conversion output using clustering of similar users with regard to conversion rate values, grouping advertisements from the same campaigns and web-pages with same category types. %the following criteria. User groups are identified from the similarity between conversion rate. The web-pages which belong to the same type publishers are clustered together. For advertisements, those are generated by the same campaign are grouped in the same cluster. 
The conversion estimation is calculated at different levels of the hierarchy made from the cross product of levels in three hierarchical structures of users, publishers and advertisers via the maximum likelihood estimation as follows.
% \vspace{-.25cm}
\begin{equation}
\small
\label{CVR_MLE}
\begin{aligned}
% P(Y=1|u\in C_{u_i}, p\in C_{p_j}, a\in C_{a_k})=\\ \begin{cases}
%       \dfrac{C_{ijk}}{I_{ijk}} & \text{if I_{ijk}>0}\\
%       unkown & \text{otherwise}\\
%     \end{cases} 
 P(Y=1|u\in C_{u_i}, p\in C_{p_j}, a\in C_{a_k}) =
    \begin{cases}
      \dfrac{C_{ijk}}{I_{ijk}} & \text{if $I_{ijk}>0$}\\
      unknown & \text{otherwise}\\
    %   0 & \text{otherwise}
    \end{cases}  
\end{aligned}
\end{equation}
where $C_{u_i}$ is the cluster that $u_i$ belongs to. $C_{p_j}$ and $C_{a_k}$ indicate the cluster of web-page $p_j$ and ad $a_k$, respectively. %In this method, conversion rate value was predicted through the maximum likelihood estimation on hierarchical structures using the combination of users, ad and web-pages clusters. %Picking user-defined number of hierarchy levels made from the combination of clusters of users, web-pages and ads, calculate  %maximum likelihood estimation value of CVR is calculated from combining clusters of users, wev-pages and ads. 
The final estimation of the conversion rate value is then modelled using logistic regression from the linear combination of MLE estimators at different hierarchical levels.

\subsection{Collaborative Filtering Based Approaches}
Collaborative filtering is an effective approach to predict online user interests. %using assumption that users presenting a similar preference in the past are likely going to have a similar interest against items in future. 
The general idea is to analyze previous behaviors of users to predict possible future user interests or to generate suggestions that may match to the preferences of the new similar users. %The similarity between users are determined through finding the neighbors of new users who have similar tastes and are close together in a latent space.
%Collaborative Filtering is a technique for personalized recommendation that learns user or/and item similarity based on their interactions and preference database: the fundamental intuition is that “similar” users will have “similar” preference, where “similar” refers to a latent notion that is learned by the model. The main goal of recommender systems is to suggest to users new items that they will probably like, but have not already consumed.

\paragraph{{Problem Definition}}
For collaborative filtering methods, the input is an incomplete sparse matrix $X\in R^{m\times n}$ of user-item preferences, which suffers from the data sparsity, \textit{i.e.} some $X_{ij}$ entries are missing. The goal is to fill in missing entries with predicted scores. The state of the art in collaborative filtering is matrix factorization~\cite{CollaborativeHu2008,ResponseMenon2011} which is based on an idea that the matrix of users' preferences \textit{w.r.t} items $X$ can be factorized into two low-rank matrices of Users $\alpha$ and Items $\beta$. It is modelled as $X\simeq\alpha^T\beta$, where $\alpha \in R^{k\times m}$ and $\beta \in R^{k\times n}$ and $k$ is the dimension of latent features. Conceptually, each $\alpha_i$ represents a user, and each $\beta_j$ represents an item. The simplest factorization model is to solve the following optimization where latent feature vector of users and items are controlled with user-defined regularization function $\sigma$ in different studies to prevent the model from the over-fitting issue~\cite{ResponseMenon2011,FactoricationKoren2008}.

\begin{equation}
\small
\label{CF_equation}
\begin{aligned}
\underset{\alpha,\beta}{\min}\dfrac{1}{|O|}\sum_{(i,j)\in{O}}(X_{ij}-\alpha_i^T\beta_j)^2+\sigma(\alpha,\beta)
\end{aligned}
\end{equation}

In general, collaborative filtering is based on the past interactions between users and products. It can be seen as the implicit feedback like click or conversion event on products or explicit feedback like product ratings. %It is applied for user response prediction task where 
Interactions between web-pages and ad banners can be shown as a matrix of web-page-by-ad feedback score (click-through or conversion rate rate or product rate values). %According to this approach, 
The correlation between web-pages and ads is captured to calculate predicted scores for missing entries which can be intuitively related to the user response prediction task. %Considering two matrices of click and impression, there are many either missing or invalid small entries. So using naive classification like MLE estimation leads to statistically high variance.\cite{EstimatingAgarwal2007}
The major studies in this category take advantage of collaborative filtering methods along with side information such as the user and item neighborhood models~\cite{FactoricationKoren2008}, the data hierarchies \cite{ResponseMenon2011,CollabLiu2014} and knowledge graph data \cite{RippleNetWang2018,MKRWang2019}. Hybrid models are used to tackle scenarios with data sparsity and cold start problems to improve the prediction performance. It is conventional that initial models resorted to apply matrix factorization and inner-product operator on latent factor vectors to establish connection between users and items. Recently, neural architecture \cite{NeuralCFHe2017,NTFWu2019} and attention mechanism \cite{AttentiveCFHe2017} are proposed as the alternative to learn higher order interactions on data. The user responses analyzed to evaluate the performance of models in the studied papers range from explicit product rate scores~\cite{FactoricationKoren2008,NTFWu2019} to implicit user feedback like CVR scores in \cite{RippleNetWang2018,MKRWang2019,ResponseMenon2011} and recommended ordered list \cite{NeuralCFHe2017,AttentiveCFHe2017}.
%the problem has a couple of challenges such as data sparsity and output rarity that majority of ads have limited historical visits and are not frequently clicked. 

% A study \cite{PredictingFruergaard2014} proposed an approach to response prediction based on matrix factorization techniques that collaborative filtering uses both latent features as well as side-information in the form of page and ad features. This extended factorization model is capable of incorporating explicit page and ad features such as the content of the ad, its
% placement on the page, known in the CF literature as side-information. They analyzed how the factorization model can be made to exploit hierarchical information about pages and ads. This extension allows the model to mitigate the challenge of data sparsity.
One of initial works in the collaborative filtering domains, developed based on latent factor models like singular value decomposition, is known as SVD++\cite{FactoricationKoren2008}. For a personalized recommendation system task, the authors improve the accuracy of system by addressing both explicit and implicit user feedback in a hybrid model. To do so, additional terms are added to optimize the loss function (\ref{CF_equation}) which is organized in three levels. In the first level, bias terms, in form of the addition of average of rating value of all items, the bias in average rating made by user $\alpha_i$, and the corresponding bias for item $\beta_i$ to control the discrepancy between actual values and predicted values in the loss function, are added. In the second level, a loss function is defined by adding a term to include implicit available feedback. They refer to all items that user had an interaction before. The implicit feedback here are considered from a series of browsing, purchasing and search user history in e-commerce systems. In the last level, a neighborhood model which addresses the effect of bias in average rating value made by neighbor users and items is added. Combining these terms together, the parameters of proposed model are updated using gradient descent optimization. It led to an improvement calculated for product rating prediction and providing top-$k$ personalized recommendation tasks.

In an extended matrix factorization design~\cite{ResponseMenon2011}, hierarchical information of web-page and ads are integrated as additional side information into latent features of collaborative filtering to tackle data sparsity and cold start problem. Explicit features from ads and web-pages in side information are linearly augmented to implicit features using a log-linear latent features model and user-interest propagation framework (FIP) to enrich input features. As a hybrid method, it includes hierarchical structural information into their factorization model using three learning ideas such as hierarchical regularization, agglomerate fitting and residual fitting.

In % the web-applications like
online advertising, interactions are generally made between multiple entities including users, items, and ads. Tensor factorization, as an extended version of matrix factorization, can use the similarity between different types to predict potential interaction between pair of instances. To address the compound similarity of entities with regard to a possible interaction, a Hierarchical Interaction Representation model~\cite{CollabLiu2014} is proposed to provide a joint representation to model mutual actions between different entities. Three dimensional tensor multiplication is used for modeling characteristics of pair of entities. %, and an iterative approach is developed to multiply tensors to form hierarchical structure of entities. % In this method, a latent vector is assigned to each entity. %Following Tensor Factorization to predict the interaction, a three dimensional multiplication of tensors is used to generate the interaction representation of pair of entities, in such a way that a hierarchical structure of entities is constructed in an iterative manner.
%The proposed latent representation provides % for multi-entity interaction described a way entities would act to capture the high-order feature interaction. 
%The 
%extracted features which are later fed into a multi-layer fully connected neural network to predict user responses.%to reveal the underlying properties of this interaction and enhance the model performance.

%Authors in \cite{RippleNetWang2018} suggested to incorporate side information like knowledge graph(KG) into collaborative filtering to address data sparsity issue and cold-start problem. Inspired by the success of applying KG in a wide variety of tasks, researchers tried to utilize KG to improve the performance of recommender systems in term of click-through rate values in such a way that it takes a user-item pair as input and outputs the probability of the user engaging (e.g., clicking, browsing) the item. The proposed method makes an extension to previous knowledge graph embedding methods by introducing preference propagation, which automatically propagates users’ potential preferences and explores their hierarchical interests in the KG.  For each user, RippleNet treats his historical interests as a seed set in the KG, then extends the user’s interests iteratively along KG links to discover his hierarchical potential interests with respect to a candidate item. The proposed method RippleNet unifies the preference propagation with regularization of KGE in a Bayesian framework for click-through rate prediction.
Recently, some studies \cite{RippleNetWang2018,MKRWang2019} propose to organize user and ad as heterogeneous information graph to improve collaborative filtering. Authors in \cite{RippleNetWang2018} suggests an end-to-end learning method to incorporate side information from knowledge graph (KG) into an item-based collaborative filtering approach for click-through rate prediction. %to address two commonly issues of the sparsity of input data and the problem of cold-start which arise in recommender systems. The model was created to evaluate user engagement through click-through prediction given a user visited an item on a web portal. 
They propose an extended knowledge graph embedding method which starts building an initial user preference sets in the knowledge graph that are originally set up from previous user click activities. An iterative propagation of user preferences along edges over the knowledge graph is used to create $k$ Ripple sets to %These sets include nodes in the knowledge graph which are in k-hop distance away from the initial set and
model potential user liking versus items.
%Following an item-based collaborative filtering, given $k$-hop Ripple set, they first calculated the embedding vector of all elements in a set of head-relation-tail triples items. Then, the relevance probability of item was calculated using a softmax function parameterized by embedding vectors of an item $v$ and head and relation entries. %The $k$-order response of each user using the click history are captured in Ripple set. 
%They were computed using a weighted summation of embedding vector of tails. 
Learning an embedding vector for each ripple set, the embedding vector of user response versus items is calculated from the sum of corresponding embedding vector of ripple sets. The click-through rate score of user $u$ versus item $v$ is modeled using dot product embedding vectors of $u$ and $v$ each of which are trained based on a Bayesian framework and gradient descent learning.

\subsection{Supervised Learning Based Approaches}
\begin{figure}[!htb]
% \captionsetup[subfigure]{aboveskip=-2pt,belowskip=-2pt}
% \centering
\subfigure[LR]{%
  \includegraphics[width=5cm]{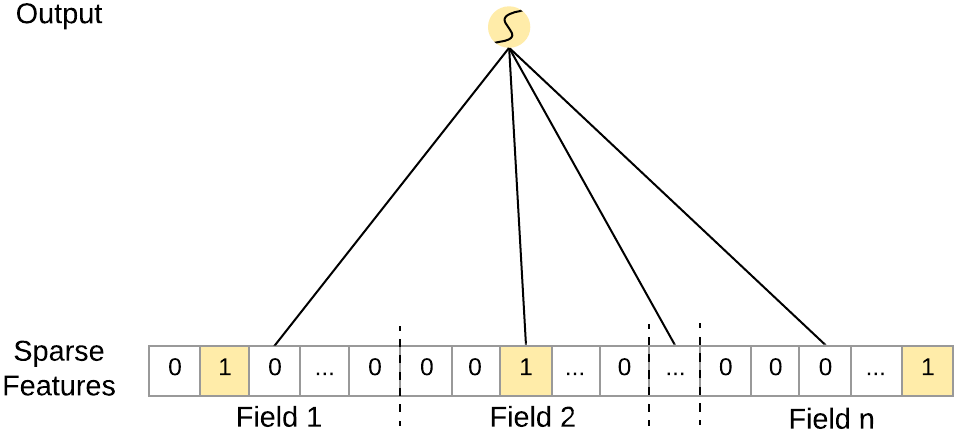}%
  \label{fig:LR}%
}
\qquad
\subfigure[Wide\&Deep]{%
  \includegraphics[width=5.5cm]{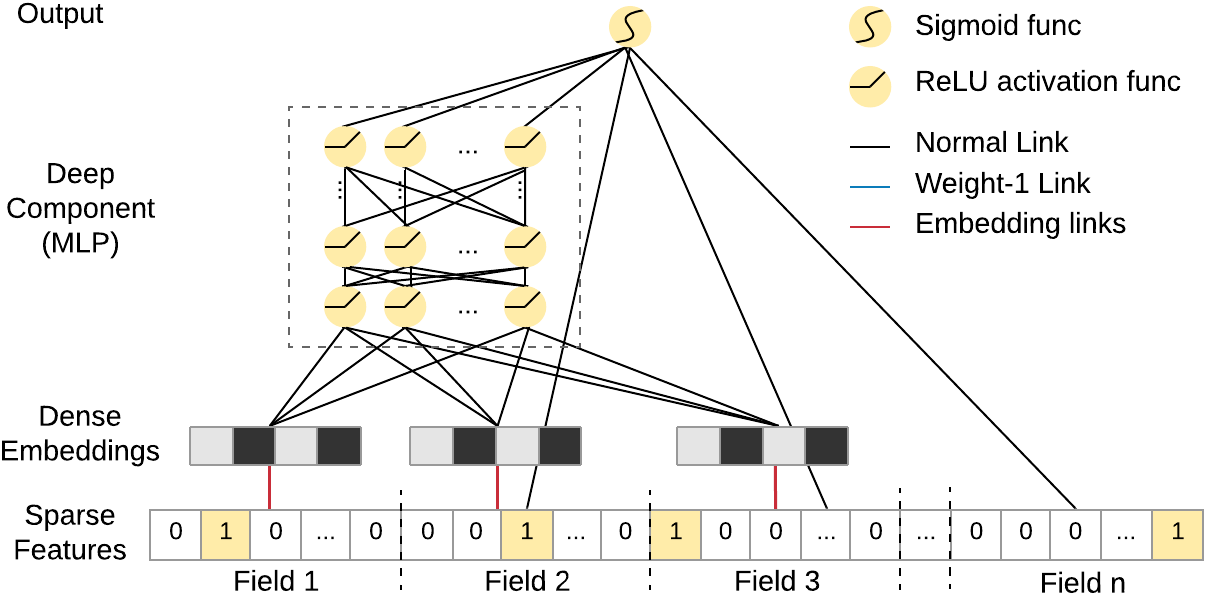}%
  \label{fig:Wide_Deep}%

\label{fig:LR_Wide_Deep}
}
\vspace{-1.3\baselineskip}
\caption{a) Logistic Regression (LR) model: linear modeling of sparse feature values b) Hybrid model (Wide \& Deep model) %\textcolor{red}{From sparse field 1 to to dense embedding, there should be a specific conversions (such as neural network). Suggest to add a layer of neural network (a bunch of circles) or use can use a receantular box marked as Feature Embedding to denote this process. The output form the feature embedding is ur current Dense Embedding. Also please add text description to briefly explain the process. }
  as the alternative to Factorization Machines to use a deep component to capture higher ($\geq$two) order feature interactions combined with the Logistic Regression addressing low order feature interactions}
% \qquad
% \subfigure[Unfolded single layer LSTM network]{%
%   \includegraphics[width=3.6cm]{figs/lstm_layer_new3.png}%
% %   \label{fig:evaluation:avgPrice}%
% }

\end{figure}

In this section, we review supervised learning based methods which formulate the prediction of user response rates as binary or multi-class classification task in online advertising platforms. These methods can be categorized into two categories including the basic and ensemble predictive methods. Following the structure in Figure \ref{fig:UserResPredict}, input features are generally considered as multiple feature fields gathered from different sources like user, advertiser and publisher. The input layer in classification methods are considered as a numeric vector from concatenation of all fields. %The concatenation of fields is straightforward approach is commonly employed in different studies \cite{EntireMa2018,DeepCovington2016,DINZhou2018,DSINFeng2019,AutoIntSong2019} as follows: 
\begin{equation}
\small
% \label{obj_function}
\begin{aligned}
x=\left[x_1 || x_2 || \cdots || x_n\right]
\end{aligned}
\end{equation}
where $n$ is the number of features and $x_i$ is the representation of field $i$. For categorical data, feature value is encoded into a numeric vector through directly one-hot-encoding. Fields with continuous values are first discretized to be encoded to binary vectors by one-hot encoding.
%vectors(one-hot-encoded)  
\paragraph{Logistic Regression based methods} 
Logistic regression is one of the first attempts to train models to predict user response from input categorical features. As it is shown in Figure \ref{fig:LR}, this method uses linear combination of coefficient values and input sparse binary feature vector to predict the binary output value. Given the input dataset with $m$ instances of ($x_i$,$y_i$) where $x_i\in\{0,1\}^n$ is an $n$-dimensional feature vector and $y_i$ is the label to represent the user response as (click:1, no-click:0). The predicted probability of $x_i$ belonging to class 1 is modeled by Sigmoid function as:
\begin{equation}
\small
% \label{obj_function}
\begin{aligned}
Pr(y=1|x_i,w)=\dfrac{1}{1+\exp(-w^Tx_i)}
\end{aligned}
\end{equation}
The model coefficient $w\in \mathbb{R}^{d}$ is achieved by minimizing the negative log likelihood  as follows:
\begin{equation}
\small
\label{obj_function}
\begin{aligned}
\min_{w}\dfrac{\lambda}{2}\left\lVert{ w }\right\rVert ^2+\sum_{j=1}^{m}\log(1+\exp(-y_i\phi_{LR}(w,x_i))) 
\end{aligned}
\end{equation}
where $\phi_{LR}(w,x)=w_0+w^Tx=w_0+\sum_{j=1}^{n}w_ix_i$ is the linear combination of coefficients along with bias value $w_0$ and features that $w_0\in\mathbb{R}$ and $w\in\mathbb{R}^n$. As it is shown in literature %\cite{}
Eq.~(\ref{obj_function}) is convex and differentiable, so gradient based optimization techniques can be applied. 
%like L-BFGS optimizer
\paragraph{\underline{{Challenges and extended methods}}}
Some studies \cite{SimpleChapelle2015} indicate that the implementation of logistic regression methods is possible with high scalability through Maximum Entropy approach and a generalized mutual information and feature hashing as the regularization. However, modeling linear interaction between feature values only address the effect of features with class label separately. Therefore, it cannot always generate an acceptable performance in user response prediction task which gets impacted by some issues such as class imbalance originated from low click and conversion rates, the cold start issue for new instances, long cycle of user purchase responses, and non-linear interaction between input features. Authors in \cite{EvaluatingDalessandro2012} use historical information of brand website visit as the proxy to model predictor using logistic regression model. A study \cite{EstimatingLee2012} suggests to create hierarchy structure from previous user performances that is captured from grouping ad campaigns and publisher pages and users. % to alleviate these issues. 
A logistic regression model is used for linear combination of local MLE estimators.

Employing the side information using transfer learning has also been studied in some work~\cite{ScalableDalessandro2014, LargeYang2016}. In \cite{ScalableDalessandro2014} a transfer learning method is developed to combine data from a model on small set of conversion data to improve post-view conversion rate for large number of ad campaigns where click event is not necessarily required. In \cite{LargeYang2016} a transfer learning approach was developed to design a natural learning processing method to capture transferable information of related campaigns. It is motivated by the fact that the similar searched content and visited web-pages by users can be indicators of their future purchase interest. In another work, a practical result from applying logistic regression for big data in social media platform demonstrated that the weakness of linear modeling could be reduced by cascading with decision tree models to implement non-linearity of input categorical data~\cite{PracticalHe2014}.

\paragraph{Factorization Based Methods} \label{FM_section}
In order to consider non-linear interaction between features values, factorization machines (FM) combine support vector machine method with factorization models~\cite{FMRendle2010}. This allows the method to carry out parameter estimation under the data sparsity using linear complexity. This can be done by modeling the feature value interactions through a product of two latent vectors $v_i,v_j\in\mathbb{R}^{k}$. The dimensionality of latent vectors is the hyper-parameter that defines the number of latent factors.
\begin{equation}
\footnotesize
\label{phi_FM}
\begin{aligned}
\phi_{FM}= w_0+w^Tx+\sum_{i=1}^{n}\sum_{j=i+1}^{n}\langle{v_i,v_j}\rangle{x_ix_j}
\end{aligned}
\end{equation}
%%%%%%%%%%%%%%%%%%%%%%%%%%%%%%%%%%
%  \begin{table}[htb!]
% 	\centering
% % 	\fontsize{9}{11}\selectfont
% % 	\begin{small}
%     \scriptsize
% % 	\scriptsize
% 	\tabcolsep 2pt
% 	\caption{{Example of multi-field categorical feature}}
% 	\renewcommand{\arraystretch}{1.2}
% 		\begin{tabular}{|l|l|l|l|l|l|} 
% 			\hline
% 			 User\_id(Field 1) & device\_type\_id(Field 2) & gender(Field 3) & good\_cate\_id(Field 4) & publisher\_id& click(Label)\\\hline
% 			  2342352& Cell Phone & Male & Sport & Adidas & 1 \\\hline
% 			  7657632& Tablet & Female & Fashion & Gucci & 0  \\\hline
% 			  1232332& Cell Phone & Female & Travel & Uber & 0  \\\hline
% % 						
% % 			User behavior features  & visited\_good\_ids  & [] \\
% 			 % & visited\_shop\_ids,visited\_cate\_ids  & [] \\	\hline		
% % 			Ad features  & goods\_id,shop\_id,cate\_id  & [] \\\hline
% % 			Context features  & time  & [] \\
% % 			\hline
% 	\end{tabular}
% 	\label{Example:Mulcat}
% % 	\end{small}
% \end{table}
%%%%%%%%%%%%%%%%%%%%%%%%%%%%%%%%%%
\vspace{-.2cm}
\begin{figure}[h]

% \centering
\subfigure[FM]{%
  \includegraphics[width=5.0cm]{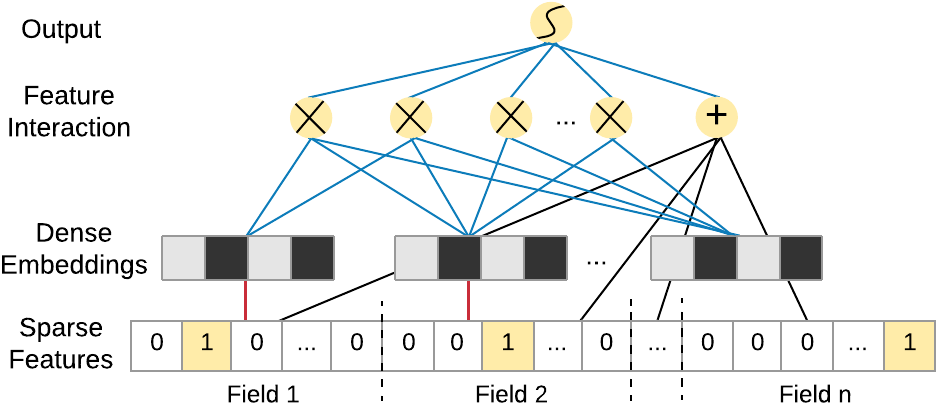}%
  \label{fig:FM}
%   \label{fig:evaluation:avgPrice}%
}
\qquad
\subfigure[FFM]{%
  \includegraphics[width=5.0cm]{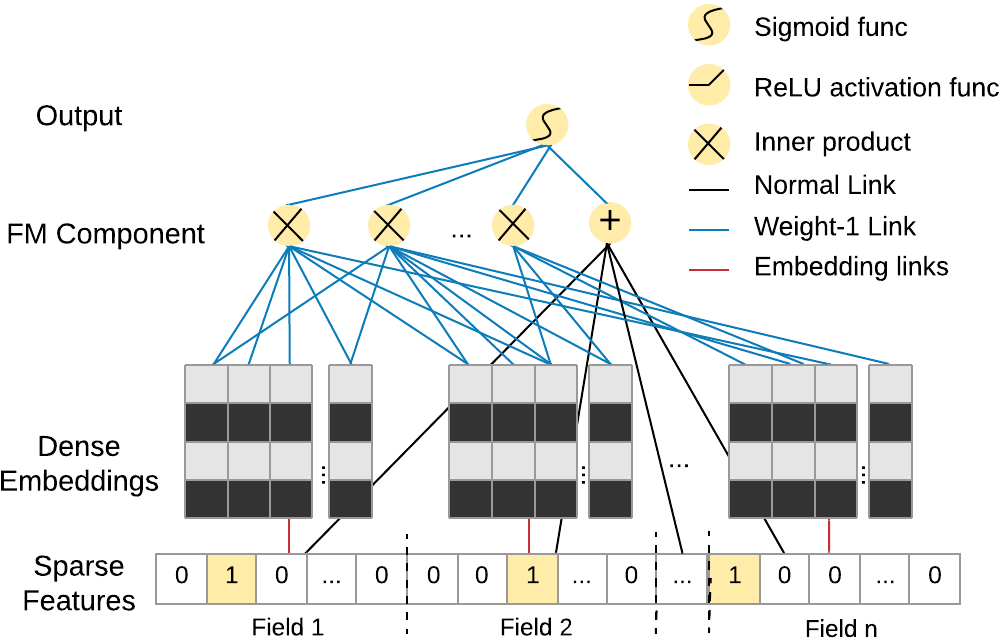}%
  \label{fig:FFM}%
}
% \qquad
% \subfigure[Unfolded single layer LSTM network]{%
%   \includegraphics[width=3.6cm]{figs/lstm_layer_new3.png}%
% %   \label{fig:evaluation:avgPrice}%
% }
\caption{a) FM Model: The architecture of Factorization Machines method %based models %\textcolor{red}{please refer to comments in Figure 5} $\Motimes$ 
  as the extension of logistic regression by using dot-product $\otimes$ operators fed by dense embedding vectors of input sparse features.
  b) Field-aware factorization machines (FFM) model: The structure is similar to FM model. The difference is that 
  %\textcolor{red}{please refer to comments in Figure 5} 
 %combines Factorization Machines cascaded by MLP neural network as the deep component. %In input layer, each categorical feature field is represented by a series of embedding vectors %for each categorical feature field as the input. 
 the sparse interaction between each feature value in the current field $i$ with another one in the other field (ex. $i+1$) is modeled by a separated embedding vector. Each feature field $i$ is represented by an embedding matrix.}
\label{fig:FM_FFM}
\end{figure}
\setlength{\abovecaptionskip}{.5cm}
% \begin{figure}[h]
%   \centering
%   \includegraphics[width=.7\linewidth]{figs/FM_v2.png}
%   \caption{The architecture of Factorization Machines based models %\textcolor{red}{please refer to comments in Figure 5} $\Motimes$ 
%   as the extension of logistic regression by using dot-product $\otimes$ operators fed by dense embedding vectors of input sparse features.}
%   \label{fig:FM_architecture}
% %   \Description{The 1907 Franklin Model D roadster.}
% \end{figure}

Figure \ref{fig:FM} shows the architecture of factorization machines as the combination of two terms including the feature interaction $\langle{v_i,v_j}\rangle$ and linear information $w_0+w^Tx$ to model click responses. The idea is that embedding vectors of features can be trained well to preserve feature interaction through dot product operation if there are enough occurrences that the features appear in the dataset. 
% \paragraph{\underline{Factorization Machines Optimization}}
\paragraph{\underline{Extended Factorization Machines models}}
Since factorization machines have a closed form equation that can be calculated in linear time, it is shown that the parameters of models can be trained using gradient based methods like stochastic gradient descend optimization (SGD)~\cite{FMRendle2010}. Some studies showed that FTRL-Proximal algorithm with $L_1$ regularization and per-coordinate learning rate, which was successfully used for logistic regression based models \cite{AdClickMcMahan2013}, can also outperform SGD algorithm for extended factorization machine models ~\cite{FMProximalTa2015}. However, this method suffers from some limitations.
% \begin{equation}
% \small
% \label{phi_FM}
% \begin{aligned}

% \end{aligned}
% \end{equation}

One of the downsides of factorization machines modeling is that for multi-field categorical data, feature values may come from different field feature that may change the interaction between feature values. But most methods deal with feature values uniformly. %Considering table \ref{Example:Mulcat}, the multi-field categorical is provided including 5 fields of user\_id, device\_type\_id, gender, good\_cate\_id, and publisher\_id. Intuitively, it can be seen that feature values of gender field have stronger interactions with features values from good\_cate\_id compared to their interaction with device\_type\_id. 
%It may indicate that there is a correlation between some users and some types of ad categories while they do not always use the same device type.
Therefore, Field-aware Factorization Machine (FFM)~\cite{FFMJuan2017} is proposed to discriminate the interaction between various feature values of different fields. To this end, it suggests to add one dimension to model parameters to allocate more than one embedding vector to features since pair of features incorporate different feature types information. %If we take the first row of table \ref{Example:Mulcat} as an example, using FM methods, the interaction model for three fields of gender, good\_cate\_id and publisher\_id would be:
% \begin{equation*}
% \small
% \label{example_FM}
% % \begin{align}
%  \langle{v_{Male},v_{Sport}}\rangle+\langle{v_{Sport},v_{Adidas}}\rangle+\langle{v_{Male},v_{Adidas}}\rangle
% % \end{align}
% \end{equation*}
% while applying the FFM method the interaction model is changed to:
% \begin{equation*}
% \small
% \label{example_FFM}
% % \begin{align}
%  \langle{v_{Male,C},v_{Sport,G}}\rangle+\langle{v_{Sport,P},v_{Adidas,C}}\rangle+\langle{v_{Male,P},v_{Adidas,G}}\rangle
% % \end{align}
% \end{equation*}
This changes the modelling of feature interaction as the following equation:
\begin{equation}
\small
\label{phi_FFM}
\begin{aligned}
\phi_{FFM}= w_0+w^Tx+\sum_{i=1}^{n}\sum_{j=i+1}^{n}\langle{v_{i,F(j)},v_{j,F(i)}}\rangle{x_ix_j}
\end{aligned}
\end{equation}
where $F(i)$ is an indicator of field name %(shown by the first letter of field name in above example)
that feature corresponding to the first entry of feature interaction while $F(j)$ is an indicator of field name that feature is related to the second entry of interaction. 
% %%%%%%%%%%%%%%%%%%%%%%%%%%%%%%%%%%%%%%%%%%%%%%
% Furthermore, %One of challenges to 
% applying factorization machines method for user response prediction tasks has the other challenges. First, using input multi-field categorical data as the input in online advertising, there are a lot of fields along with millions of feature values should be coded by one-hot-encoding. This leads to a severe data sparsity. It usually comes along with the cold-start problem in recommendation systems where historical data about samples are often limited for prediction. %Recent years have seen a growth in Internet 
% In addition, the reliability of data can get affected
% with the diversity in using online devices. The increase in the number of customer access points in online content providers can lead to high rate of churn rate where many online user interaction through different devices are considered independent and unrelated against each other. It leads to the performance of factorization machines methods that is not acceptable as the user response data is dependant on many factors like time and is subject to these issues. Some studies consider to use methods to evaluate the importance of feature interaction and sample~\cite{AttentionalXiao2017,FwFMPan2018}, and others propose a robust prediction in the case of noise data~\cite{RobustFMPunjabi2018}.
% %%%%%%%%%%%%%%%%%%%%%%%%%%%%%%%%%%%%%%%%%%%%%%
Lack of consideration into the importance of features and the limitation of inner-product to model feature interaction are two issues in baseline methods of factorization methods have been studied in many other works as follows.

%It should be considered that, 
The baseline factorization machine methods usually consider all combinations of feature values in different fields with the same weight. But interactions between features often vary and do not have equal values. %Given multi-field categorical data in Table \ref{Example:Mulcat} including user profile and advertiser features, the interaction between layout\_id and user gender level compared to ad\_id and category\_id is not the same. 
So there is a chance that using less important features in the training set, the noise is actually learnt by the model which can have the adverse effect on the performance. This motivates studies to impose weights on the interactions~\cite{FwFMPan2018,AttentionalXiao2017,FIBINETHuang2019}
To this end, authors in \cite{FwFMPan2018} proposed a weighted version of field-aware factorization machine (FwFM) that can use the memory efficiently for model parameters. It adds more information through a weight matrix to consider the difference in the strength of interaction between feature values originating from different pair of fields. In \cite{AttentionalXiao2017} follows a deep learning study in which the importance of feature interaction were studied using an attention network through a layer to learn corresponding weights. A study \cite{PredictingOentaryo2014} centered the work on a different aspect to use a cost sensitive approach to address the cold-start issue and using the data hierarchy for the data sparsity. They designed an importance-aware loss function to assign the more importance weights and penalty values for ad samples which are shown more to users but their user response predicted wrongly. %in the case they have seen wrong user response prediction. 
Authors in \cite{RobustFMPunjabi2018} also proposed a robust factorization machines method considering user response prediction as a classification problem under the noise. The uncertainty within input samples is modeled by an optimization through an uncertainty vector with each dimension corresponding to independent noise value.

Aside from the above mentioned points, the capability of factorization machines to address data sparsity issue using inner-product operation can be limited when confronting high dimensional data. Modeling only 2nd-order feature interactions is not expressive enough for implicit higher order feature conjunctions. This stimulated motivations to propose high order variants of factorization machines method \cite{LearningXu2019,HigherOrderBlondel2016,FIBINETHuang2019}. In \cite{FIBINETHuang2019} authors extended the feature interaction modeling in factorization machines using a bilinear interaction method which combines inner product and Hadamard product together to generate a fine-grained feature interactions.
Very recently, \cite{LearningXu2019} %argued that although deep learning methods have a potential to capture arbitrary complex interactions, but they need the large number of parameters for training within layers. Taking advantage of geometric properties for hyperbolic triangles, they 
proposed a score function to replace inner-product operation between embedding vectors of input features. They discussed that using Lorentz distance, the triangle inequality principle between two points with regard to the origin point is not always consistent. They suggested to use the sign of triangle inequality to learn feature interactions through a proposed Lorentz embedding layer. To this end, a novel triangle pooling layer is proposed to substitute for the typical factorization machines structure. %that is consisting of inner-products ended by a Sigmoid function. 

\paragraph{Hybrid Approaches}
% Hybrid methods follow a classification technique that involves a usually small number of heterogeneous methods, which act complementarily to each other. Each method solves a different task and the classification decision is reached by one method. This idea makes an motivation for attempts to develop more complex machine learning methods which are able to predict user interaction in online advertising systems. 
Hybrid methods follow a classification technique that involves a number of heterogeneous methods each of which acts complementary to each other. Each method solves a different task and the classification decision is reached by the one method at the end. The distinction between ensemble methods and hybrid methods is that the former models are trained separately to generate the predictions at the inference time. On the contrary, hybrid models follow a joint training which optimizes all parameters simultaneously. This idea makes a motivation for attempts at developing more complex machine learning methods which are able to model a non-linear user interaction in online advertising systems. Regarding the user response prediction, the primitive predictors based on the logistic regression or factorization machines have weakness to capture limited range of feature interaction by addressing linear relations or dot product interactions between input features. Their performance is suffered from the data sparsity, class imbalance problem and cold-start problems. To address these issues,% and also leverage the benefits of these methods, 
hybrid architecture of classifiers has been proposed in many studies. The categorization of hybrid models are presented as follows:

\paragraph{\underline{Logistic regression based methods}}
One of the first studies to improve logistic regression performance was the addition of decision trees to the structure of model~\cite{PracticalHe2014}. In order to address the data sparsity in input data consisting of multi-field categorical data, they use a cascade of decision trees structured by boosting ensemble paradigm to provide a non-linear transformation of categorical features. Following a gradient boosting machine, the boosted decision trees generate a feature vector with the user-defined dimension $k$ that is passed to logistic regression classifier for prediction. 

The success of deep learning methods in capturing higher order interactions motivates research to include deep neural networks to improve the Logistic Regression in different studies~\cite{WideAndDeepCheng2016,DeepCrossingShan2016}.

Although the logistic regression models have shown a good scalability and interpretability to handle the massive data in the online advertising industry~\cite{SimpleChapelle2015}, the generalization of model for predicting new samples is limited and highly dependant to whether high quality features can be obtained through feature engineering. Using the polynomial regression applied, the logistic regression model can only capture low-order feature interactions. This drives authors in \cite{WideAndDeepCheng2016} to approach a hybrid structure of logistic regression and deep neural networks which are trained jointly to consider low and high order feature interactions when there is the data sparsity issue and dealing with massive data. As shown in Figure \ref{fig:Wide_Deep}, the framework includes two components \textit{i.e.} wide and deep. Wide linear component is modeled by the logistic regression classifier. It analyzes two sets of input features including raw categorical features and transformed features which are designed to memorize sparse feature interactions using a cross-product feature transformation. Following the feature engineering approach on the training data, the transformation function is designed to represent the frequent co-occurrence of features to explore the possible correlation with user responses. The deep neural network component is trained to generalize the prediction for unseen inputs through low-dimensional embeddings. In this model, the final output is calculated from the combination of wide and deep components using the logistic loss function.
%trained based on the embedding layer 
%followed by fully connected neural network.
In initial studies, the embedding vectors are generated from a embedding dictionary by feature hashing method \cite{SimpleChapelle2015} where most frequent categorical values are transformed by projecting into pre-defined fixed-size numerical vectors~\cite{DeepCovington2016,WideAndDeepCheng2016,DeepCrossingShan2016}. The other models covered in the next section fix this issue using trainable embedding vectors.

Some recent studies in this category extend different elements in the hybrid design like embedding vectors \cite{NFMHe2017,StructuredNiu2018,ResembeddingZhou2019} and neural network architectures \cite{DeepCrossingShan2016,DINZhou2018,DIENZhou2019} to improve prediction performance. Although it is typically expected that stacking of multi-layer fully connected neural networks can capture arbitrary non-linear relations between input features, dealing with a lot of parameters can cause different issues such as the degradation and over-fitting. A study \cite{DeepCrossingShan2016} proposes to use a residual neural network for a deep component, where five hidden layers of residual units combined with original input features are added to the result of two layers of ReLU transformations. The effect of aggregation of embedding vectors on the prediction performance is also studied in \cite{NFMHe2017}. The baseline methods \cite{WideAndDeepCheng2016,DeepCrossingShan2016} follow a simple concatenation of embedding vectors in Figure \ref{fig:Wide_Deep} to be fed in a deep component to capture feature conjunctions. They demonstrate that it may carry less non-linear information in the low-level. Therefore, they suggest a Bi-Interaction pooling encoder to capture more informative feature interactions. Considering an embedding vector for each feature value, the Bi-Interaction pooling operation is designed to generate the aggregated vector as follows:
\begin{equation}
\footnotesize
\label{bi-interaction}
% \begin{aligned}
f_{BI}(V_x)= \sum_{i=1}^{n}\sum_{j=i+1}^{n} x_i\mathbf{v}_i \odot x_j\mathbf{v}_j
% \end{aligned}
\end{equation}
where $V_x=\{x_1\mathbf{v_1},...,x_n\mathbf{v_n}\}$ is the set of embedding vectors, $x_i$ is binary feature value in sparse input vector, $\mathbf{v_i}$ is embedding vector and $\odot$ operator makes element-wise product of two vectors. 

Further, authors in \cite{StructuredNiu2018} pinpointed that the semantic intrinsic relations between embedding vectors of user and ads can be captured through their proposed structured semantic models. They propose a series of orthogonal convolution and pooling operators rather than trainable convolutional operators which can be applied as embedding vectors to address semantic relations in input features. %The result of the flatten layer includes extracted features that are passed through the fully connected deep neural component for the optimization. 
Experiments reported in the above studies show that applying hybrid methods can improve logistic regression, which highly depends on the quality of features prepared by using feature engineering. This encourages further studies to develop extensions of factorization machines with better generalization ability.

\paragraph{\underline{Factorization based hybrid methods}}

\begin{figure}[h]

% \centering
\subfigure[DeepFM]{%
  \includegraphics[width=5cm]{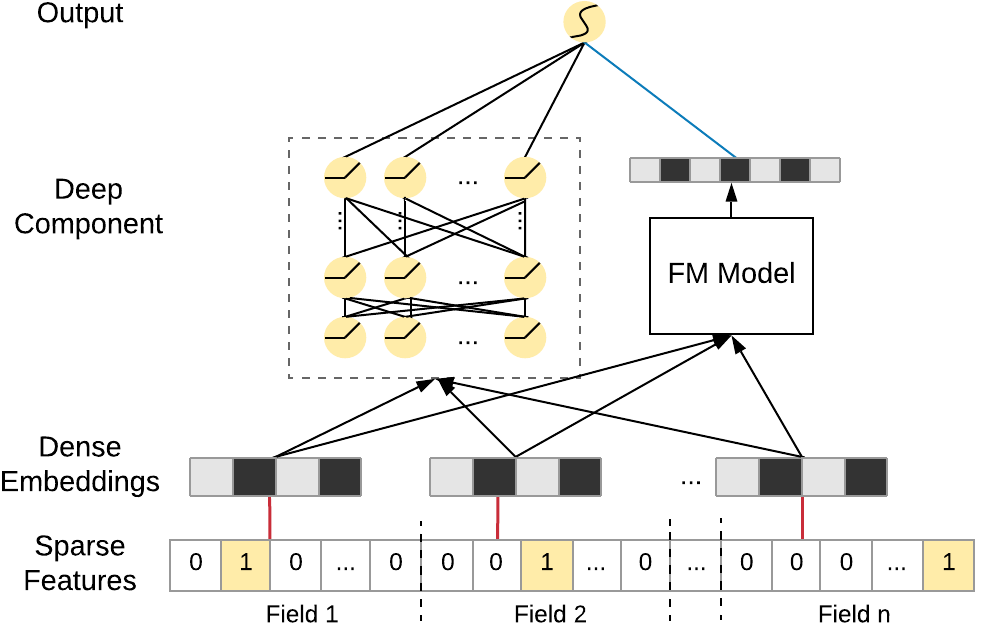}%
  \label{fig:DeepFM}%
}
\qquad
\subfigure[DeepFFM]{%
  \includegraphics[width=5cm]{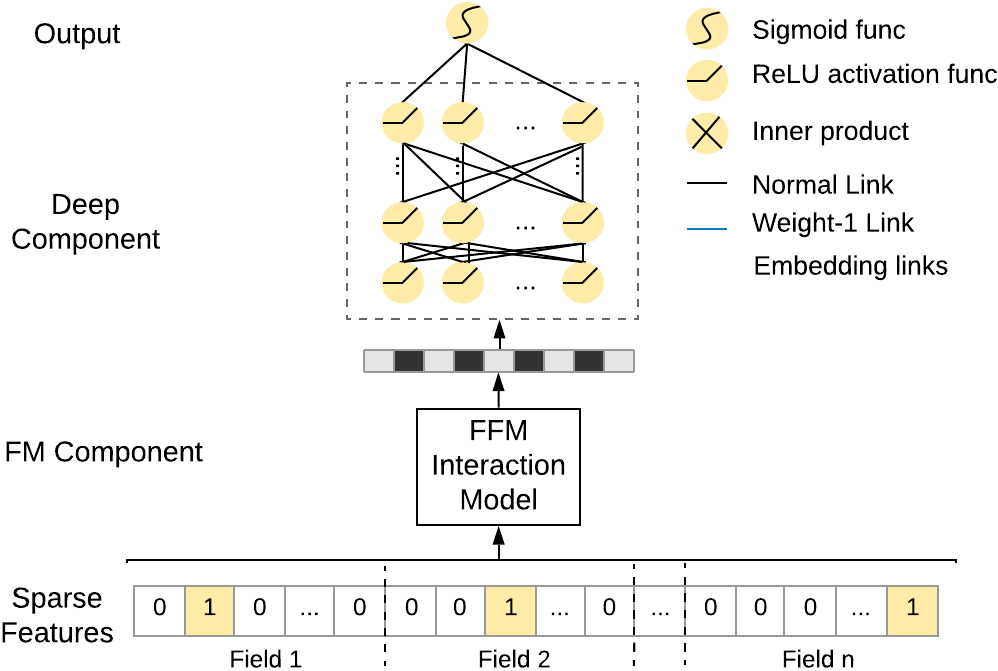}%
  \label{fig:DeepFFM}%
}
% \qquad
% \subfigure[Unfolded single layer LSTM network]{%
%   \includegraphics[width=3.6cm]{figs/lstm_layer_new3.png}%
% %   \label{fig:evaluation:avgPrice}%
% }
\vspace{-1.3\baselineskip}
\caption{a)Hybrid model (DeepFM):  %\textcolor{red}{please refer to comments in Figure 5}
  It is combined by a deep component (fully connected neural network) with a Factorization Machines method(Figure \ref{fig:FM}) using a shared embedding layer to feed dense embedding vectors as the input to the structure b)Hybrid model (DeepFFM): It 
  %\textcolor{red}{please refer to comments in Figure 5} 
 cascades the FFM interaction model in Figure \ref{fig:FFM} to a MLP network as the deep component. %In input layer, each categorical feature field is represented by a series of embedding vectors %for each categorical feature field as the input. 
 %The sparse interaction between each feature value in the current field with another one in the other field is modeled by a separated embedding vector. Each feature field $i$ is represented by an embedding matrix.
 }
\label{fig:hybrid_DeepFM_DeepFFM}
\end{figure}

In \cite{DeepFMGuo2017}, the authors provided a successful version of hybrid methods as the stack of factorization machines and fully-connected neural networks. The success of this design later led to employ this structure as a base for developing many extensions~\cite{NewWang2018,FAT-DeepFFMZhang2019}. The study pinpointed that Wide \& Deep method \cite{WideAndDeepCheng2016} has some challenges in modeling of feature interactions, since the wide component includes the logistic regression model trained using a feature engineering. It can cause a poor generalization. They investigated to use a factorization machine to automatically capture feature interactions from one-hot-encoded features. Following the structure shown in Figure \ref{fig:DeepFM}, the  proposed model, DeepFM, combines the power of factorization machines and deep learning for the feature learning in a recommendation application. The new neural network architecture models linear and 2nd order feature interactions through FM and models high-order feature interactions by fully connected neural network. Replacing the logistic regression with factorization machines and using a shared embedding layer between these two components, they build a model in an end-to-end manner without a feature engineering. % being trained using the same input and embedding layer. 

\begin{figure}[h]
  \centering
  \includegraphics[width=.4\linewidth]{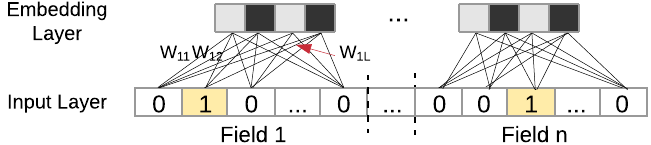}
\vspace{-1\baselineskip}  
  \caption{The structure of embedding layer to generate dense embedding vectors. It includes a linear mapping from discrete categorical features represented by one-hot-embedded vectors to dense numerical embedding vectors. It equals to a layer of linear neurons above input layer whose weights are getting trained using gradient descent optimization. The weights are formed in a embedding matrix(lookup table) to accomplish the linear transformation. %The embedding matrix is considered as a lookup table whose 
  The rows in the matrix represent embedding vectors for discrete values in the categorical fields}
  \label{fig:embedding}
%   \Description{The 1907 Franklin Model D roadster.}
\end{figure}

Figure \ref{fig:embedding} shows the embedding layer structure, which  is designed to project discrete feature values to a dense numerical vector space. This projection is modelled by a layer of linear neurons defined on the top of one-hot-encoded input vectors~\cite{EntityGuo2016}. It includes an embedding matrix of parameters learned for each feature field. Embedding vector representing each categorical field can be shown as follows:
\begin{equation}
\small
\label{eq:emb}
\begin{aligned}
e_i=W_ix_i
\end{aligned}
\end{equation}
where $e_i$ is the dense embedding vector and $x_i$ is the sparse binary representation. $W_i$ is the embedding matrix for i-field with the dimension of $m_i \times d_i$.  $m_i$ denotes the number of discrete values for categorical field $i$ and $d_i$ is the user-defined dimension of dense embeddings. In practice, the functionality of embedding layer is identical to one layer of densely connected neurons without considering bias links and activation functions. It is shown that the embedding matrix $W_i$ can be considered a lookup table for each field. This is because in the case of one-hot-encoded input, the multiplication of input with embedding vectors in Eq. \eqref{eq:emb} can be replaced by corresponding embedding vectors at referred indices in the embedding matrix. Randomly initialized, the weights $w_{ij}$ in the embedding matrix are trained during the optimization of the target value in different models.

% \begin{figure}[h]
%   \centering
%   \includegraphics[width=.7\linewidth]{figs/DeepFM_v3.png}
%   \caption{Hybrid model (DeepFM)  %\textcolor{red}{please refer to comments in Figure 5}
%   to combine a deep component including a fully connected neural network with Factorization Machines using the shared embedding layer to feed dense embedding vectors as the input to this structure}
%   \label{fig:DeepFM}
% %   \Description{The 1907 Franklin Model D roadster.}
% \end{figure}  
% \vspace{-0.5cm}

% \vspace{-.5cm}

%In the different network architecture 
In ~\cite{ProductBasedNNQu2016} a new hybrid is proposed through combining embedding vectors and a cascade of factorization machines and a MLP network. This method takes advantage of learning ability of neural networks and discriminative power of latent patterns in a more effective way than MLPs, through adding a product layer between the embedding layer and the first layer of the fully connected neural network. The model is examined with inner and outer product operations in the product layer, to examine different methods to model the feature interactions, combined with  Stochastic Gradient Descent training (using L2 regularization) and a dropout mechanism to address the over-fitting issue. 

%Taking a look on other work, a study in \cite{ProductBasedNNQu2016} proposed a new hybrid method using embedding vectors combined with a cascade of factorization machines methods with MLP network. They presented a method which takes advantage of learning ability of neural networks and mine the latent patterns of data in a more effective way than MLPs, through adding a product layer between the embedding layer and the first layer of the fully connected neural network. The model is examined with inner and outer product operations in the product layer, to examine different methods to model the feature interactions. The model is trained with Stochastic Gradient Descent by L2 regularization and a dropout mechanism to address the over-fitting issue. 

As introduced in Section \ref{FM_section}, in FFM\cite{FFMJuan2017} method as the field-aware factorization machines, each feature value is represented by more than one embedding vectors to model combinatorial features in input space. It addresses different weights for interactions occurring between different feature types. %Authors later extended this study to model feature interaction using a kernel product method \cite{QuProductBasedNN2018} to not only model higher level feature interaction is modeled by applying, inner-product of embedding vectors in kernel mode, it proposed a micro network as the special kernel to address the large number of features.

The large number of features in latent vectors generally cause  space complexity problem and memory bottleneck~\cite{FFMJuan2017}. In addition, DNN based models may run into insensitive gradient issue when dealing with multi-field categorical data which deter the progress of gradient based optimization. To tackle these challenges, a net-in-net architecture is proposed as the generalization of kernel product~\cite{ProductMultiQu2018} to model feature interactions. So a micro network including one layer of the fully connected neural network cascaded by dot-product feature latent features is used as the special kernel function to alternate a simple inner-product function in factorization machines.
% \begin{figure}[h]
%   \centering
%   \includegraphics[width=.55\linewidth]{figs/DeepFFM_OFM_v2.png}
%   \caption{Hybrid model (DeepFFM) 
%   %\textcolor{red}{please refer to comments in Figure 5} 
%  combines Factorization Machines cascaded by MLP neural network as the deep component. %In input layer, each categorical feature field is represented by a series of embedding vectors %for each categorical feature field as the input. 
%  The sparse interaction between each feature value in the current field with another one in the other field is modeled by a separated embedding vector. Each feature field $i$ is represented by an embedding matrix.}
%   \label{fig:DeepFFM}
% %   \Description{The 1907 Franklin Model D roadster.}
% \end{figure}

Following the success of field-aware factorization machines~\cite{FFMJuan2017} in capturing feature interaction with regard to feature fields information, a study \cite{OperationYang2019} extends this idea to provide a hybrid model of FFM and a fully connected deep neural network to learn feature conjunctions in the input data, as shown in Figure \ref{fig:DeepFFM}. % demonstrates the architecture of combining FFM model with MLP neural network. 
In this case, each sparse input feature is represented by multiple embedding vectors to address the effect of feature with regard to the feature field in inner (dot-product) feature interactions. The embedding vector are organized as a 2-dimensional matrix with size of $k \times n$ where $k$ is the dimension of embedding and $n$ is the number of feature fields. Applying $n(n-1)/2$ inner-product calculations between pair of embedding vectors to generate intermediary input vectors, the predicted click-through rate value is generated from the output of Deep component. Field-aware Neural Factorization Machines~\cite{Field-AwareZhang2019} further extends this method, by introducing a Bi-Interaction operator with a wide concatenation based on Hadmard product operator as the alternative to the inner product layer in Figure \ref{fig:DeepFFM}. 
Using Bi-interaction operator to calculate feature interaction, the dimension of input vector of deep component is changed to $n(n-1)/2 \times k$. The increase of parameters in the embedding vectors of field-aware based factorization machines methods can decrease the prediction performance because of the over-fitting issue. Therefore, it demands to select features before the feature interaction procedure in factorization machines. Compared to Attentional Factorization Machines method \cite{AttentionalXiao2017} which captures important cross features interaction step in FM model, a recent study \cite{FAT-DeepFFMZhang2019} evaluates the importance of features before applying feature interaction step using Squeeze-Excitation network \cite{FIBINETHuang2019,FAT-DeepFFMZhang2019}. The authors~\cite{FAT-DeepFFMZhang2019} introduced an attention-based method to selectively use more informative features in embedding vectors. They propose to apply Compose-Excitation network as the extension of Squeeze-Excitation Networks to select important feature representations.

\paragraph{Generic hybrid methods}
Some studies \cite{DCNWang2017,xDeepFMLian2018} develop a hybrid method to generalize the idea of factorization machines method. The authors in \cite{DCNWang2017} extended the second order feature interaction in factorization machines to higher order levels through a multi-layer network structure where the maximum level of interaction order is determined by the number of layers. Considering the structure of DeepFM method in Figure \ref{fig:DeepFM}, a cross network is employed for feature crossing operation using equation \eqref{eq:FeatureCross} through weighted dot product of current output vectors at subsequent layers. 
\begin{equation}
\small
\label{eq:FeatureCross}
\begin{aligned}
\mathbf{x}_{l+1}=\mathbf{x}_0\cdotp \mathbf{x}_l^\intercal\cdotp w_l+b_l+\mathbf{x}_l
\end{aligned}
\end{equation}
where $\mathbf{x}_l \in \mathbb{R}^d$ denotes the output vector calculated at level $l$ and $w_l \in \mathbb{R}^d$ and $b_l \in \mathbb{R}^d$ constitute system parameters in this structure. In the first layer, the dot-product of concatenated embedding vectors $\mathbf{x_0}$ are used to generate the first output.
For input data as a multi-field categorical data, the proposed component explains the main difference between this model and factorization machines methods in which dot product interaction here is applied on the concatenation of feature fields rather than between pair of feature fields. The study \cite{xDeepFMLian2018} is the other extended work to create a new cross network in which feature interaction is modeled at a vector-wise level through outer product instead. This leads to generating an embedding matrix in which the operation in each layer has intuitively connection to convolution neural networks by considering the weights as filters.

\begin{figure}[h]
  \centering
  \includegraphics[width=.4\linewidth]{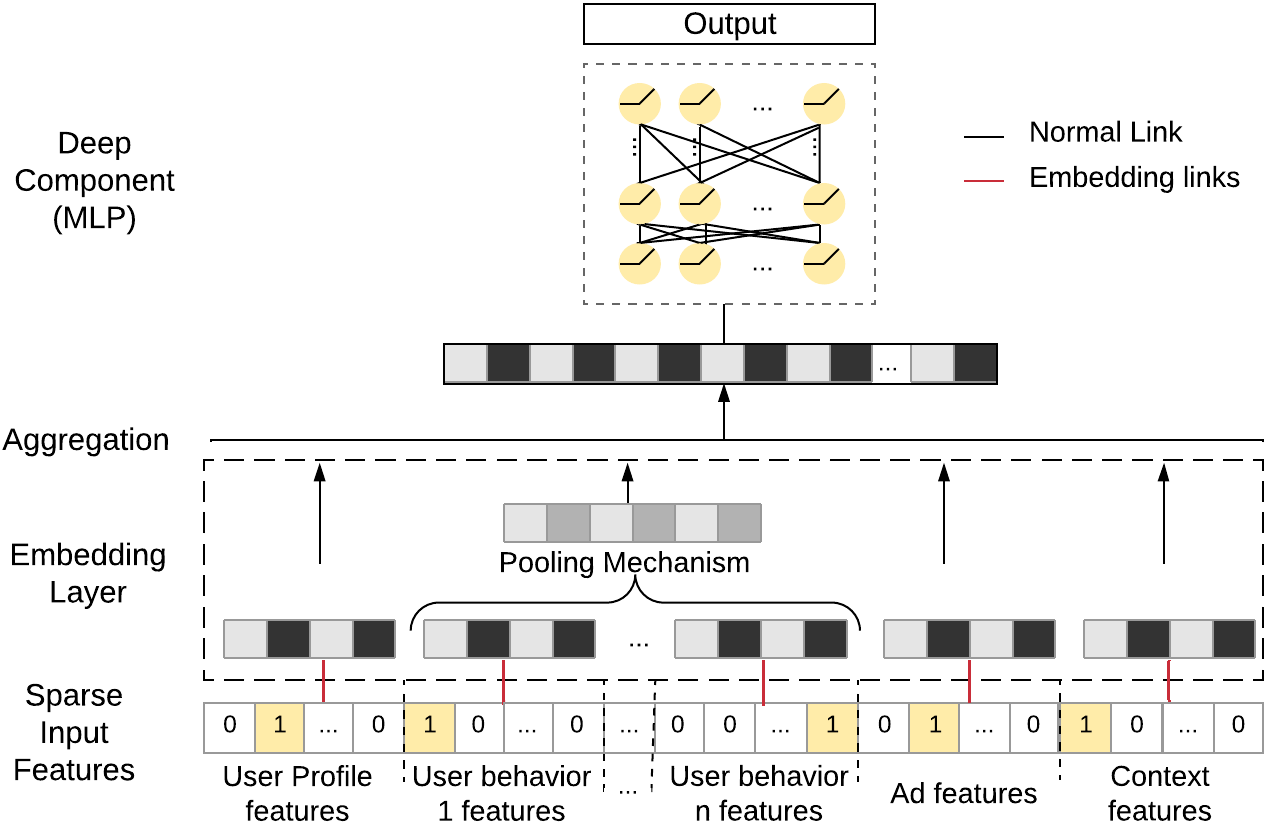}
\vspace{-.5\baselineskip}  
  \caption{DNN based architecture (Embedding \& MLP) 
   including two main steps embedding layer followed by the aggregation operation to generate a merged vector. The vector is used as the input for second step using MLP to generate predicted user response value(CTR, CVR, ...) developed in different studies.
   The selected input features are chosen user profile and online user behaviour sequence in addition to those selected for Ads and context
  %DNN based architecture to address online user behavior sequence
  }
  \label{fig:DNN}
%   \Description{The 1907 Franklin Model D roadster.}
\end{figure}

\paragraph{Deep learning based methods}
In literature, various deep learning techniques have been studied for the user response prediction. The majority of previous work following a deep network structure are typically based on two components of embedding and interaction basically designed through deep neural networks to capture non-linear feature interaction in sparse input data. Figure \ref{fig:DNN} demonstrates the structure of this paradigm. Embedding component is designed to transform the sparse input data into a low-dimensional dense latent space. The embedding vectors are then processed by applying an aggregation mechanism to produce a fixed length vector for the deep component. The high-order interactions between features are addressed through feeding a fixed-length vector into the deep neural network component generally implemented by the multi-layer perceptron \cite{DeepCovington2016,DeepCrossingShan2016}. Gradient based training is adopted to learn the non-linear correlation between user features and user responses. %The schema of these model is shown in figure \ref{fig:DNN}.
In this regard, there are lots of studies conducted to improve the performance of each component. Table \ref{tab:dnn_summary} demonstrates a summary of representative methods mainly developed based on multi-layer perceptron, recurrent neural networks and convolutional neural networks some of which are combined with attention mechanism design their proposed models.

% \begin{figure}[h]
%   \centering
%   \includegraphics[width=.6\linewidth]{figs/DNN_v3.png}
%   \caption{DNN based architecture (Embedding \& MLP) 
%   including two main steps embedding layer followed by the aggregation operation to generate a merged vector. The resultant vector is used as the input for the fully-connected neural network to generate predicted user response proxy value(CTR, CVR, ...) developed in different studies}  
%   %\textcolor{red}{please refer to comments in Figure 5}
%   \label{fig:EmbeddingMLP2}  
% %   \Description{The 1907 Franklin Model D roadster.}
% \end{figure}
In online advertising, input features can be gathered from different sources. In Figure \ref{fig:DNN}, sparse binary features representing the input data can be grouped into multiple categories like Table \ref{tab:cate_features} regarding users, advertisers and context. Deep neural network can process input data vectors with a fixed length dimension. However, using a fixed-length vector for users with diverse interests against advertisements can be  bottlenecks for prediction, since each user and web-page can have different labels and diversities at the same time. Addressing the variety of user interest generally needs the expansion in the dimension of embedding vector for user features in the aggregation step which increase the risk of over-fitting and the cost of computation. 

Dealing with this issue, two sub categories of features such as user profiles and user behaviors~\cite{DINZhou2018} are proposed for click-through rate prediction, where an array of user behavior in a period of time attributed by categorical data \cite{DINZhou2018,DeepOuyang2019} like visited good, shop and web-page category ids, are used to describe users and their interests. % task\cite{DINZhou2018} %is one of the first studies organizing an input feature set consisting of four categories for user advertisements and context like table \ref{tab:cate_features}. In this way, they 
%described users and their interests with regard to advertisement using two sub categories of features such as user profiles and user behaviors for the click-through rate prediction task. In this regard, they considered an array of user behavior in a period of time attributed by categorical data \cite{DINZhou2018,DeepOuyang2019} like visited good, shop and web-page category ids. 
The idea is further followed by \cite{ImageGe2018} to model user behaviors from the continuous image data. %like figure~\ref{EmbeddingMLP2}
In the case of categorical features, since the category of a shop and web-page visited by users may be shown with multiple values, the binary representations are modeled by multi-hot-encoding. %They argued that following figure \ref{} to use fixed-length vector as the aggregated vector of user interest cannot reflect the diversity user interest. Visiting a website, user may have different tendency to various items at the same time. 
They also address the importance of feature interaction in modeling of user behaviors. To this end, they design a local activation unit to provide an adaptive feature representation with regard to different ads, and assign weights to the relevant pair of a visited page and advertisement in the user behavior sequence with regard to the targeted ad. The output vectors are later passed to a weighted sum pooling to generate a fixed length user behavior embedding vector, and then passed into deep component to generate the predicted output value.

% \vspace{-0.5cm}
The application of attention units to model user behavior history has also been studied for click-through rate and conversion rate prediction~\cite{ AutoIntSong2019,DeepOuyang2019,PrivilegedXu2019,ImageGe2018,DKNWang2018,InterpretableLi2020,AttentionGao2018}. A multi-head self-attentive networks~\cite{ AutoIntSong2019} is proposed where features are mapped to multiple subspaces through multi-head mechanism. This would help the model to consider different orders of feature interaction with adaptive weights. In addition, this method proposes a residual neural network rather than the conventional MLP network to model high order combinatorial feature interactions. The study \cite{DeepOuyang2019} extends the analysis of user behavior from two temporal and spatial aspects. Because a web-page can be filled with more than one ad, they model user behavior history not only by ads clicked by users but also those not clicked by users. They consider ads shown in the same page above or below the targeted in both spatial and temporal order. Their interactions to targeted ads are modeled by adding an attention based factorization machines layer followed by a fully connected neural network. In \cite{PrivilegedXu2019} the multi-head attention mechanism is adopted to model the user behavior history from a sequence of clicked/purchased advertisement information by adding discriminative features likes dwell time on landing page for a conversion rate prediction task.

Some studies also extend the deep component using different structures of MLP based networks~\cite{ClickThroughOuyang2019,EntireMa2018,ConversionWen2019,RepresentationOuyang2019}. In order to tackle the data sparsity, researchers~\cite{RepresentationOuyang2019} propose to consider input features to be fed into a couple of sub-nets built based on MLP network for a feature interaction modeling, using features of users, query and ads entities. The sub-nets are created to model the interactions between user-ad, the correlation between ads. These models are then combined with the third one designed for the prediction in a joint optimization.
For conversion events followed after clicking on ads, a deep learning based model is developed~\cite{EntireMa2018, ConversionWen2019} to consider not only all clicked ad impressions, but also include all impressions and further user actions like ``add to card'' (DAction) and ``add to wish list'' (OAction) taking place before conversion events. Following a multi-task framework, multiple deep components are trained for each event to generate the prediction of conversion post-click rates. 

For the embedding, some studies propose to handle categorical and continuous input features at the same time~\cite{ResembeddingZhou2019,ImageGe2018}. For images, convolutional neural network have been developed in several studies \cite{ ImageGe2018, DeepChen2016,LearningYang2019}. In the scale of industrial applications, authors in \cite{ ImageGe2018} suggest to generate embedding vectors of images using a pre-trained very deep convolutional neural network rather than employing an end-2-end training model. The convolutional networks are also adopted in \cite{FeatureGenLiu2019} to extract implicit features from the sparse input data and deal with the overfitting issue in fully connected based networks~\cite{DeepCovington2016}. In the proposed model, the convolutional neural network structure designed based on shared weights followed by pooling module can considerably reduce the number of parameters. Considering these embedding vectors along with raw features result aggregated vector to be processed by a deep multi layer perceptron component. 
%Another study \cite{ResembeddingZhou2019} focuses on a new approach to improve embedding mechanism for the user response prediction. They considered input features as the combination of user and ad features when user features modeled from the sequence of user behaviors. The analysis indicated that user behaviors can follow an interest delay model that embedding vectors of visited ad related to a similar user interest category have proportionally a small distance against others. They generate an interest graph basically made from ads captured in user behavior sequences to describe the similarity of items with regard to different user interest categories. They proposed that the combination of ads and their neighbors can play a role as an auxiliary information. It is added to initial ad embedding vectors for training DNN based user response prediction models. In their modeling, they evaluated the effect of several fusion methods, like weighted average, GCN, and attention, to produce auxiliary embedding vectors.

\paragraph{\underline{Recurrent neural network based methods}}  \label{sec:rnn}

Deep neural networks is typically made of multi layer fully connected neurons. Following a stateless structure for neurons, the independent features incorporate the data that flow through multi-layer perceptrons to generate the output without backward links. Considering independently visited or clicked advertisements to model the user behavior history fail to extract efficient useful user interests with regard to user response prediction. Therefore, recurrent neural networks are developed to process the sequence of input data sorted by time %They are of hidden states for neurons following by backward links to act as a memory to persist temporal information from the past input values.This form of neural networks have been studied widely in many work to model user interests and 
to improve user response prediction performance \cite{FiGNNLi2019,DIENZhou2019,DSINFeng2019,PracticePi2019,OnlineChao2019,SequentialZhang2014}.

In online advertising or e-commerce platforms, user intention is often not explicitly expressed through their behavior history. Therefore, it is hard to identify real interested users only based on captured online user behaviors. In addition, living in a dynamic life environment, people's knowledge increase and their latent interest and behavior may change over time \cite{OrderPan2019}, indicating that temporal correlation in online user behaviors can show the evolution in the user interest to have more tendency to advanced items compared to previous converted ones. It leads to studies of developing LSTM models based on user behavior sequence from previous bought and clicked items to predict conversion rates~\cite{OrderPan2019}.
%For industrial applications, considering user behavior history created from previous bought items\cite{OrderPan2019}, authors reported a successful study to build a LSTM network% over word2vec embedding vectors 
%to predict the next item with the higher conversion rate. The study in \cite{PredictionBigon2019} similarly used the sequence of click events to represent user behavior histories. They were used to model a LSTM model for conversion rate prediction. 
Following Figure \ref{fig:DNN}, to aggregate embedding vectors, two categories of aggregation modules are generally discussed in different recurrent neural network based studies. % to aggregate embedding vectors of user behavioral data. 
The approaches like min-max or sum pooling based aggregation are generally applied to model user behavior from independent input features. In the body of neural network structure, GRU/LSTM neural units were commonly used in many studies %The sequence based methods based on GRU/LSTM networks are used
to model latent user interests~\cite{FiGNNLi2019,DIENZhou2019,DSINFeng2019,PracticePi2019}.

%For example, to generate embedding vectors for user behavior history, \cite{DIENZhou2019} proposed an embedding layer to capture temporal information from user behavior histories. They used GRU neurons to model the temporal dependency in a sequence of user behaviors. It followed an attentive update gate to represent the user interests by controlling the drifting effect. This phenomenon describes the case that the diversity of user interest may be dependant to activity long time ago rather than the one in the recent past. This model was used to track the evolution of user interest over time and their effect through attention weights with regard to user click events.%%to model the effect the history of click event on different items are modelled using attention units.
%Here, the embedding vectors are aggregated by concatenation, and are fed into fully connected network to predict the output through joint gradient descent optimization. Similarly, in \cite{DSINFeng2019} the authors decreased a granularity of user interest sequence into a series of sessions based on their observations in an e-commerce platform that user activities can be organized within separate sessions to have related online behavior regarding one type of items. A Bi-LSTM neural network layer followed by self-attentive mechanism was proposed to model user interest in each session and its evolution over time. 

Comparing to deep neural network based methods, recurrent neural networks suffer from computational and storage overheads. They include hidden states in the structure to capture user interests from sequence of user behavior data. Therefore, it makes it difficult to use these network for industrial applications visiting numerous users and ads everyday. It causes limitations to apply these methods to model long term user interests based on long sequential user history records. To tackle these challenges, some studies introduced memory based architectures \cite{ClickThroughOuyang2019,PracticePi2019,OnlineChao2019}. %A study \cite{PracticePi2019} addressed the storage and computational bottlenecks to deal with long sequential user behavior data through designing a serving system. This system acts as an external memory to keep the latest version of user interest representations and tackle resource-consuming parts of optimization using in recurrent neural networks. Borrowing the idea from Neural Turing Machines, they designed a multi-channel user interest memory network which utilizes an attentive based GRU neural network in the design to adaptively learn the change of user interests over time.

%Although experiments in above discussed methods provide a considerable performance for user response prediction task, but RNN based methods developed based on LSTM or GRU units have a downside to only model one dimensional temporal dependency from captured user behavior data. There are higher order patterns like the seasonality and trend in time for user responses which is not addressed well. In addition, ad category hierarchies can be an important to identify correlated user responses with regard to related ads. These challenges encouraged a study \cite{OnlineChao2019} to design a model to deal with input data from both hierarchical and temporal perspectives. Considering together both temporal info and ad hierarchies, a multi-scale pyramid network was designed for modeling user behavior data. A two-folded predictive model was proposed to produce latent features from a combined version of convolutional and recurrent neural networks. For the aggregation step, they designed a proposed resolution-wise gating mechanism followed by average pooling to represent most related user purchase features from different temporal patterns and resolutions in hierarchies. The final embedding vector was fed into MLP network to generate the predicted conversion rate.

\paragraph{\underline{Convolutional neural network based methods}}
Studying to design deep learning network structures are not limited to above discussed ones, since the input space suffer from the data sparsity which makes it hard to learn directly using simple gradient descent methods. Although the deep neural network including multi-layer perceptrons in theory is known as an universal approximator which has a capacity to capture almost all non-linear feature interaction in input space, but the order of magnitude of parameters used a fully connected neural network deters to capture feature interaction in a sparse feature space and leads to over-fitting issue. This encourages to apply convolutional neural networks (CNN) which benefits from parameter sharing and pooling mechanism to work with a feasible number of parameters~\cite{FeatureGenLiu2019,ConvolutionalLiu2015,DeepEdizel2017,AttentionGao2018,ConvolutionalChan2018}. Dealing with image data along with multi field categorical data, CNN networks are used to extract non-linear latent features in the form of embedding vectors for raw pixel image data \cite{ImageGe2018,DeepChen2016,DeeplyGligorijevic2018,StructuredNiu2018}. As one of the primitive studies, authors in \cite{ConvolutionalLiu2015} conducted experiments to apply convolution filters followed by a flexible max pooling in a CNN network on two datasets including multi-field categorical data and a series of impressions in an e-commerce platform to capture neighbor patterns in input data. The downside of this method was that they applied the convolution for neighbors field feature while the feature interaction between non-neighbor fields is neglected. However, for user response prediction tasks, any order of feature fields are possible. The order of feature fields in the certain alignment in input data does not have meaningful inference like images or texts. Therefore, the other studies developed methods to take advantage of both CNN and deep multi layer perceptron to address high order and low order feature interactions. 

%A study \cite{FeatureGenLiu2019} suggested to use CNN, followed by a recombination layer, to model both local and global feature interactions. It is stacked by a deep component including a multi layer perceptron network for the click-through rate prediction. 
For news recommendation, a knowledge-aware model~\cite{DKNWang2018} proposes to use knowledge graph to represent news items, with each news article being attributed by word, contexts, and entity embeddings. For user response prediction,  CNN network, previously proposed for sentence representation learning, is used to generate the final embedding vector of user and ad features. Attentive neural network is applied to address user interest diversity to estimate click-through rate values.
%the input data including user behavior histories and targeted ad were described in the form of news items in~\cite{DKNWang2018}. Each news item incorporates a sentence of words as the title. Following knowledge graph representation modeled over news items, each news article was attributed by word, contexts and entity embeddings. With regard to user response prediction task, authors applied typical CNN network previously proposed for sentence representation learning to generate the final embedding vector of user and ad features. Attentive neural network is applied to address user interest diversity to estimate click-through rate values. %. It was fed into a deep MLP component to generate estimated click-through rate values.   

% First, using linear model like logistic regression, the interaction should be obtained from feature engineering. Thus it can be designed by experts like examples is provided in above method. Some feature interactions can be easily understood, However, most other feature interactions are hidden in data and difficult to identify beforehand. Moreover, although they are easy to detect, manually addition of pairwise feature interaction into the input feature vector is generally computationally expensive and lacks generalization capability for unseen interaction in training data. 
% So 
% \setlength{\rotFPtop}{0pt plus 1fil}
% \begin{landscape}
%%%%%%%%%%%%%%%%%%%%%%%%%%%%%%%%%%%%%%%%
% {\renewcommand{\arraystretch}{.5}
\begin{table*}[!htbp]
\setlength{\tabcolsep}{.4em}

% \begin{table*}[H]
% \begin{table*}[H]
% \begin{table*}[t]
% \small
 \tiny
\centering
\caption{Summary of selected DNN based user response prediction methods with a hybrid structure.}
% \vspace{-1\baselineskip}
\label{tab:dnn_summary}
% \begin{minipage}{\textwidth}
% \begin{threeparttable}
\setlength\tabcolsep{2pt}
\begin{tabular}{|p{.06\linewidth}|>{\raggedright}p{3.8cm}|>{\raggedright}p{0.09\linewidth}|>{\raggedright}p{1.2cm}|>{\raggedright}p{.07\linewidth}|>{\raggedright}p{2.6cm}|>{\raggedright}p{.11\linewidth}|p{.05\linewidth}|} %{\linewidth}{|L|p{4cm}|L|L|L|p{3cm}|L|L|} 
    \hline
% \textbf{{\multirow{3}{1.2cm}{\raggedright Main DNN structure}}} & \multicolumn{3}{c|}{\textbf{{\multirow{3}{3cm}{\centering System Framework}}}} & \textbf{{\multirow{3}{1.2cm}{\raggedright Method}}} & \textbf{{\multirow{3}{1.2cm}{\raggedright Features}}} & \textbf{{\multirow{3}{1.2cm}{\raggedright Examined\\advertising\\type}}}& \textbf{{\multirow{3}{1.2cm}{\raggedright Prediction Task}}} \\
% &&&&&&&\\
% \cline{2-4}
% & \textbf{{\multirow{2}{3cm}{Embedding Component}}} & \textbf{Deep Component} & \textbf{{\multirow{2}{1.2cm}{\centering Aggregation}}} &  & & & \\
%     \hline
% Advertising \\ Type
 \textbf{\multirow{3}{*}{\centering{\shortstack[l]{Main \\ DNN \\ Structure}}}} & \multicolumn{3}{c|}{\textbf{\multirow{1}{4cm}{\centering \multirow{1}{*}{System Framework}}}} & \textbf{\multirow{3}{*}{Method}} & \textbf{\multirow{3}{*}{Features}} & \textbf{\multirow{3}{*}{\shortstack[l]{Application \\ Domain}}}&
 \textbf{\multirow{3}{*}{\shortstack[l]{Predict. \\ Task}}} \\
 & \textbf{\multirow{2}{3cm}{Embedding Component}} & \textbf{Deep \newline Component} & \textbf{\multirow{2}{1.2cm}{\centering Aggregation}} &  & & & \\
     \hline
\multirow{21}{*}{MLP} &\textbf{All features}:RE{\tnote{a}}
 & MLP&
FM/Pooling Layer & DSTN
\cite{DeepOuyang2019}& User, Query, Ad, Context, 
User clicked ad history
%User unclicked ad history
  & E-commerce\tnote{b}&CTR \\%\hline
    \cline{2-8}
& \textbf{All features}: RE\tnote{a}& MLP
 &Concatenation& MA-DNN
\cite{ClickThroughOuyang2019}
& User, Query, Ad, Memorized user interest (memory link)
  &
  E-commerce\tnote{b}&CTR \\%\hline
    \cline{2-8}
& \textbf{All features}: RE\tnote{a} cascaded by FM& MLP
 &Concatenation& PNN
\cite{ProductBasedNNQu2016}
& 
User behavior ,item , context information
  &
  Display advertising&CTR \\%\hline
    \cline{2-8}    
& \textbf{All features}: RE\tnote{a}& MLP alongside FM
 &Concatenation& DeepFM
\cite{DeepFMGuo2017}
& 
User behavior ,item , context information
  &
  Display advertising&CTR \\%\hline
    \cline{2-8}        
& \textbf{All features}: RE\tnote{a}& Multiple MLP Stacking &Concatenation& $ESMM$ ($ESM^2$)
\cite{ ConversionWen2019, EntireMa2018}
& User, Item, users’ historical preference scores
  
 &E-commerce\tnote{b}&CVR \\%\hline
    \cline{2-8}
& \textbf{All features}: RE\tnote{a}& Multiple MLP Stacking
 &Concatenation &DeepMCP
\cite{RepresentationOuyang2019}
& User, Query, Ad, Context, negative ad sample features &E-commerce\tnote{b}&CTR\\ %\hline
     \cline{2-8}
& \textbf{All features}: RE\tnote{a}& MLP alongside Hadamard product layer
 &Concatenation &NCF
\cite{NeuralCFHe2017}
& The identity of Users and Items

 &Movie/Image recommendation&Ranking %\hline
 \\\cline{2-8}
 & \textbf{All features}: via Auto- feature grouping and high-order feature interaction selection  & MLP alongside FM
 &Concatenation& AutoGroup
\cite{AutoGroupLiu2020}
& 
User behavior ,item , context information
  &
  Display advertising&CTR %\hline
 \\\hline
\multirow{10}{*}{CNN} & CNN(Pre-trained CNN model using orthogonal base convolutions)& 
W\&D\cite{WideAndDeepCheng2016} & Concatenation & W\&D SSM
\cite{StructuredNiu2018}
& User behavior ,item , context information
 &Display Advertising&CTR \\\cline{2-8}
& \textbf{Image ad features}: CNN subnet
\newline\textbf{Other Ad features}: fully connected subnet
 & MLP&Batch\newline Normalization & DeepCTR
\cite{DeepChen2016}
& Image,categorical (Impression)
  &Display Advertising&CTR \\\cline{2-8}
& \textbf{All features}:RE\tnote{a} augmented to
 CNN subnet
 & MLP&Concatenation & FGCNN
\cite{FeatureGenLiu2019}
& Categorical+continuous features in display advertising
  &Display Advertising&CTR \\\cline{2-8}
% &\newline regular embedding\footref{1sttablefoot}
%  & CNN &-& CCPM
% \cite{ConvolutionalLiu2015}
% & Item(Impression), Item(Impression) sequence
%   & Display Advertising, E-commerce&CTR, CVR \\\cline{2-8} 
  & \textbf{Ad,Query features (under character-level)}: 1d CNN subnet,
Ad,Query features (word-level)
%pre-trained word embeddings& 
& MLP & Cross-convolutional pooling & DCP/DWP
\cite{DeepEdizel2017}
& Ad, Query

 &Sponsored search&CTR \\\hline
 \multirow{8}{*}{RNN}%\newline(LSTM, GRU)}
%  RNN(LSTM,GRU)

  & \textbf{Ad, context features:}RE\tnote{a}\newline
\textbf{User behavior features:} Hierarchical GRU based memory network& MLP & Concatenation &HPMN
\cite{LifelongRen2019}
& User behavior sequence,item
context info
 &E-commerce\tnote{b}&CTR \\\cline{2-8}
  & \textbf{All features}: RE\tnote{a}& MLP\newline LSTM & Concatenation &NTF
\cite{NTFWu2019}
& User behavior sequence,item, time &Recommendation&Product rate  \\\cline{2-8} 
  & \textbf{Ad, context features:}RE\tnote{a}\newline
\textbf{User behavior features:} Memory induction GRU based unit& MLP & Concatenation &MIMN
\cite{PracticePi2019}
& User behavior sequence,item
context info

 &Display advertising&CTR \\\hline
\multirow{30}{*}{\shortstack[l]{Neural \\ Attention}}    & RE\tnote{a}& Multi-head ResNet%Self-Attentive 
%ResNet
&Concatenation & AutoInt
\cite{ AutoIntSong2019}
& User profile, item attributes&Display advertising&CTR\\\cline{2-8}
& \textbf{user profile, Ad features}: RE\tnote{a} \newline
\textbf{User behavior features:} Self Attentive Bi-LSTM
 & 
MLP & Concatenation & DSIN
\cite{DSINFeng2019}
& User profile, 
User behavior sequence,
item,
  &Display advertising&CTR \\\cline{2-8}
&\textbf{User behavior (event) sequence:} RE\tnote{a}\newline
\textbf{User behavior (timestep) sequence:}
%Attentive trainable embeddings
Attentive embeddings
  followed Bi-GRU&
MLP& Attention Mechanism & DTAIN
\cite{TimeGligorijevic2019}
& Event, Timestep information
&E-commerce\tnote{b}&CTR \\\cline{2-8}  
  & \textbf{user profile, Ad, context features:}RE\tnote{a}\newline
\textbf{User behavior features:} self Attentive GRU relative to target ad& MLP& Concatenation
 & DIEN
\cite{DIENZhou2019}

& User profile, 
User behavior sequence,
item,

 &E-commerce\tnote{b}\newline Display advertising&CTR \\
    \cline{2-8}  
& \textbf{User behavior sequence features:} 
  RE\tnote{a} controlled by multi-head self-attention structure
\newline\textbf{Other features}: RE\tnote{a}
& Jointly training two MLP stacks for CTR and CVR
%– knowledge Transferring using joint optimization
 &Concatenation& PFD+MD
\cite{PrivilegedXu2019}
& User, Item, Post- click info, User Clicked/Purchased sequence,
User-item interaction statistical info
  &Display Advertising	&CVR,\newline CTR
\\\cline{2-8}
& \textbf{User profile features}: RE\tnote{a}
User behavior sequence features based on Ad image:
Pre-trained embeddings
 & 
MLP & Concatenation,\newline Sum/Max/\newline Pooling & DICM& User, Ad (image), user behavior sequence(image)
\cite{ImageGe2018}
&E-commerce\tnote{b}&CTR \\\cline{2-8}
& Pre-trained knowledge graph Word embeddings combined with, entity and context embeddings via CNN
 &
MLP & Attentive pooling,\newline Concatenation & DKN
\cite{DKNWang2018}
& User (clicked news item), News item
  &News recommendation&CTR \\\cline{2-8}
 & \textbf{Query \& ad under word-level:} 1)Pre-trained word embedding, 2) regular embedding\tnote{a}
Followed by bi-LSTM and MLP& CNN \newline MLP& %Attentive pooling, Query-Ad tensor matching & DSM
Pooling, Query-Ad tensor matching & DSM
\cite{DeeplyGligorijevic2018}
& Ad(title, URL, description), query words
  &Sponsored search&CTR \\ 
 \hline
\end{tabular}
% \begin{tablenotes}[flushleft]\footnotesize
% \renewcommand{\TPTtagStyle}[1]{\makebox[.6em][l]{#1}}
%%%%%%%%%%%%%%%%%%%%%%%%%%%
\begin{tabular}{l}
\tnote{a} Regular Embedding using trainable look-up table parameters
(matrix embedding per feature) following the structure shown in Figure \ref{fig:embedding}\\
\tnote{b} In the e-commerce scenarios, the prediction task is defined as the the probability that user clicks or makes an conversion on the recommended items(ads) 
\end{tabular}
%%%%%%%%%%%%%%%%%%%%%%%%%%
% \end{tablenotes}
% \end{threeparttable}
% \end{minipage}
%\label{tab:dnnsummary}
% \end{table*} 

% \label{}

\end{table*}
% }
%%%%%%%%%%%%%%%%%%%%%%%%%%%%%%%%%%%%%%%%
% \vspace{1.5cm}
% \end{landscape}
% \end{adjustbox*}

% \begin{table*}[t]
%  \footnotesize
%     \centering
% \caption{Summary of selected DNN based user response prediction methods. (Continued)}
% \label{tab:datasummary}
% \begin{minipage}{\textwidth}
% % \begin{tabularx}{\linewidth}{|L|L|L|L|L|L|L|L|} 
% \begin{tabularx}{\linewidth}{|L|p{4cm}|L|L|L|p{3cm}|L|L|} 
%     \hline
% \textbf{Main DNN structure} & \multicolumn{3}{c|}{\textbf{System Framework}} & \textbf{Method} & \textbf{Features} & \textbf{Experimented Advertising type}& \textbf{Task} \\
% & \textbf{Embedding Comp.} & \textbf{Deep Comp.} & \textbf{Aggregation} &  & & & \\
%     \hline
% CNN(Cont.)& \textbf{Query and ad under word-level:}1)Pre-trained word embedding,
% \newline 2) regular embedding\footref{1sttablefoot}
% Followed by bi-LSTM and MLP& CNN \newline MLP& Attentive pooling, Query-Ad tensor matching & DSM \cite{DeeplyGligorijevic2018}
% & Ad(title, URL, description), query words
%   &Sponsored search&CTR \\\hline

% Autoencoder   & \textbf{Contextual features}:K-means Clustering and tensor dimension reduction
% ,
% \textbf{Other features}:
% \newline regular embedding\footref{1sttablefoot}& Autoencoder\newline
% FM & Attentive pooling & ASAE\cite{ NewWang2018}
% & User profile,
% Contextual features(AdID, position, and the number of advertisementson the return page), Statistical features

%  &E-commerce&CTR \\\hline

% \end{tabularx}
% \end{minipage}

% % \label{}
% \end{table*} 
\paragraph{Other methods}%{\underline{Other methods}}
In the previous section, we have provided an review on classification methods ranging from linear logistic regression based methods to advanced deep learning based methods. A few other classification based methods which may gain attention are generative adversarial network (GAN) based models \cite{AdversarialLi2020,DisguiseDeng2017}, transfer learning \cite{ImplicitZhang2016,ImprovingSu2017}, fuzzy design \cite{ImprovedJiang2018}, decision trees \cite{MultiWen2019,ClickThroughGutpa2018} and multi task learning \cite{MKRWang2019,PredictingPan2019}  approaches.

% https://quinonero.net/Publications/predicting-clicks-facebook.pdf
\subsubsection{Ensemble Based Approaches}
\begin{figure}[H]
  \centering
  \includegraphics[width=0.8\linewidth,height=4.5cm]{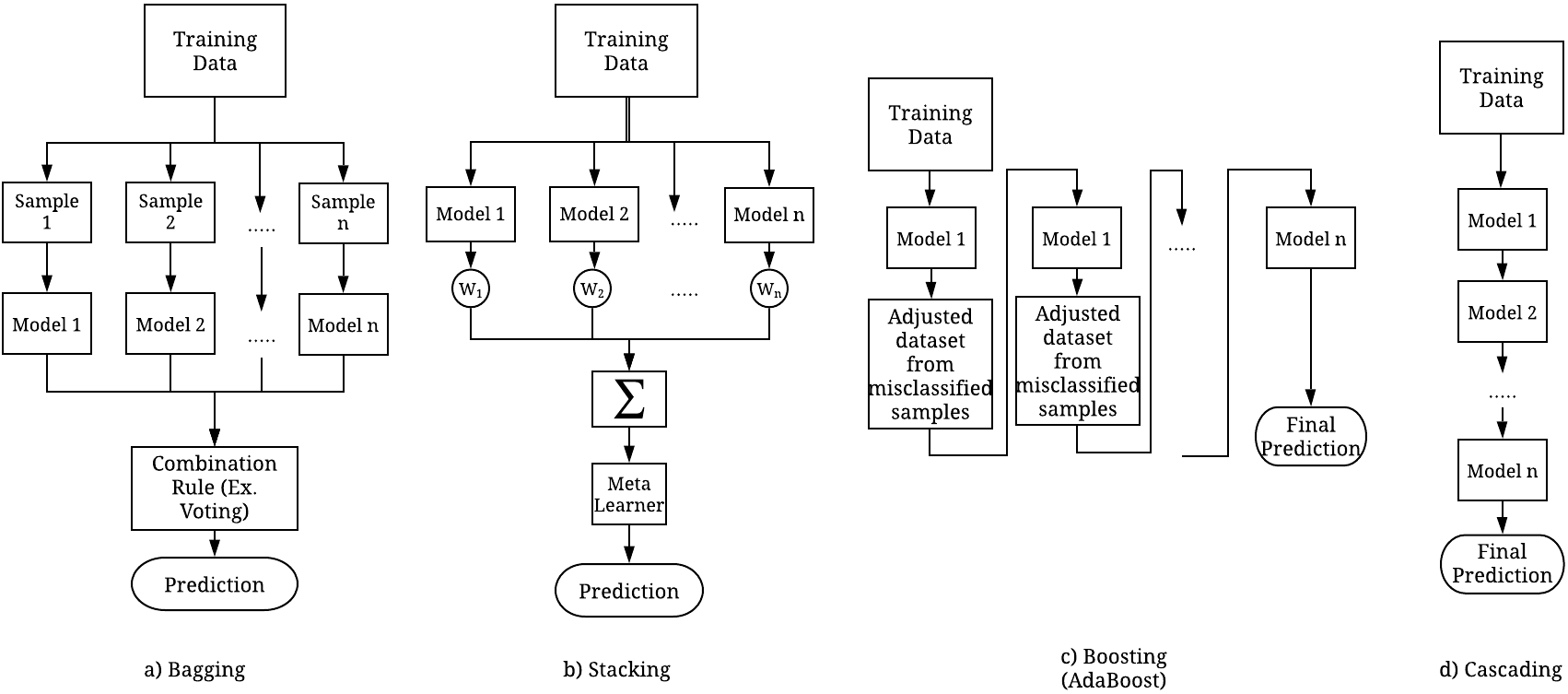}
  \vspace{-.75\baselineskip}
  \caption{Ensemble structure types: a) Bagging randomly samples training data, with replacement, to generate N subset data and train N predictor models. The final result is the combination of N classifier outputs; b) stacking: N models are trained in parallel based on the same training data. The final output is combined through a meta-classifier. This classifier is fitted on the output of base classifiers; c) Boosting(AdaBoost): a series of predictor models are trained using a subset of training data sequentially. The subsets of data are created adaptively using misclassified samples in previous model; d) Cascading: is based on concatenation of multiple classifiers}
  \label{fig:entypes}
%   \Description{The 1907 Franklin Model D roadster.}
\end{figure}

While early proposals for user response prediction mainly use linear logistic regression classifiers, which provide the simplicity along with the scalability, modern approaches are developed to address non-linear interactions in data using methods likes factorization machines, generalized version of decision trees, and neural networks. Some studies show that using a single machine learning method may lead to non-optimal results, and propose a new aspect of development to design a model structured from an ensemble of machine learning models. These models can bring more improvement in the level of accuracy for the prediction task. Generally, the design of ensemble models are mainly categorized into four sections like Bagging and Boosting, Stacked Generalization and Cascading shown in Figure \ref{fig:entypes} ~\cite{IntroAlpaydin2010}. 

The combination of different classification methods in the form of ensemble structures are utilized in different studies. %The cascading version of ensemble models can be considered as a hybrid model to combine two or more learners. 
%Among them 
The study in \cite{FNNZhang2016} followed a cascading version of an ensemble model which includes two learners. They investigated the performance of combining factorization machines with a fully connected neural network to predict CTR values for the digital advertising. Because of the data sparsity in the categorical input data, the feature interaction cannot be easily detected directly using deep neural networks which generally lead to the overfitting issue. They propose to cascade factorization machines to a deep neural network in order to address this issue.  

In the context of e-commerce websites, \cite{EnsembleAryafar2017} suggested multi-modal ensemble learning to consider texts and images of posts as different modalities. They separately built a logistic regression model for historical CTR values and another model for embedding vectors of images and textual information. Following the multi-modal learning approach, the stacking ensemble model is used to combine linearly their results by passing to the final logistic regression classifier.

\begin{figure}[h]
  \centering
  \includegraphics[width=.55\linewidth]{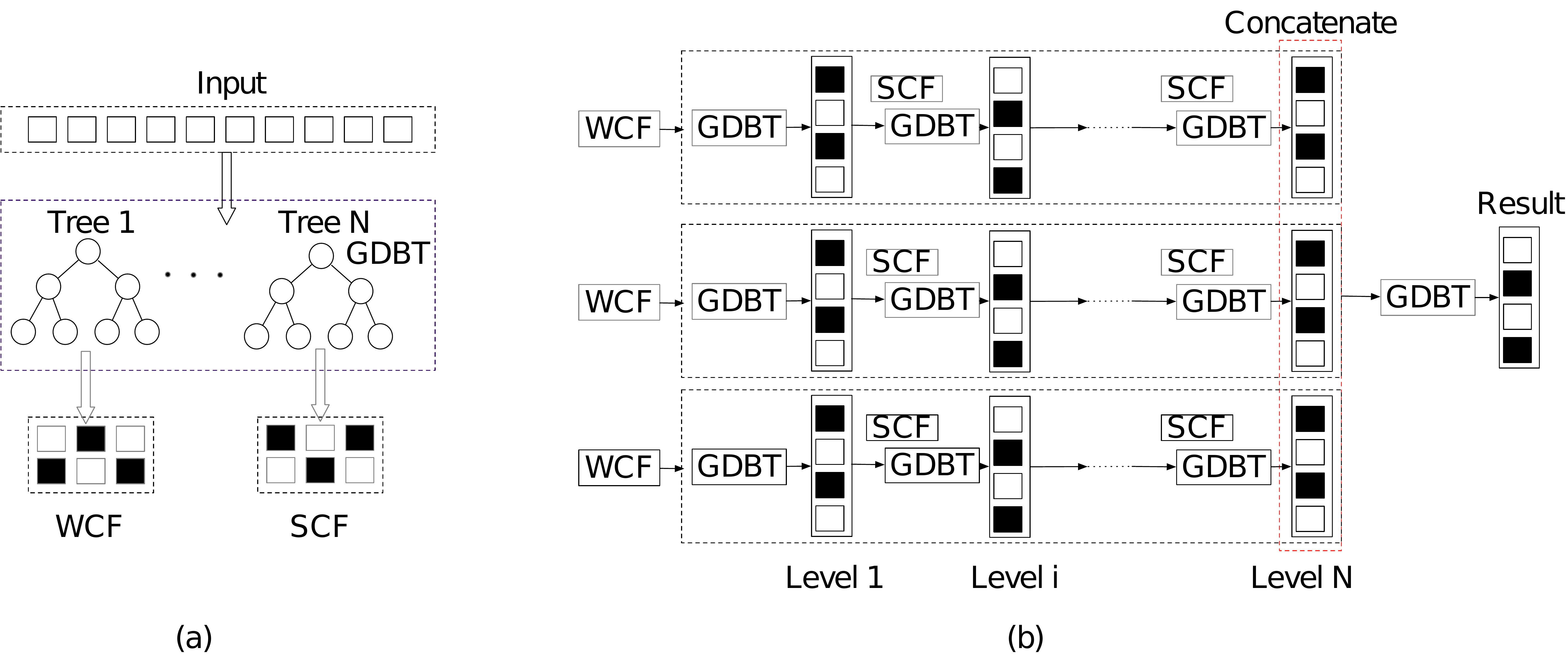}
  \vspace{-1\baselineskip}
  \caption{Ensemble GBDT model: a) Input features randomly selected from the feature pool fed into GDBT model. Their importance for prediction is evaluated based on their correlation to class labels by traversing from the root node into leaf in a tree. They are sorted into two categories according to the explanatory power into WCF(Weak correlation features) and SCF(Strong correlation features) b)The structure of proposed multi-level deep cascade trees including stacking multiple sequence of GDBTs}
  \label{fig:entree}
%   \Description{The 1907 Franklin Model D roadster.}
\end{figure}

In another study, authors in \cite{MultiWen2019} propose to develop an ensemble model for conversion rate prediction which is mainly based on GDBT learners. Following the Cascading and stacking techniques, they used multi-level cascade of GBDT models to extract features which are coming for values received from the previous model. To improve the diversity of extracted features, multiple cascade of decision trees are aggregated like Figure \ref{fig:entree} through the concatenation to be passed to the conclusive GDBT to generate the final features for the classification. As a part of the contribution of this work, in order to improve the prediction performance, the importance of input features is also considered. They use a separate GBDT model to pre-process the input raw features and generate two class of features that have weak and strong correlations (WCF and SCF) with regard to the prediction result. These class of features are used relatively as the input to train the model.
% \vspace{-.5cm}

In literature, some works also comparatively study the effect of ensemble techniques~\cite{EnsembleLing2017,EnsemebleKing2015}. Following the ensemble techniques introduced in the beginning part and the goal to improve click-through rate values, the study in \cite{EnsembleLing2017} examined two ensemble techniques Boosting and Cascading with GBDT, LR and a fully connected deep neural network classifiers. They compared the performance of corresponding single learners with the cascaded and boosted version of pair of models in the click-through rate prediction for the sponsored search advertising. %They used three source of features from query, ad, and user with a format of multi-field and statistic numerical features to form a training set. 
For the sponsored search advertising application again, the study in \cite{EnsemebleKing2015} made a comparison between the effect of four different structures of ensemble learning such as the majority voting, bagging, boosting and stacking for pay-per-click classification. The features are selected from 
different information sources like the attributes describing ad impression,  click-through rate value, conversion rate value, and the position of ads in addition to the textual features captured from the title and the body of ads and campaign categories. They are joined together to train ensemble learners such as Naïve Bayes, Logistic Regression, Decision tree and SVM to estimate the pay-per-click value of campaigns.

\subsection{Unsupervised and Semi-supervised Approaches}
In this section, we review two categories of methods in the literature that do not fully rely on labeled data. In this case, the predictive models are designed based on the implicit and explicit pattern in data. Semi-supervised models refers to approaches like graph neural network based models that involve designing a user feedback estimation model using both labeled and un-labeled sampled data. Two categories of these methods are represented in the following sub-sections.
\subsubsection{Clustering Based Approaches}
%Many studies consider user response prediction for online advertising as a supervised prediction problem to deal with issues like the data sparsity and class imbalance problem in the input feature space which are mostly based on categorical data. However, 
Clustering methods have also been investigated in the literature for online advertising. As an unsupervised approach, clustering involves grouping sample data into related clusters based on similarity among data points. 

%The increasing need to group categorical data into groups led to many studies to develop new clustering methods, since traditional clustering approach can only handle the input with numerical features and cannot be directly applied to the categorical data. This originates from the point that similarity of different clusters cannot be well understood for categorical data. There are 
Some studies develop statistical clustering methods categorical data in different contexts, such as $k$-modes \cite{KmodesSharma2015} as an alternative to the popular $k$-means method which uses hamming distance as the distance metric, COOLCAT made by \cite{COOLCATBarbara2002} uses the notion of entropy for the similarity metric, or CLOPE clustering approach in \cite{CLOPEYang2002} research which develops a scalable method to leverage the trend of a height-width ratio of the cluster histogram as the similarity criterion. But these categorical clustering methods are not well studied in the online advertising with the multi-field categorical data.

On the other hand, some clustering based studies proposed the idea to improve classification based methods. As the initial study in this category, to supplement the logistic regression \cite{SimpleChapelle2015} and gradient boosting decision trees (GBDT) \cite{ImprovingHillard2010} methods for user response prediction task , authors in \cite{PredictingRegelson2006} propose to use feedback features which are prepared from historical user behaviors. Considering advertiser and publisher come from with different hierarchical granularity, they incorporated the combination of publisher page and advertisement along with user-publisher-creative which are created based on the hierarchical structure of user as the new features. The extra features are quantized using the $k$-means clustering to be added to input features for training.

%In this regard, understanding a user historical behavior is the main goal in many clustering methods. 
Some methods~\cite{PredictingRegelson2006,PersonalisingReps2014,ClickWang2011} organize the user input, such as keywords used in search engine or pages visited by users, by using clustering to reduce the severity of above mentioned issues and improve the correlation with user responses. In \cite{PredictingRegelson2006}, the authors suggest that since a different click response probability can be assumed for different query keywords, the topic of ads and query keywords can be used to organize data into clusters where more closely related terms have more similar click-through rate values. They propose to generate groups of terms using hierarchical clustering and keyword-advertiser matrix. The similarity of samples intra and between clusters are evaluated by the textual similarity of terms in ads. Therefore, assuming the fixed CTR value for clusters, the estimated value of click-through rate for new samples is determined by the nearest neighbor clusters.

\subsubsection{Network Based Framework} \label{sec:network_based}
With development on the Internet, information networks are the common element of online businesses. In this subsection, we will go over approaches addressing the network structure in the input data to develop predicting models for the user response prediction. 
\paragraph{Graph embedding based methods}
Recent years have seen a lot of studies which focused on the application of network representation learning methods for recommender systems and the user response prediction. Motivated by the success of CNN and RNN, there has been an interest in developing neural network based models for the graph structured data. Considering three major challenges in recommender systems, scalability, data sparsity, and cold start, many methods have been proposed in the literature using graph embedding \cite{BillionWang2018} and graph neural network~\cite{TripartiteKim2019,FiGNNLi2019,MetapathFan2019}.

% \textcolor{red}{suggest to add a figure to show how graph/network is used to model data. For example, you may consider Figure 2 for user behavior embedding: https://arxiv.org/pdf/1803.02349.pdf}
\begin{figure*}[htb!]
  \centering
  \includegraphics[width=.72\linewidth]{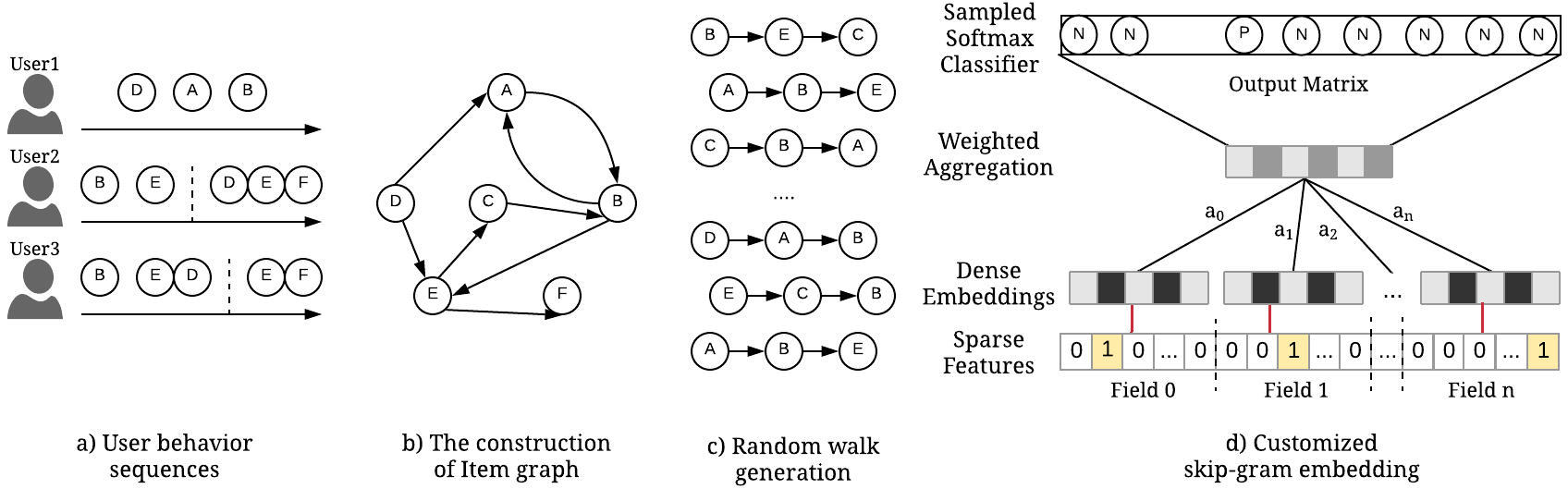}
%   \vspace{-.75\baselineskip}
  \caption{The  billion-scale commodity embedding proposed by \cite{BillionWang2018} for E-commerce recommendation in Alibaba: a) user behavior sequences including items(Ads) visited in one or more session specified by dash lines. b) item(ad) graph generated user behavior sequences that a direct edge representing two subsequent items(ad) in user history. c) generating sequence of nodes using random walk in graph following Deepwalk method. d) Proposed graph embedding algorithm to use side information to reduce data sparsity. Field 0 specifies a node in random walk and field 1 to field $n$ are the one-hot-encoded vectors for side information corresponding to the ad node in the graph. Hidden layer is a weighted average aggregation of dense embedding vectors}
  \label{fig:graph_embed}
%   \Description{The 1907 Franklin Model D roadster.}
\end{figure*}
Authors in \cite{BillionWang2018} designed a recommender model based on the graph embedding which takes advantage of side information to cope with the three challenges. The model includes two sections such as matching and ranking. Focusing on the first section with the network of users interacting with items in an e-commerce website, they applied DeepWalk \cite{DeepWalkPerozzi2014} to generate embedding vectors of items in the directed graph of items formed using user online behavioral history. Because of the data sparsity, there is a lack in the number of interaction in the graph. Therefore, as the part of the contribution, side information such as the price value, shop, category and brand, are included as the one-hot encoded vectors in network representation learning procedure. %The embedding vectors of items are updated from the linear weighted combination of the side information vectors and their embedding vectors in the training step.
Graph Neural Networks (GNN) are known as one the most effective solutions to develop predictive models for network based data. Extended from recurrent neural networks and convolutional neural networks, GNNs have a unique capability to generalize neural networks to cope with directed and un-directed input graphs, by using an iterative process to propagate node status over their neighborhoods. After optimization, it can provide an embedding output of nodes in the graph in which the feature information are aggregated using their neighbor nodes. Recently, GNN based methods have received more attention for online advertising and recommendation systems applications \cite{TripartiteKim2019,FiGNNLi2019,MetapathFan2019,NGCFWang2019,HetGANWang2019,HetGNNZhang2019}. Like deep learning based models in Table \ref{tab:dnn_summary}, Embedding and Interaction components are two major components in the design of these predictive models in the literature. The embedding component is dedicated to map information associated to node and edges in the graph-structured data into numeric vectors. However, the interaction component tries to make the reconstruction of user feedback in the system. This part can be designed by different ways through a MLP component in deep learning based models \cite{HetGANWang2019} or inner-product between embedding vectors \cite{NGCFWang2019} or element-wise multiplication of embeddings \cite{HetGNNZhang2019} to represent collaboration between users and items to rank and predict user preferences in recommendation and online advertising systems. The embedding component is developed using recent advance in graph neural networks. Authors in \cite{NGCFWang2019,LightGCNHe2020} propose an propagation layer in their embedding component to refine embedding vectors via aggregation of embedding vectors of neighboring nodes in the graph. The model in \cite{NGCFWang2019} is built using the message-passing architecture and defining a model via layer-wise message constructing and aggregating operations.  %Authors in \cite{TripartiteKim2019} discussed issues to apply GNNs to social recommendation tasks. They pointed out to the cold-start problem, complex noisy connections and high scale heterogeneity in addition to the over-smoothing where the structural information is ignored and embedding vectors of different nodes are not separate and recognizable. Although the extra network of user-user interactions can alleviate the cold-start problem in recommender systems, but it is not tractable in the social media scale recommendation. In this study, the concept of group is introduced in order to have two bipartite graphs of group-user and user-items network. To address the over-smoothing issue, they proposed an embedding approach that uses the personalized PageRank idea for the propagation. They applied this embedding separately for the group-user graph and user-item graph. Node embeddings from each graph are aggregated at the end of using attention mechanism to generate embedding vector of users and items. The conclusive node in the model computes predicted CTR value through dot-product operation on the user and item embedding vectors. 
A recent study \cite{FiGNNLi2019} addresses the limitation of methods like DeepFM~\cite{DeepFMGuo2017}, by considering a graph structure for feature fields in the input data in which nodes corresponding to feature fields interact with others through weighted edges to reflect the importance of field interactions. A GNN based model is developed to model complex interaction between input field features. In this model, field feature as the nodes in graph are attributed by hidden state vector which is updated using a recurrent approach. An interaction step parameter is defined to consider higher and lower interactions between nodes and their neighbor nodes which are located in one or more hops away. The ending point of the model includes an attention layer to predict CTR values. 
One essential challenge of network embedding for the user response prediction is that the embedding learning might not be directly optimized towards the underlying user response prediction. %Regarding the graph embedding methods, experiments in different studies demonstrate that they could achieve considerable improvement. 
Following an unsupervised learning, nodes are represented using embedding vectors, however, they may not be optimized for downstream tasks like the click-through rate prediction. This issue can be considered as the bottleneck to improve the task. Therefore, the user intention modeling is considered as an alternative~\cite{DINZhou2018,DIENZhou2019}.
% Considering some challenges in
Considering the sequence of the user behavior from the interaction between users and ads in the user intent modeling still have some challenges like the data sparsity and weak generalization. A study \cite{GINLi2019} showed that sequence of user behavior can be organized as a co-occurrence commodity graph with node representing clicked commodities and weighted edges describing number of co-occurrence times. To address the sparsity problem, a multi-layered neighbor diffusion is performed on the commodity graph. The preceding result is combined with using an attention layer to generate user intention features. These features combine with other ones, such as user profiles, query keywords, and context in fully connected network, for the click-through prediction.

\paragraph{Knowledge graph based methods}
Knowledge graphs (KG) are semantic heterogeneous networks including a collection of entities with attributes that are inter-connected together through edges. %Here relations corresponding to edges may have different types and functionalities. 
They are usually described through a triplet with relation connecting head and tail entities like (Head, Relation, Tail). This structure of data has been studied for different applications like link prediction \cite{TemporalDunlavy2011} and Web search analysis \cite{WebAnalysisTorres2014}. The advantages of the knowledge graph for applications suffering from data sparsity and cold start problems, such as recommendation systems, have been observed from different perspectives. First, networks can provide additional semantic relationship information to improve the recommendation performance. Moreover, the diversity of information in the knowledge graph can extend the information for matching with user interests. In addition, the historical information linked to items in recommender systems can provide implication ability for the system. In summary, we separate knowledge graph based methods for recommender systems and online advertising into three categories: embedding-based, path-based, and hybrid methods.

\subparagraph{\textit{\underline{Network embedding methods}}}
This category of methods %consist of embedding methods for the knowledge graphs. They use a specified random walk to capture the structural and topological information~\cite{KnowledgeWang2017}. The main idea is 
aim to map the components of the knowledge graph, including entities and relations, into low-dimensional embedding space to preserve network structural information~\cite{KnowledgeWang2017}. Some jointly incorporate heterogeneous attributes and content that are assigned to nodes in the graph for modeling~\cite{HetGNNZhang2019}. %The application of this idea for recommendation systems have been studied in several papers.

%For example, authors in \cite{DKNWang2018} studied knowledge graph representation for news recommendation. They proposed a network embedding method to address the challenge of topic and time sensitivity for the news items selected and visited by users. 
For example, knowledge graph based representation for news recommendation~\cite{DKNWang2018} has been studied to address the challenge of topic and time sensitivity for news items selected and visited by users. It means that users generally visit selected news at a short specific of time which may not happen later. In addition, news content usually has brief words and diverse topics. %To handle these challenges, they provide a deep neural network which basically takes advantage of a customized CNN module as the key component. This component is designed to model user interest through multiple channels which consider both word semantics information and corresponding information from a generated knowledge graph data. 
To handle these challenges, a deep neural network is proposed to take advantage of a customized CNN module as the key component to model user interest through multiple channels which consider both word semantics information and corresponding information from a generated knowledge graph data. This leads to generating three categories of embedding vectors for words in the body of news, the associated entity and immediate neighbors in knowledge graph. In the design, an attention mechanism is used to aggregate embedding vectors of user behavior sequences. They avoid using concatenation strategy for the aggregation in this step since entity and word embeddings may have different dimension generated from different contexts. The output are fed into a fully connected neural network to learn the probability of user's click for a selected news piece. Likewise, a study~\cite{HetGNNZhang2019} presents a heterogeneous graph neural network model which adopts the aggregation of feature information with regard to a sampled neighboring nodes. %As part of contribution, %a procedure to sample neighbor nodes
A node sampling procedure is suggested to aggregate selected neighbors grouped by their types and their frequency in a designed random walks. %This data is combined with information encoded from heterogeneous contents associated to nodes in a heterogeneous graph-structured data. 
Using attention mechanism, the content embedding of neighbor nodes with the same type are first aggregated. It is then followed by another attention round to aggregate embedding vectors of neighbor nodes from different types in the graph. They train embeddings using heterogeneous skip-gram learning. To compare the performance of proposed model, element-wise multiplication and inner-product operations of user and items embedding vectors are used to simulate user response for link prediction and recommendation experiments.% and generate samples for recommendation experiments based on element-wise multiplication of trained embeddings to simulate user responses.
%by taking the embedding vectors of the news and the history of user’s clicked news as the input through a fully connected network.
%For example, authors \cite{DKNWang2018} centered their study on knowledge graph representation called DKN for news recommendation. Generally, news recommendation is quite difficult as it poses three major challenges. First, unlike other items such as movies [9]news articles are highly time-sensitive and their relevance expires quickly within a short period.  Second, people are topic-sensitive in news reading as they are usually interested in multiple specific news categories  Third,news language is usually highly condensed and comprised of a large amount of knowledge entities and common sense. Addressing these challenges, a content-based model for click-through rate (CTR) prediction is proposed which takes one piece of candidate news and one user’s click history as input, and outputs the probability of the user clicking the news. A CNN module is designed to jointly learn from semantic-level and knowledge-level representations of news. The multiple channels and alignment of words and entities enable the method to combine information from heterogeneous sources and maintain the correspondence of different embeddings for each word. To model the different impacts of a user’s diverse historical interests on current candidate news, DKN uses an attention module to dynamically calculate a user’s aggregated historical representation. 
\subparagraph{\textit{\underline{Meta-path based methods}}} 
This category contains knowledge graph embedding methods which employ meta-path schemes as the guideline to generate random walks and in turn embedding vectors. Although many studies in the category have shown a decent performance for recommender systems \cite{MetapathFan2019,SHNEZhang2019}, current methods heavily rely on manually building random walk corpus for further processes. The selection of meta-path schemes are generally considered as the hyper-parameters set differently by researchers in experiments. So this can be an issue in practice. To tackling this problem, attention mechanism has been employed in recent studies. Authors in \cite{HetGANWang2019} design a heterogeneous graph neural network %that leverages attention mechanism 
to automatically address the effect of different neighboring nodes and meta-paths using two-level attention layers. In the first level, node-level attention is applied to train the weights for meta-path guided neighbors of each node in the graph. It is then fed to a semantic attention step to calculate weighted combination of different meta-paths for the node embeddings. The predicted interaction between different node types in heterogeneous graph is modelled through training a fully connected neural network at the end. %The category of methods explore various patterns of connections among items in a KG (a.k.a meta-path or meta-graph) to provide additional guidance for recommendations. Path-based methods make use of KGs in a more intuitive way, but they rely heavily on manually designed meta-paths/meta-graphs, which are hard to tune in practice. 

\subparagraph{\textit{\underline{Other knowledge based methods}}} 
In this section, we present hybrid knowledge based methods %which combine above two categories 
which learn user/item embeddings by exploiting structural information in the knowledge graphs~\cite{Knowledge-AwareWang2019,End-to-EndQu2019}. Recently, a study \cite{Knowledge-AwareWang2019} discusses the extension of GNN method made for a knowledge graph where the edge weights between user and item nodes are not available beforehand. So a personalized scoring function is proposed for training to determine the edge weights %are determined by training  
via a supervised approach following a relational heterogeneity principle in the knowledge graph. To address the data sparsity issue in recommendation systems, a leave-one-out loss function is used as a label smoothness regularization to calculate predicted weight values. It leads to calculating node embeddings through aggregating node's feature information over the local neighborhood of the item node with different weights.

%Recently, %research~\cite{End-to-EndQu2019} investigates %that knowledge graph embedding methods provides side information to tackle the data sparsity and cold start problems in recommender systems, and identify
A common approach to model user response in knowledge graph based methods is to apply aggregation mechanism to combine embedding vectors of user and items entities via average pooling or attention units over their neighbors. Authors in ~\cite{End-to-EndQu2019} consider this as early summarization problem. %common weakness of modeling user response prediction using user/item embedding vectors has been investigated in~\cite{End-to-EndQu2019}. It is named as an early summarization problem. The authors elaborate the problem regarding the method used in current knowledge graph-based recommendation methods to create embedding vectors of user and items from applying aggregation mechanism (average pooling or attention units) over their neighbors. 
They argue that modeling user response using the inner product of embedding vectors of user and item can have a limitation for user response prediction. Accordingly, a neighborhood interaction model is proposed to integrate a higher order neighbor-neighbor interactions through %a graph neural network. This model takes advantage of 
a bi-attention network in the aggregation step to improve user click through rate prediction.

\subsection{Stream Based Framework}
Online advertising is essentially a streaming platform, where users, auctions, and ads are continuously and dynamically changing~\cite{DisplayWang2016}. In this context, data stream refers to continuous feeds of news and information generated by users in an interactive way~\cite{leong:2014:Advertising}. Social media platforms are examples of these systems in which millions of users generate data continuously being uploaded. The stream environment provides an opportunity to emerge in-stream advertising with commercials in stream of data. Also known as native advertising, in-stream ads look similar to regular feeds. They are differentiated by an assigned tag indicating a commercial target or the content of feed.

%In this context, using the same terminology, displayed ads within data stream are known as ad creatives. They receive user responses in the form of the click events to direct users to promoting websites. 
The performance of advertising strategies for stream data has been studied from different aspects according to the condition and policies in online platforms. Click-through rate value is not only a metric to evaluate user experiences. Many studies have developed methods to address pre-click and post-click user experiences~\cite{PromotingLalmas2015,ImprovingBarbieri2016,PredictingZhou2016}. From a different perspective, user response prediction was cast to evaluate ad quality. The high rate of quality value is considered as the positive influence for users to use the platform more and produce even more click responses for the long run. There are some studies experimenting a model to address the impact of ads quality for predicting user response and user engagement, based on in-app advertising such as Yahoo Gemini platform~\cite{PromotingLalmas2015,ImprovingBarbieri2016}. To this end, a post-click experience instead of click-through rate value is used to evaluate the user experience on the landing page of advertising web-sites. %Post-click experience is usually described using different metrics to evaluate the user engagement like 
Post-click experience is attributed by metrics like dwell time and the bounce rate. The former measures the spending time in the landing page where the latter indicates the percentage of short and momentary dwell times. The level of user engagement with ads is considered to have a natural connection to the time length users spend in landing websites. %Lack of interactive features like query keywords being available in sponsored search advertising, longer stay is connected to higher%the higher the length of staying in considered to give more chance to see further activities regarding ads conversion. 

Aside from ad quality metric, in the context of social media, CTR prediction for stream data in Twitter is first studied \cite{TwitterLi2015}, where positive use responses are defined as retweet, reply, and actual click on promoted tweets. %proposes to study stream data in Twitter for the click-through rate prediction. %This platform is equipped with different capabilities to circulate commercials among users. %The facilities are seen as an invite to visit user accounts,  hashtag\footnote{Hashtags are prefixed expressions using symbol $\#$ to be used for marking specific topic in Tweeter} or promoted tweets in the stream of daily tweets. 
%To calculate the prediction of click-through rate values, authors expand the definition of click event to three forms of positive responses like retweet, reply, and actual click on promoted tweets. 
They also use a dismiss feature in Twitter to identify %The additional feature in this platform is the ability to dismiss tweets which can provide the 
explicit negative instances for the analysis. According to the fact that the number of spots dedicated for promoted tweets are limited,  
%Considering the limit in the number of dedicated spots for promoted tweets among the stream of tweets, 
in this study a learning-to-rank method with a calibration mechanism is proposed to combine traditional classification with pair-wise learning to address data sparsity and scalability issues.
%and the rank of tweets with a favor of high scalability. %As the contribution to consider the stream of input, 
They formalize two problems of classification and ranking in the framework. 

In an alternative work~\cite{Real-timeDiaz2012}, time-sensitivity of streaming data in Twitter and the short memory issue for online learning are studied to exclude obsolete tweets from being considered. Therefore, authors propose to analyze hashtags\footnote{Hashtags are prefixed expressions using the symbol of $\#$ to be used for marking a specific topic in Twitter} in social media as the indicators of user interests to provide a personalized ranking of topics. They present an online collaborative filtering method following pairwise ranking approach for matrix factorization (Stream Ranking Matrix Factorization), and propose a pairwise learning to optimize an ordinal loss and a selective negative sampling based on a selective active learning, %and using a pre-defined small buffer, 
using three objective losses, including hinge loss, SVM, and RankSVM %are examined 
for training.

Recently, authors in \cite{AddressingKtena2019} have centered their work on delayed positive feedback at stream media to study the effect of two factors, such as the trend and seasonality, in online advertising. In live streams, the predicting models are dealt with the cold start issue. This is because in online real-time scenarios, fresh data lack enough label information and the few appearance of the positive response of users, which leads to the underestimation %does not fix the underestimation 
of CTR values. They conduct experiments to estimate %In the problem of  %estimation of 
CTR values for video ads in Twitter platform, and examine predicting models with logistic regression and Wide\& Deep~\cite{WideAndDeepCheng2016} models using five loss function designed for delayed positive samples to identify the combination of learners and loss functions for continuous stream data. 

%To monitor the ad quality, some online platforms put in place an ad feedback mechanism to give users a chance to provide the negative feedback on the ads serve \cite{PredictingZhou2016}. With ability to express negative feedback by dismissing an ad in many online platform, in the study conducted by \cite{PredictingZhou2016}, the quality of ads in native advertising was discussed from a different aspect called pre-click experience. They indicated that the study of correlation between the rate of offensive feedback and click-through rate values does not show an explicit relation. Therefore, the offensive feedback rate was considered as the main learning target. A feature engineering approach through the crowd sourcing is used to select the combination of relevant features like the page rank describing trustworthiness of ads, visual features about text and image of ads. They are complimented by historical user behavior features regarding landing page like the dwell time and bounce rate. To predict the offensiveness of ads, a learning framework based on Logistic Regression is developed for experiments. 

\subsection{Summary}
%In this section, we try to summarize the studied approaches and provide the main ideas along with merit and demerit of different type of methods. 
%Predicting user response in online advertising and recommender system is essential problem to assess the performance and prosperity of systems. That is why in last decade different solutions have been studied by the research community in this field.
To summarize different framework covered in above sections, Table \ref{tbl:adv_disadv} outlines main learning strategies used by different methods. %It then compares methods based on their related positive and negative points. 
In Table \ref{tbl:model_InOut_comparison}, we also outline studied methods from a different aspects, including feature engineering, downstream tasks used for the evaluation of models, and domain applications. Recent years have witnessed a significant growth in networking technologies and a larger number of online users across the world. As a result, scalability is a major challenge for recommender and online advertising. %In the literature, we are mainly focused on models to predict potential set of candidate advertisements or promoted products evaluated using offline metrics in an isolated environment. But the perspective can be different when it comes to deploy these complex models for online service provider systems. 
In Table \ref{tbl:industry_summary} we overview different efforts made to provide technical solutions for user response prediction in real world applications. Comparing with academic scale solutions, models deployed for production system need massive resources to store and execute internal processes. %So they are different in terms of three constraints such as application-level requirements, memory and computation complexities. 
To address these requirements, industry attempts to devise paralleled model and data architectures that data can be processed with high throughput and remarkably low latency. Recently some work \cite{MLPrefMattson2019,DeepNaumov2019} focus on developing  benchmark framework suites to provide adequate flexibility along with good test results to make fair comparisons between academic and industrial models. In Appendix \ref{future_sec}, we also outline some potential directions for future studies.%However, there is still need to have more open-source tools and workloads to facilitate research in the domain of high-scale recommendation and online systems.
%Although, the current studies around developing commercial recommendation systems and online advertising established a solid background in this field, we noticed that there are still some open issues. 

%In the following, we outline some potential directions which can drive the future studies.
%%%%%%%%%%%%%%%%%
%future directions
%%%%%%%%%%%%%%%%%

%with different multiple business companies working on campaigns to get competitive profits. So this may lead to gradual shift in user preference

\begin{table*}[h]
\tiny
\centering
\caption{Overview of main ideas of user response prediction methods along with pros and cons}
\label{tbl:adv_disadv}
% \arrayrulecolor{black}
\begin{tabular}{|>{\raggedright}p{0.12\textwidth}|>{\raggedright}p{0.07\textwidth}|p{0.4\textwidth}|p{0.3\textwidth}|} 
\hline
\textbf{Learning Strategy}                       & \textbf{Algorithm}                    & \textbf{Advantages}                                                                   & \textbf{Disadvantages}                                                                                                                                   \\ 
\hline
Data Hierarchy Analysis                                        & Clustering based      &+ Using clusters as an auxiliary information for samples with insufficient observations& - May have a variation in user response rates  \\ 
\hline
\multirow{4}{*}{Matrix Factorization}                                        & Collaborative Filtering      &+ Good scalability along with simplicity\newline~  + It can provide robust performance against sparse data                                                                        &-Explore all historical data\par
-Weak on anonymous behaviour sequences\par - No promised performance in case of lack of user info due to privacy issues  \\ 
\hline
\multirow{2}{*}{\shortstack[l]{Training a classifier }}                                          & LR                           & + Scalability                                                                  & - Needs feature engineering                                                                                                                       \\ 
\cline{2-4}
                                        & FM based                           &                 + Have a closed form equation that can be calculated in a linear time                                                          & - Limited to model 2nd order feature interactions%; higher learning time than linear models                                                                                                  \\ 
                                        \\
\hline
\multirow{10}{*}{\shortstack[l]{Feature Learning +\\ Training a classifier }}  & DNN based & +  end-to-end interface with representation learning and non-linear transformation\newline
+ High flexibility using a modular implementation via open-source frameworks&- Interpretability\par
- Prone to over-fitting due to requirement of large amount of input data\newline - Hyper-parameter tuning issue                                                         \\ 
\cline{2-4}
                                        & RNN based                    & + Can learn from sequential data with variable lengths \newline + Robust performance with regard to data sparsity                                                                             & - Rely on linear sequential structure;\par 
                                - Hard to take full advantages of GPU/TPU computing architectures;\par - Long training time   \\ 
\cline{2-4}
& GNN based~& + Addressing the network structure in the input data\newline to aggregate feature information of neighboring nodes \newline
+ Joining with attention mechanism to provide good interpretability &- A model trained cannot be directly applied to an input graph with different structure \par- Computational cost                                                                                                                                           \\ 
\hline
\multicolumn{2}{|l|}{\multirow{3}{*}{Stream based  framework}}                           & + Adjust prediction in user preferences over time \newline
                        + Joining with external memory network for increment updates
                        \newline + Reservoir technique to use more samples to update the model  
                        & - Unpractical to stack up the training data for modeling                                                                                   \\
\hline
\end{tabular}
% \arrayrulecolor{black}
\end{table*}
\begin{table*}[h]
\tiny
\centering
\caption{Comparing user response prediction methods in terms of feature characteristics, application domains, and downstream tasks.}
\label{tbl:model_InOut_comparison}
\begin{tabular}{|>{\raggedright}p{.06\linewidth}|>{\raggedright}p{0.1\textwidth}@{\hskip1pt}|p{0.09\textwidth}@{\hskip1pt}|p{0.085\textwidth}@{\hskip1pt}|p{0.07\textwidth}@{\hskip1pt}|p{0.47\textwidth}|} 
% \begin{tabular}{|>{\raggedright}p{.12\linewidth}|p{0.1\textwidth}|p{0.1\textwidth}|p{0.08\textwidth}|p{0.5\textwidth}|p{0.5\textwidth}|} 
\hline
\multicolumn{2}{|l|}{\multirow{3}{*}{\textbf{\shortstack[l]{Feature Types\\/Organization}}}} & \multicolumn{3}{c|}{\textbf{Application Domain}}                                  & \multirow{3}{*}{\textbf{Prediction Task (Publications)}}                                                                                                                                                                                            \\ 
\cline{3-5}
\multicolumn{2}{|l|}{}                                            & \multirow{2}{*}{\tiny{\textbf{E-commerce}}}                & \tiny{\textbf{Display Advertising}}       & \tiny{\textbf{Recomm. Systems}}           &                                                                                                                                                                                                                                           \\ 
\hline
\multicolumn{2}{|l|}{\multirow{1}{*}{Feature Engineering}}                         &                           & \checkmark &                           & (1) CVR (\cite{EstimatingLee2012,CostVasile2017,FaceLiu2020}); (2) CTR (\cite{SimpleChapelle2015,EvaluatingDalessandro2012,PracticalHe2014,EffectsKumar2016,UserRen2016,ZhangBidAware2016,ZhuSEM2017}); ~(3) Ranking(\cite{DeepCovington2016}  \\ 
\hline
\multirow{12}{*}{\shortstack[l]{Feature\\Learning}} & Collaborative Filtering Based               &      \checkmark                     & \checkmark & \checkmark & (1) CTR (\cite{CollabLiu2014,ResponseMenon2011}); (2) Ranking(\cite{CollabLiu2014});\newline~ (3) Product Rating(\cite{NTFWu2019})\\ 
\cline{2-6}
&Multi-field\\(categorical)                    &                           &                           & \checkmark & (1) CTR(\cite{WideAndDeepCheng2016,FIBINETHuang2019}), \newline~
(2) Ranking(\cite{NeuralCFHe2017})\\ 
\cline{2-6}
&Textual                                     &                           & \checkmark & \checkmark & (1) CTR (\cite{UsingBaqapuri2014,ClickEffendi2016}, Deep(Char) WordMatch \cite{DeepEdizel2017}, DSM \cite{DeeplyGligorijevic2018}); ~(2) Ranking (\cite{DeeplyGligorijevic2018})                                                                      \\ 
\cline{2-6}
&Visual                                      & \checkmark                          &                           &                           &(1) CTR(DICM\cite{ImageGe2018},\cite{DeepChen2016}), 
(2) Ranking(ACF\cite{AttentiveCFHe2017}, PinSage\cite{GraphYing2018})\\ 
\cline{2-6}
&Sequential                                  &   \checkmark                        &   \checkmark                        &   \checkmark                        &(1) CTR(DSIN\cite{DSINFeng2019}, DIEN\cite{DIENZhou2019}, DIN\cite{DINZhou2018}, RNN\cite{SequentialZhang2014}, MIMN\cite{PracticePi2019}),\newline
(2) CVR(GMP\cite{OnlineChao2019}, DTAIN\cite{TimeGligorijevic2019}),\newline
(3) Ranking(FGNN\cite{ExploitingQui2020}, GAG\cite{GAGQui2020}, LGSR\cite{JointXu2020})
\\ 
\cline{2-6}
&Network based                               &         \checkmark                  &          \checkmark                 &                           &     (1) CTR(FiGNN\cite{FiGNNLi2019}, GIN\cite{GINLi2019},\cite{BillionWang2018}, KNI\cite{End-to-EndQu2019});  (2) Ranking(KGNN-LS\cite{Knowledge-AwareWang2019})\\
\cline{2-6}
&Hybrid                                      & \checkmark &                           & \checkmark & (1) CTR(\cite{EnsembleAryafar2017},  RippleNet\cite{RippleNetWang2018}, MKR\cite{MKRWang2019},  DKN\cite{DKNWang2018}),\newline~ (2) Ranking(RippleNet\cite{RippleNetWang2018}, MKR\cite{MKRWang2019})                                                          \\
\hline
\end{tabular}
\end{table*}

\begin{table*}[!htbp]
\tiny
\setlength{\tabcolsep}{3pt}
\caption{Summary of selected practical solution applied in industrial environments}
\label{tbl:industry_summary}
\centering
\begin{tabular}{|>{\raggedright}p{.09\linewidth}|>{\raggedright}p{0.3\textwidth}|>{\raggedright}p{0.39\textwidth}%|>{\raggedright}p{0.055\textwidth}
|>{\raggedright}p{0.08\textwidth}|p{0.06\textwidth}|} 
\hline
\textbf{Algorithm}                       & \textbf{Challenge}                                  & \textbf{Introduced Strategy}%y&A/B Test Improv. Rate                        
& \textbf{Application Domain}               & \textbf{Provider}  \\ 
\hline
AdPredictor\cite{WebScaleGraepel2010} & Scalability                  & Bayesian probit regression model, Weight pruning, \\Parallel training       %&
& Sponsored search ads           & Microsoft Bing            \\ 
\hline
EtsyCTR\cite{EnsembleAryafar2017} & Dealing with image data~~                  & Transfer learning, Feature Hashing, Ensemble model % ,Progressive Validation       %&
& E-commerce           & Etsy            \\ 
\hline
FBCTR\cite{PracticalHe2014}   & Massive data                               & Uniform sub-sampling, Cascade of classifiers, Ensemble model %&N/A
& Display ads  & Facebook        \\ 
\hline
DLRM\cite{DeepNaumov2019}   & Memory constraints in embeddings and computational costs of DL components                               &  Using PyTorch and Caffe2 for model and data parallelism   %&N/A
& Recomm. Sys & Facebook        \\ 
\hline
DeepFM\cite{DeepFMGuo2018}\\PIN\cite{ProductMultiQu2018} & Insensitive gradient issue in DNN based models and space complexity of FFM-based models& Shared embedding vectors, an end-to-end prediction model\cite{DeepFMGuo2018}, Net-in-Net architecture to combine FM and DNN units\cite{ProductMultiQu2018} & Recomm. Sys           & Huawei               \\ 
\hline
DIN\cite{DINZhou2018}\\DIEN\cite{DIENZhou2019} & Large number of DNN parameters, Addressing temporal drifts in user interests representation%Fixed-length embedding vectors to represent user interests, Temporal drift in user interests
& Mini-batch aware regularization and local adaptive activation function\cite{DINZhou2018}, Attention based user interest extractor layer\cite{DIENZhou2019} & Display ads           & Alibaba               \\ 
\hline
MIMN\cite{PracticePi2019}    & Handling long user behavior sequences  & Multi-channel memory network                                                %&7.5\%(CTR)
& Display ads & Alibaba         \\ 
\hline
HPMN\cite{LifelongRen2019}\\UBR4CTR\cite{UBR4CTRQin2020}\\SIM\cite{SIMQi2020} & Tackling long sequential user behaviours & Memory network model along with a GRU network\cite{LifelongRen2019}. Self-attentive retrieval module to select relevant user behaviors\cite{UBR4CTRQin2020}. Cascaded two-stage  search model\cite{SIMQi2020}.                                                       %&N/A
& E-commerce           & Alibaba               \\ 
\hline
EGES\cite{BillionWang2018}    & Scalability                                & Graph Embedding Using XTensorflow~                                          %&
& Recomm. Sys       & Taobao          \\
\hline
DICM\cite{ImageGe2018}    & Dealing with images to represent user behaviours                                &A distributed model server to handle image data embedding and reduce the communication latency                                           %&9.2\%(CTR)
& Display ads       & Taobao          \\
\hline
RAM\cite{Jointly2020}    & Balance immediate advertising revenue and long-run user experience                                & Joint optimization using two-level reinforcement learning~                                          %&N/A
& Recomm. Sys       & ByteDance          \\
\hline
HPS-4\cite{DistributedZhao2020}    &       Massive model with large number of parameters                          & A distributed hierarchical GPU parameter server  %                                        &N/A
& Display ads       & Baidu          \\
\hline
PinSage\cite{GraphYing2018}    &       Massive input graph with billions of links                          & Highly scalable graph convolutional network                                  %&10-30\%
&Recomm. Sys       & Pinterest          \\
\hline
DCN\_V2\cite{DCNV2Wang2020}    &       Controlling the number of model parameters to learn feature interactions for real-time data                         & Mixture of low-rank approximation of DCN method\cite{DCNWang2017} organized in stacked and parallel structures                                 %&10-30\%
&Recomm. Sys       & Google          \\
\hline
\end{tabular}
\end{table*}

\section{Conclusion} \label{sec:conclusion}
This survey provides a comprehensive overview of computational methods for user response prediction in online advertising. Our goal is to provide a detailed review and categorization of the online advertising ecosystem, stakeholders, data sources, and technical solutions. To achieve the goal, we review and categorize online advertising platforms, type of user responses, data sources and features, and propose a taxonomy to characterize main stream approaches for user response prediction. %The survey provides direct answers to some important quesltions like main advertising platforms? What type of use response can be modeled/predicted using computational approaches? What are the features and the source of features useful for use response prediction? How to utilize features for use response prediction? What are the main types of technical solutions for user response prediction? Are there anyone benchmark and online resources (datasets/software) for evaluation purposes?
%We propose taxonomies to categorize existing methods into different groups based on embedding approaches and algorithmic designs. 
For each type of user response prediction methods, we also briefly study technical details of representative methods, with a focus on machine learning, especially deep learning, based approaches. % It follows with reviews and comparisons between representative methods within each categories. 
In addition to the algorithms, we also review user response prediction applications, benchmark data, and open source codes. The survey delivers a first-hand guideline for industry and academia to comprehend the state-of-the-art. It also serves as a technical reference for practitioners and developers to design their own computational approaches for user response prediction.

\section{Acknowledgments}
This work is partially sponsored by the U.S. National Science Foundation through Grant Nos. IIS-1763452 and CNS-1828181 and by the Bidtellect Inc. through a sponsorship agreement.

\bibliographystyle{ACM-Reference-Format}
\bibliography{acmart}

% % %%
% % %% If your work has an appendix, this is the place to put it.
% \newpage
\appendix
\counterwithin{table}{section}
\counterwithin{figure}{section}

\section{Appendix}
%\subsection{Additional Figures}
%\begin{figure}[H]
%  \centering
%  \includegraphics[width=0.8\linewidth]{figs/Ensemble_models_new.png}
%  \vspace{-.75\baselineskip}
%  \caption{Ensemble structure types: a) Bootstrap Aggregating or Bagging that randomly sample with replacement of training data to generate N subset data. N predictor models are learnt using each subset. The final result is the combination of M classifier outputs b) stacking: N models are trained in parallel based on the same training data. The final output is combined through a meta-classifier. This classifier is fitted on the output of base classifiers. c)Boosting(AdaBoost): a series of predictor models are trained using a subset of training data sequentially. The subsets of data are created adaptively using misclassified samples in previous model. d) Cascading: is based on concatenation of multiple classifiers}
%  \label{fig:entypes}
%   \Description{The 1907 Franklin Model D roadster.}
%\end{figure}
\counterwithin{table}{section}
\subsection{Term Definitions} 
In Table  \ref{tab:notations}, we introduce some important terms used in the context of online advertising, and also summarize common terminologies used in the paper.

%In this subsection, we introduce some important terms used in the context of online advertising. The list of common terminologies used in entire of this paper is presented in Table  \ref{tab:notations}.

\begin{table*}[!htbp]
\begin{scriptsize}
\centering
\caption{Summary of key terms and notations in online advertising}
\label{tab:notations}
% \vspace{-.75\baselineskip}  
\begin{tabular}{p{0.21\textwidth}|p{0.75\textwidth}}
\toprule
Notation  &  Descriptions             \\ 
\midrule
Display Ads & The form of ads containing rich media content that are displayed in reserved spaces in websites\\
Search Ads & The form of ads that are displayed in the search result page triggered based on user search query. They are sorted and displayed to users to attract their attention to commercial products. \\
Native Ads & The form of ads which are similar to regular posts in stream media platforms including video and images commercial contents.\\ 
Advertiser & The stakeholders which promote products and services in online advertising, by serving ads to online users\\
\hline
Publisher & The stakeholders which run websites with potential placements to show ads to online ads\\
Audience & The online users who are exposed to ads in a respective online context\\
Ad Exchange & The marketplace where advertisers and publishers are connected, through DSP and SSP, to negotiate selling and buying price using real-time bidding (auctions)\\
Auction & A virtual real-time sale being held upon SSP ad request to gather bids of advertisers for an ad impression. \\
Demand Side Platform (DSP) & A software platform in online advertising eco-system working on the behalf of advertisers to manage campaigns and submit a relevant bid price to bid requests\\
\hline
Supplier Side Platform (SSP) & A software platform in online advertising working on the behalf of publishers to send a bid request to Ad exchange network. It loads the ad creative by calling ad server which has a winning ad in an auction\\
Ad Campaign & A set of advertisements with a common theme (or objective) in targeting similar group of users \\
Ad Creative & The message or artwork (advertising object), designed by advertisers, to be served to the audience' devices\\
Ad Placement & It refers to the location where an ad is displayed on the web-page. The common places that are available for advertiser to run an ads are the footer, the header and sidebars of website or anywhere in the beginning or middle of article and videos.  \\
\hline
Banner Ad  &  A rectangular graphic window, such as an iFrame, dedicated to show ad creative (image or text content) in the publisher's web-page \\ 
Impression   & The rendering/presence of an ad on the user's device \\ 
Landing page   & A web-page used to show to viewers, after they click on the ad\\
Click  & A user mouse click event or user tap event on the ad when they visiting an ad on desktop or mobile devices\\ 
Conversion & The user actions, such as purchasing a product or subscribing to a service, after clicking an ad creative and being directed to advertiser's landing web-pages\\
Gross Merchandise Volume & A e-commerce metric considered by businesses to indicate the amount of sales made by users over a specific period before deducting any expenses like those associated with online marketing\\
\hline
Data Management Platform\newline (DMP) & A software platform in online advertising designed to collect and analyze data for both advertisers and publishers. DMPs provide services to  DSPs or SSPs to improve ad campaign efficiency\\
First party Data & First party is referred to as the stakeholder itself, so first party data is referred to as data collected from the activities of business users of respective stakeholders. The data ranges from user profile information like demography user historical behaviors such as visited pages and purchase history, user subscription data or their activities in social media\\
Second party Data & Second party is referred to as the other party of each stakeholder. The first party data of a company is referred to as the second party data of other companies\\%. Relying on quality of data made by others, the data can be used to enrich user profile information to improve ad campaign and identify potential customers through DMP nodes.\\
Third party Data & Data gathered from outside sources to be packaged and sold to others. The data is organized to clusters and segments in terms of page information, user characteristics, and audience interest to be chosen by buyers. Third party data are typically gathered from DMP by analyzing user cookie information\\

\bottomrule
\end{tabular}

\end{scriptsize}
\end{table*}
%In this section, we introduce some important terms that are used in the context of online advertising. The list of common terminologies used in entire of this paper is presented in Table  \ref{tab:notations}.
% \subsection{Additional Figures}
% \counterwithin{figure}{section}

\section{Media Types, Devices, and Platforms}
Driven by the communication and networking technologies, online advertising has been continuously evolving in the past two decades. Starting from static banners on websites, the industry is now dynamically serving ads based on types of media platforms, user devices, and media types~\cite{DisplayWang2016}. In this section, we briefly summarize types of media platforms, user devices, and ad types in the online advertising eco-system. % Many types of media platforms are supporting online advertising, and we briefly summarize media types as follo leads to introduce a wide range of advertising types can be used in different type of online platforms. 

\subsection{Media Platforms} There are various advertising platforms that have been proposed to serve advertisements to users, depending on the context of users accessing to the network. % users are accessing to the  via different platforms. 
\subsubsection{Sponsored Search Marketing}
Sponsored search is a search engine based advertising platform that uses user query as the context and returns a list of ads related to user queries for advertising. %query issued by users developed by search engine platforms like Google that provides the list of relevant ads related to query issued by users. 
The result of ads are usually displayed as a sorted list of commercial hyperlinks among the search result page on the basis of similarity to the search result. To differentiate ads from regular list of search results, an ``Ad'' tag is usually marked to promoted records. Following the pay-per-click mechanism, the advertisers pay the search engine platform when user click on ads. Then the potential revenue for publishers is calculated using click-through rate values. 

\subsubsection{Display Advertising}
Display advertising is one of latest forms of online advertising working on the basis of real-time bidding to select and show personalized ads in the dedicated areas of web pages. Different from search advertising where user queries provide clear context of user preference, the context of display advertising is often limited to the pages visited/requested by users. Given a user requesting to visit a URL in the web browser, ads are delivered from real-time auctions set up by ad exchange network. The main approach in this advertising is targeting individuals using information available from them in different internet sources (like cookies for browsing history). The main goal for advertisers is to find strategies to reach to right customers to engage them to take desirable action. The performance of marketing is evaluated based on user responses like click or conversion which can determine display-related advertising revenues and cost for publishers and advertisers.

\subsubsection{Social Media Advertising}
Social media networks are becoming common elements of daily life, and providing an opportunity for people to interact and share information. The popularity of these platforms motivate businesses and companies to target their potential customers among online users. To this end, advertisers aim to provide personalized posts or tweets in social media platforms, like Twitter, Facebook, and Instagram, to attract user responses, which are usually measured through click-through rate or conversion rate. To address the quality of ads, some metrics like post-click-experience are also introduced to analyze the dwell time users may take on a landing page following the click event. Some studies~\cite{PromotingLalmas2015,ImprovingBarbieri2016} show that post-click-experience can be used to gain additional knowledge about user preferences and modeling user responses.

\subsection{Ad Types}
%Related to the context of advertising, different 
In order to promote products and services in online advertising, different types of advertisements have been designed as the means of advertising. %to promote products and services by advertisers.
\subsubsection{Banner Ads}
Banner ads, which have existed since the very beginning of online advertising in 1990s, incorporate the main standard media type in the Internet for advertising. Organized by compounds of text, image, or animated contents placed in the specific area of web pages, banner ads present the advertisers' message to users to attract their attention about the promoted content. Clicking on the areas generally indicate some sort of matching between ad content and user preferences. As a result, a click leads to a transition of users from the publisher web pages to advertisers' websites for marketing purposes.
\subsubsection{Textual Ads}
Text ads are the most well-known type of ads in many advertising  platforms like sponsored search advertising, short message (SMS) marketing, email advertising, or display advertising~\cite{ResearchJiang2016}. A text ad includes a textual creative ad shown alongside search results \cite{DeepEdizel2017} or part of the emails or text message sent to subscribed users to show promotional messages. The elements of ads are organized to lead to a click response directing users to promoted pages~\cite{EffectsKumar2016}.

\subsubsection{Video Ads}
With the increase of broadband Internet, the usage of video ads to deliver promotional content has gained increasing popularity and becoming an effective way of interact with audience. Today the striking amount of video ads are employed to transfer commercial messages to online users. In the form of live videos or downloadable video content, this type of advertisements can be presented to users like banners in websites, tweets, or feeds in social media platforms.
%\url{https://dl.acm.org/doi/pdf/10.1145/1291233.1291467?download=true}

\paragraph{Roll-out Ads} Roll-out ad is a specific form of video ads that are joined to the other video content in streaming websites like YouTube or Vudu. These videos automatically appear before playing the original content can be seen in either skippable or non-skippable formats. According to the place, the video ads being attached to the beginning or the end of original one are called pre-roll or post-roll advertisements. The mid-roll version of video ads is defined to augment to some points in the middle of original videos which is contextually relevant and less intrusive to users~\cite{VideoSenseMei2007}.

\subsubsection{In-App Ads}
This type of ads represents the version of advertisements which appear within applications like video games in desktop systems or smartphones. Like digital advertising, mobile apps are designed to reserve some space for ads. There are generally two ways used to display and place ads in apps. The first is considered as static brands involved in the background and integrated as the part of app. This type follows the guaranteed contract setting and similar to the primitive version of ads in the past which cannot directly receive user responses. Therefore, they can be considered  superficial or even not recognizable by users~\cite{AdvertisingBidmon2018}. The second type, which is considered more interactive ones, places ads in transitions. They usually appear in two forms of full screen or regular rich media ads. Following ad real-time bidding model, relevant ads are selected from negotiation between DSP and SSP to be displayed across the screen~\cite{IntegratedTruong2019}.
\subsection{Source of Features}
The source of features for user response prediction task is related to information transferred in online advertising ecosystem. In the ad network workflow in Figure \ref{fig:ecosystem}, actors, such as advertiser and publisher, play a role together with intermediary nodes to provide users with relevant commercial content. In order to acquire positive user feedback, various data features are used to represent users and describe advertisers and online content providers. The data sources for feature extraction are information from demand side or supplier side, or even external third party sources. The representative list of features regarding supplier-side platform, demand-side platform and third-party groups are shown in Table~\ref{feature_sources}.
% In online advertising, actors such as advertiser, and publisher are contributing together to induce a positive feedback from users being served by commercial contents. %by displaying a commercial content to the customers. 
% The typical ecosystem of online advertising can be seen in Figure \ref{fig:ecosystem}. %Aside from the major aforementioned actors,
% Ad exchange network along with intermediary nodes are working in background to provide users with relevant advertisement in different online content providers. These nodes deal with contextual features and information which have seen useful in user response prediction tasks \cite{DeepCovington2016}. The representative list of features regarding supplier-side platform, demand-side platform and third-party groups are shown in Table \ref{feature_sources}. % for advertising. 
\subsubsection{Suppler Side Features}
As showing in Figure~\ref{fig:ecosystem}, SSP or supplier side platforms are intermediary nodes in the ad network which work on behalf of publisher to manage the inventory of available ad placements in web pages. As soon as users submitting a keyword in search engine or visiting a website in display advertising, an auction is triggered to find ad to be served to the users. In this case, information regarding visiting user and ad slots are transferred to relevant DSP nodes through ad exchange network. As shown in Table \ref{feature_sources}, this information is characterized via features describing ad creatives and their appearance in the publisher web-pages, such as features about the content of web-pages, placement id, size, width and height, visibility status and format, as well as user information such as device types, user agent, browser information \textit{etc.} %to data about the format of ad along with 
%features about the content of web-pages.   
% \begin{figure}[h]
%   \centering
%   \includegraphics[width=0.5\linewidth]{figs/publisher_features.png}
%   \caption{SSP(Publisher) features}
%   \label{fig:ecosystem}
% %   \Description{The 1907 Franklin Model D roadster.}
% \end{figure}
\begin{table}[H]
\footnotesize
\begin{tabular}{|l|p{.75\linewidth}|}%{\linewidth}{|l|L|}
 \hline
 Source & Features\\
%  \multicolumn{2}{|l|}{\centering{features}} \\
%  \hline
%  Country Name     or Area Name& ISO ALPHA 2 Code &ISO ALPHA 3 Code&ISO numeric Code\\
 \hline
 Supplier Side Platform & Page URL, Device type, Devide Id, HTTP cookie, OS Version, Browser type/Version, User agent, Geo location, Ad slot (ID, width, height, visibility, format), placement ID, Publisher ID, User ID\\%Creative ID, Advertiser ID, user ID\\
 Demand-Side Platform & Bidding price, Paying price, Campaign Category, creative ID, Advertiser ID \\ %Page Category \\
 Third-party & User segmentation, User demography, Site information, Page information, Page categorization \\%Device-id, HTTP cookie, OS Version, Browser Version, %IP address, WiFi network, GPS, Gyroscope\\
%  Algeria    &DZ 012\\
%  American Samoa&   AS 016\\
%  Andorra& AD 020\\
%  Angola& AO 024\\
 \hline
 
\end{tabular}
\caption{A summary of major sources of categorical features in online advertising}
\label{feature_sources}
\end{table}
\raggedbottom
% \vspace{-0.75cm}
\subsubsection{Demand Side Features}
Another important party in the online advertising ecosystem is DSP (Demand side platform) which intervenes in connections between advertisers and ad exchange network. An ad exchange casts auction for bid request triggered by SSPs to DSPs to select the display of ad on the publisher’s website.% is selected through an auction process. 
The Ad exchange network collects bid prices offered by connected advertisers through their DSPs, and selects ads corresponding to higher bid to present in publisher websites. The bid price calculated by advertisers %using efficient algorithms by DSPs 
depending on three set of main factors such as information available about users, the constraints of publisher pages, and ad campaigns. The information is characterized using features such as ID and the floor price of ad creatives, the URL of landing page, the segmented user profiles, and ad campaign categories to describe the campaign and content of ads. They are augmented with the online user browsing history captured by DMP(Data Management Platform) nodes to facilitate decision making to choose matching ads for users. % This additional information about user behavior history is generally processed by DMP nodes in ecosystem 
%related to publishers ad requests.  

\subsubsection{Third-Party Side Features}
Features used to select the best matching ads to users' interests are not only limited to those from by DSPs and SSPs. Instead, external sources can also provide compliment and valuable information needed to describe objects involved in advertising scenarios. In this context, they are known as %third party companies which offer 
third party data which are captured in various websites and social media platforms, using data aggregation, cookie, or machine learning approaches. % by people are not directly interact with users. 
These data can include different information exploited ranging from user web cookies, to meta-data of devices and geographic information. For example, White Ops%~\cite{whiteops:2020}
, an online ad verification and fraud detection company, provides third-party side services allowing users to query whether the traffic (\textit{i.e.} the page visit) is initialized from a genuine human user or a bot in real-time, using data collected from ``trillions of transactions''. Such services allow an advertiser to determine whether an auction is potentially fraudulent and stop bidding on fraud auctions~\cite{zhu:2017:fraud}.  %To have receptive targeted users and obtain higher performance, advertisers use further information from different sources. 
 %To address session management on the web, internet cookie
%cookie data in form of textual data of http header 
%can be used to access to the status of individual users. These features are prepared from the corresponding third party data produced in different platforms. 
%Purchasing from third-party sellers, 
Online advertising systems can leverage third party data to organize new features used to target users for their campaigns. %using different data about the user agent application, browser, Device, the location in form of IP and geographic region and city ids.  
% \begin{figure}[h]
%   \centering
%   \includegraphics[width=0.6\linewidth]{figs/Third-party.png}
%   \caption{third party features}
%   \label{fig:ecosystem}
% %   \Description{The 1907 Franklin Model D roadster.}
% \end{figure}

%https://link.springer.com/chapter/10.1007/978-1-4302-4603-9_9

%https://link.springer.com/chapter/10.1007/978-3-658-22681-7_6

%https://ieeexplore.ieee.org/abstract/document/8802632

%https://journal.acs.org.au/index.php/ajis/article/view/1971

\subsection{Device Platforms}
\subsubsection{Desktop Advertising}
Since the very beginning of banner ads on the web, desktop was and still is the dominate device platform for advertising. %which got started with displaying banner ads on the web. %can be seen though desktop system on the web are banner ads. 
Available through desktop systems, desktop advertising entails expanded version of ads including text-based advertisement, roll-out video ads, and in-app advertisements appearing in search engine results, streaming web services or software. The capabilities of smart phones make these devices as the predominant opponent of desktop systems since it can be used for same purposes. However, it does not discount the value of desktop branding as long as the desktop and laptop systems stays on.

\subsubsection{Mobile Advertising}
As smartphones and mobile devices are becoming essential tools for communication, online advertising also quickly adapts to mobile devices for marketing. %  Today, no one can picture the world without smartphones and mobile devices. 
In early days of mobile phones, the common advertising form was SMS advertising in which advertisers send the textual ads to customers. The rise in popularity of multi-purpose mobile phones and wearable devices made an opportunity for online companies to use a new way for advertising and targeting audience. Nowadays, mobile advertising makes up a significant portion of online advertising \cite{PredictingOentaryo2014,YouTolomei2018} which can be roughly classified into the following two types:

% \noindent\vspace{0.1cm}\textbf
\paragraph{Mobile Web Advertising}
Like desktop workstations, one of the major advantages of smartphones is web browsing, which relies on search engines to get relevant information for user needs. There is no wonder that users take advantage of their phones to look for services and products located near them. In era of virtual assistants, voice search has seen significant growth using smartphones. Now, mobile web advertising is being developed to leverage new ways, like natural language based optimization, to provide sponsored search for verbal personalized ads for users. Using cross-platform compatibility followed in mobile devices make it possible for advertisers to focus on the content of their ads for potential customers with low cost about the ways that are published in different devices~\cite{RobustFMPunjabi2018}. But advertising is not limited to voice data. The transcript of verbal or written conversations between customer-service agents and people consist of valuable information about users implicit and explicit behavior which can be analyzed to predict user life events and campaign relevant products \cite{ImplicitEbadi2019}. 
% \noindent\vspace{0.1cm}\textbf

\paragraph{Mobile App Advertising}
With the development of smartphone devices, there is an continually increase in the production of mobile applications. The bulk of research confirmed the fact that users today spend more time in online applications than web browsing on cellphones~\cite{TargetAliannejadi2018}. Mobile app advertising, known as in-app advertising, is the specific type of advertising where ads are served within smartphone app eco-system. In this way, there is a trend that smartphone app developers prefer more to produce free applications by supporting an ad model to gain revenue from presenting ads on their apps~\cite{ComprehensionSaborido2017}. This type of advertising deals with different advertisement units like banners ads which are shown at top or bottom of apps and transitional apps that are places for transitions inside applications.

\subsubsection{Tablet Advertising}
Tablet advertising is designed to present relevant ads for tablet users. Although both tablet and smartphones are essentially mobile devices, they are quite different in terms of quality of display and goals of usages. Some studies highlighted the difference in user behavior regarding advertisement when using tablet devices \cite{TabletsMcCreery2017}. The recent years have seen a decline in using tablets since brands continue to create bigger screen smartphones. But there are still many people use this device for entertainment activities\footnote{http://googlemobileads.blogspot.com/2011/11/consumers-on-tablet-devices-having-fun.html} like playing online video games and second screen viewing\footnote{https://www.statista.com/topics/2531/second-screen-usage/} to watch the main media stream or live event and use tablets at the same time. Business users also use it to complement office activities. It can lead to generate plenty of opportunities to design distinctive tablet-based business-themed approaches for ad campaigns. In these devices, advertisements take advantage of high resolutions banners and videos to provide a better user experience and grab more user engagement.
% https://search.proquest.com/docview/1621400331?pq-origsite=gscholar

% https://www.tandfonline.com/doi/full/10.1080/10641734.2017.1291386

% https://digitalcommons.calpoly.edu/grcsp/70/

% \url{http://virtual.vtt.fi/virtual/nextmedia/Deliverables-2012/D1.1.1.6_eReading_Tablet%20advertising_the%20look%20on%20best%20practices_design%20patterns%20and%20methods.pdf}

% \vspace{-0.5cm}
\section{Evaluation procedure of user response prediction models}
%In this section, we review the common procedures proposed by researchers in this domain to conduct experiments and provide evaluation results.
%The evaluation of user feedback in recommendation systems and online advertising systems comes with two perspectives from academia and industry. 
The evaluation of user response prediction can be categorized into offline (simulation based) evaluation and online test, which are from academia and industry perspectives, respectively. 
\subsection{Offline Evaluation}
Majority papers published in this domain are developed by researchers from academia. Common approaches used by them to assess model performance are to use a simulation of the real-world environment. In this condition, sample data from different stakeholders in recommender and online advertising systems are gathered to follow offline trials. As shown in Figure \ref{fig:UserResPredict}, the studies conducted about recommendation systems and online advertising typically aim to provide two types of outputs corresponding to different stages in the system. These studies are categorized into two type of point-wise and list-wise methods. In the former, the models predict the single numeric estimated value of user feedback score such as product rates, click-through rate, or conversion rate values. The latter prepares the ranked list of products ordered by predicted user interaction scores. In point-wise models, models are evaluated using different metrics such as root mean square error for regression problems (product rating), Area over ROC curve (AUC) and Accuracy metrics for binary classification methods i.e. CTR prediction, CVR (purchase) prediction. The output in list-wise scenarios are optimized via ranking metrics like Precision (Precision@K) and Recall (Recall@K) over top $k$ recommended list. The performance of these systems are also evaluated from other aspects such as Normalized Discounted Cumulative Gain (NDCG) and Mean Reciprocal Rank (MRR). A common approach to evaluate the quality of a recommended item list with $k$ elements is evaluating the predicted score of user interaction for test samples located among random selected set of unvisited items\cite{NeuralCFHe2017}. Table \ref{tab:auc} compares the performance of different methods for a binary classification task to predict click events. %According to reported results, Criteo and MovieLens are two frequently used public datasets which are examined in different studies. The table also shows that methods like DeepFM, xDeepFM and FiGNN and AutoInt which are the ones taking advantage of deep learning components, therefore could achieve better performance. AutoInt uses the attention mechanism, and outperforms the remaining approached by getting the higher AUC value. For MovieLens dataset, methods like KNI, RippleNet consider knowledge network as the side information to organize input categorical features obtained the top two performances. 
%These information can prove the effectiveness of graph data as side information and attention mechanism to create effective embedding vectors for modeling implicit user feedback. 
Although the success of model designed based on deep learning make it cornerstone to develop models in academia and industry for user response prediction, it cannot be neglected that in some scenarios like recommendation system contests (RecSys conference)~\cite{WhyJannach2020}~\cite{MethodologicalDacrema2020} which evaluate the methodologies for offline experiments and indicate that winning solutions could be selected from the old techniques like SVMs, KNN, Logisitic Regression and Decision using ensemble based models. Preparing proper environments to assess baselines is also important~\cite{OnRendle2019}, and the research has shown that early methods could outperform recent proposed algorithms as long as they are well set up for experiments. These results provide some indications about the evaluation procedures. It first can show the significance of feature engineering and knowledge about of the application domain to use appropriate features in modeling. It also raises the reliability issue in experiments. Although many reported results are based on cross-validation, statistical significance tests and availability of their code for reproducibility (Some of which are gathered in Table \ref{tab:sourceLinks}), several arbitrariness in experiment designs should be addressed by research community. In the following sub-section, we discuss evaluation methodologies followed in the reviewed papers covering common approaches to set up experiments.  

\subsubsection{Experimental Setup} 
The evaluation step in prediction and classification tasks aim to assess how well model can cope with new unseen samples. Typical data splitting is performed by randomly selecting samples without replacement from datasets to create three partitions: training, validation, and test. %  fraction of data  for training, validation and test set%The standard way to follow-up this procedure is preparing separate data-sets for training and test steps. This can be performed by randomly selecting the portion of of samples in the original dataset. The common proportion which is used in academic papers is 80:20 \cite{} i.e., 80 percent of samples randomly selected without replacement in the dataset are assigned to the training set while the remaining samples are in the test set.
However, in recommendation and online advertising systems, the dataset include data logs of user interactions with online systems in which each sample come with a timestamp. So considering the sequence of samples in term of time and applying a chronological order constraint in the data splitting in datasets is a rational expectation. The data-splitting in this case is followed by choosing a arbitrary cut-off point in the dataset to prepare training and test subsets. A typical approach is leave-one-out method which assigns the latest data to the test-set while the reminder data are dedicated for training and validation sets~\cite{NeuralCFHe2017}. % the sequence of samples in term of time is considered, each sample  
To address short-term and long-term sequential data, there are two form of event-based and session based datasets. %So the chronological order constraint in the data splitting is imposed. 
In the event-based dataset, the common idea is selecting randomly a couple of time intervals where samples before the split point are assigned for training set whereas the one after splitting time point are for test set (ex. in a dataset including 7 subsequent days, the first subsequent days are assigned to training while the last day data is for test set \cite{DeepFMGuo2017,ProductBasedNNQu2016}). In the session-based datasets, the same procedure is applied with a difference that the samples before or after the split point include the session of events. There are different approaches to define the session notion to represent short-term user interaction logs. In \cite{DSINFeng2019}, the cut-off to split data logs into session is determined as the gap of 30 minutes between subsequent interactions. %All in all, there is no standard way to create datasets, in 
Section \ref{app:benchmark_dataset} summarizes a number of benchmark datasets presented in different existing academic and industrial experiments.

\subsection{Online A/B Test}
A/B test is an evaluation mechanism which provides controlled environment to compare models %. Before deploying any models to real online system, A/B test experiments are conducted 
by splitting user traffic into two different portions, A \textit{vs.} B, to compare model performance. So models are assessed using real-world system users to drive desired user feedback signals. To avoid misleading evaluation results using the knowledge about the field, the statistical significance tools are applied to create variation of data, and make statistical test to evaluate confidence intervals. The rate of improvement compared to the baselines and rate of error are central in results based on A/B test. A common metric used to present online comparison is to calculate the relative improvement like Normalized Log Loss (NLL) metric as $\frac{{LL(\overline{p})-LL(p)}}{LL(\overline{p})}$ where $LL(\overline{p})$ is the log loss value of the best predictor on the test dataset while $LL(p)$ is for any baseline predictor $p$. %Authors in different work provide an online comparison between the performance of their proposed models with the baseline ranking and user response predictor models. 
Typically baseline models are built based on logistic Regression (LR) models that are highly engineered with rich features in production simulations~\cite{FFMJuan2017,WideAndDeepCheng2016,DeepFMGuo2018,ProductMultiQu2018}. In recent developed models, the baseline models are chosen from successful models like DeepFM~\cite{DeepFMGuo2017} or DIN\cite{DINZhou2018} for online experiments~\cite{AutoFISLiu2020,AutoGroupLiu2020,DIENZhou2019}. However, previously, authors in \cite{FFMJuan2017} compare the performance of the proposed model, \textit{i.e.} FFM with the baseline LR models via calculating Return on Investment (ROI) relative improvement in display advertising. Some other works like ~\cite{DeepFMGuo2018} calculate the relative improvement in recommendation metric values like Coverage, Popularity and Personalization to compare DeepFM model with a baseline LR model in Huawei app market environment. Because of complexities in online businesses, there is no standardized plan to follow online experiments. So the number of research papers conducting this evaluation procedure are relatively limited in the literature.  
\section{Applications, Resources \& Future Directions} \label{sec:app}

For the online advertising, the click-through rate and conversion rate are common user response metrics to evaluate the marketing performance and determine the revenue of advertisers and publishers in online advertising and recommender systems.     %It can be used for analyzing further user engagement applications. 
In this section, we review some applications and tasks related to user interactions in online content providers. %In the appendix, 
We will introduce the benchmark user-log datasets
% , evaluation methods 
and the publicly available source-codes provided by researchers. Such information is useful in a case we need to deal with baseline methods in the domain of user response prediction.

\subsection{Benchmark datasets for the user response prediction} \label{app:benchmark_dataset} %In order to compare the performance of new proposed methods with current ones for user response prediction, the access to benchmark dataset can provide a standard evaluation studies. 
Benchmark datasets are helpful type of data sources to carry out a fair comparison between different methods. A list of publicly available datasets that were examined in previous studies are summarized in Table \ref{tab:datasummary}. Several benchmarking datasets have been prepared and made publicly available to conduct studies for the user response prediction.  Table \ref{tab:auc} demonstrates the performance of different user response prediction methods on selected datasets. We describe two representative datasets in the following.
\paragraph{Criteo Dataset} This is one of the important benchmark datasets gathered from the seven days real data logs by Criteo company. It was initially prepared for a competition held by Kaggle in 2014 to encourage the development of approaches for click-through rate prediction task. Criteo dataset includes 39 high cardinal features which are consisting of 26 categorical features and 13 continuous features for each pair of ad and user describing the event that ad visited by the user. Each row corresponds to one impression. Each instance of user and ad has a label to indicate whether the impressed ad receives ad click response or not. It have a full and condensed version containing around 500 and 45 million samples respectively.
\paragraph{Avazu Dataset} The second dataset prepared by Kaggle for a competition in 2014 including user click behavior gathered from Avazu mobile advertising platform. Each row in this dataset describes an impression (an ad displayed to users). Each impression event is attributed by a set of features such as categorical features for user, device and advertisement like hour of day, banner position, site id and device model. Avazu dataset contains around ten days click-through data of mobile ads following the chronological order. For experiments, two subsets of dataset including 24 fields from the first 9 days of logs for training and remaining for test and evaluations are also available.

\begin{table*}[!htb]
\centering
\scriptsize
\caption{Summary of benchmark datasets for online advertising user response prediction.}
% \vspace{-.75\baselineskip}
\label{tab:datasummary}
\begin{tabular}{|l|l|l|l|l|l|l|l|p{.13\linewidth}|} 
\hline
Category                                      & Dataset                            & \textbar{}Feature\textbar{} & \textbar{}Impression\textbar{} & \textbar{}User\textbar{} & \textbar{}Item\textbar{} & \textbar{}Click\textbar{} & \textbar{}Conversion\textbar{} & Citation                                                                     \\ 
\hline
\multirow{9}{1.3cm}{Display Advertising}          & Criteo\tnote{a}                             & 39                          & 45,840,617                     & -                        & -                        & -                         & -                              & \cite{SimpleChapelle2015,FFMJuan2017,AddressingKtena2019,FiGNNLi2019,xDeepFMLian2018,FeatureGenLiu2019,FwFMPan2018,RobustFMPunjabi2018,ProductBasedNNQu2016,ProductMultiQu2018,TowardsSong2020,AutoIntSong2019,ModelingChapelle2014,CostVasile2017,DCNWang2017,DCNV2Wang2020,OperationYang2019,FAT-DeepFFMZhang2019,DeepFMGuo2017,DeepFMGuo2018,FIBINETHuang2019,InterpretableLi2020,AutoGroupLiu2020}  \\ 
\cline{2-9}
                                              & Avazu(In-app)\tnote{b}                      & 33                          & 40,428,967                     & -                        & -                        & -                         & -                              & \cite{ClickThroughGutpa2018,FIBINETHuang2019,ConvolutionalChan2018,FiGNNLi2019,FeatureGenLiu2019,ConvolutionalLiu2015,RobustFMPunjabi2018,ProductMultiQu2018,TowardsSong2020,AutoIntSong2019,LearningXu2019,FAT-DeepFFMZhang2019,Field-AwareZhang2019,FLENChen2019,DisguiseDeng2017,InterpretableLi2020,AutoGroupLiu2020,AutoFISLiu2020,CollabLiu2014,PBODLLiu2017,IterativeLevine2020,StructuredNiu2018}                           \\ 
\cline{2-9}
                                              & iPinYou\tnote{c}& 16                           & 19.5M                              & -                        & -                        & 14.79K                         & -                              & \cite{SparseFMPan2016,ProductBasedNNQu2016,ProductMultiQu2018,UserRen2016,FNNZhang2016,ZhangBidAware2016,ZhuSEM2017,AutoGroupLiu2020,AutoFISLiu2020}                                               \\ 
\cline{2-9}
                                              & Taobao\tnote{d}                             & -                          & -                      & 987K                        & 4.1M                        & -                         & -                              & \cite{PracticePi2019,LifelongRen2019,UBR4CTRQin2020,BillionWang2018}                                                          \\ 
\hline
\multirow{3}{1.3cm}{Conversion
  post-click logs} & Yoochoose\tnote{e}                          & 9                           & -                              & -                        & -                        & -                         & -                              & \cite{ConvolutionalLiu2015,TimeGligorijevic2019,ExploitingQui2020,JointXu2020}                                                              \\ 
\cline{2-9}
                                              & Taobao\tnote{f}                             & -                           & 84M                            & 0.4M                     & 4.3M                     & 3.4M                      & 18K                            & \cite{EntireMa2018}                                                                        \\ 
\cline{2-9}
                                              & Tencent\tnote{g}                            & 12                          & -                              & -                        & -                        & 50M                       & -                              & \cite{OperationYang2019,WarmupPan2019}                                                              \\ 
\hline
\multirow{3}{1.3cm}{E-commerce
  Platform}        & Amazon (Review logs)\tnote{h}    & -                           & -                              & -                        & -                        & -                         & -                              & \cite{LifelongRen2019,SIMQi2020,PracticePi2019,End-to-EndQu2019,DeepShi2020,BillionWang2018,NGCFWang2019,DIENZhou2019,DINZhou2018,ResembeddingZhou2019,LightGCNHe2020,AdversarialLi2020}                                                    \\ 
\cline{2-9}
                                              & Avito\tnote{i}
  (Classified
  Ad
  portal) & 27                          & 170,588,667                    & -                        & -                        & -                         & -                              & \cite{ClickThroughOuyang2019,DeepOuyang2019,RepresentationOuyang2019}                                                                       \\ 
\cline{2-9}
                                              & Frappe\tnote{j}
  (mobile app)               & 10                          & 288609                         & 957                      & 4082                     & -                         & -                              & \cite{NewWang2018,AttentionalXiao2017,InterpretableLi2020}                                                              \\
\hline
\end{tabular}
\begin{tabularx}{\textwidth}{>{\raggedright}X}
$^{\text{a}}$ \url{http://labs.criteo.com/2014/02/kaggle-display-advertising-challenge-dataset/}\\
$^{\text{b}}$ \url{https://www.kaggle.com/c/avazu-ctr-prediction}\\
$^{\text{c}}$ \url{https://contest.ipinyou.com/}\\
$^{\text{d}}$ \url{https://tianchi.aliyun.com/dataset/dataDetail?dataId=649&userId=1}\\
$^{\text{e}}$ \url{https://recsys.yoochoose.net/challenge.html}\\
$^{\text{f}}$ \url{https://tianchi.aliyun.com/dataset/dataDetail?dataId=408&userId=1}\\
$^{\text{g}}$ \url{https://algo.qq.com/}\\
$^{\text{h}}$ \url{https://nijianmo.github.io/amazon/index.html}\\
$^{\text{i}}$ \url{https://www.kaggle.com/c/avito-context-ad-clicks/data}\\
$^{\text{j}}$ \url{https://github.com/hexiangnan/neural_factorization_machine/tree/master/data/frappe}\\
\end{tabularx}
\end{table*}

\begin{table*}[!ht]
 \footnotesize
    \centering
\caption{Reported user response prediction (click-through rate prediction) results in different experimets  for eight commonly used datasets. The performances were evaluated using AUC-ROC score metric.}
% \vspace{-.75\baselineskip}
% \begin{small}
\scriptsize
% \caption {Input field feature type}
% \cline{2-3}
% \begin{tabular}{|l|l|l|l|l|l|l|l|l|l|l|}
\label{tab:auc}
% \begin{threeparttable}
\begin{tabular}{|l|l|l|l|l|l|l|l|l|}
\hline
% \toprule[1pt]
%\multicolumn{1}{|l}{{\multirow{2}{*}{\centering Category}}} &
\multicolumn{1}{|l|}{{\multirow{3}{1.2cm}{\centering \textbf{Algorithm}}}} & \multicolumn{8}{c|}{\textbf{Dataset(AUC-score)}}                      \\ 

% \multicolumn{1}{|l|}{}& & & &\multicolumn{1}{l|}{{\multirow{2}{1cm}{\centering majority\\voting}}} \\

%\multicolumn{1}{|l}{}&
\multicolumn{1}{|l|}{}&Criteo &Avazu
% &Yoochose
&Avito&Amazon&\shortstack{Taobao \\ Alibaba}& MovieLens&iPinYou&\shortstack{Bing\\News}\\%&Tencent\\
\hline     
%(no-sampling)
% MA-W\&D\cite{WideAndDeepCheng2016}&&&&0.7976(full)&&&&&&\\\hline
MA-W\&D\cite{ClickThroughOuyang2019}&&&0.7976\tnote{a}&&&&&\\\hline
W\&D-SSM\cite{StructuredNiu2018}&&&&&&0.7754\tnote{d}&&\\\hline
% DeepMCP\cite{RepresentationOuyang2019}&&&&0.7927&&&&&&\\\hline
DeepMCP\cite{RepresentationOuyang2019}&&&0.7927&&&&&\\\hline
% DTSN-I\cite{DeepOuyang2019}&&&&0.8395&&&&&&\\\hline
DTSN-I\cite{DeepOuyang2019}&&&0.8395&&&&&\\\hline
% DIEN\cite{DIENZhou2019}&&&&&\shortstack{Elec.(0.7792) \\ Book(0.8453)}&&&&&\\\hline
DIEN\cite{DIENZhou2019}&&&&0.7792\tnote{b},\space    0.8453\tnote{c}&&&&\\\hline
% \multirow{2}{*}{\centering DIEN}&&&&&\multirow{2}{*}{ \shortstack{Elec.(0.7792) \\ Book(0.8453)} }&&&&&\\\hline
% &&&&&&&&&&&\\\hline
% DIN\cite{DINZhou2018}&&&&&Elec.(0.8818)&0.6031&0.7337&&&\\\hline
DIN\cite{DINZhou2018}&&&&0.8818\tnote{b}&0.6031&0.7337\tnote{d}&&\\\hline
DIN-ATT\cite{ResembeddingZhou2019}&&&&0.9106\tnote{b},\space    \textbf{0.9404}\tnote{c}&&0.7429\tnote{d}&&\\\hline
% DSIN\cite{DSINFeng2019}&&&&&&0.6375&&&&\\\hline
DSIN\cite{DSINFeng2019}&&&&&0.6375&&&\\\hline
DeepFM\cite{DeepFMGuo2017}&0.8016&&&&&&&\\\hline
% DeepFM\cite{DeepFMGuo2017}&0.8007&&&&&&&&&\\\hline
% xDeepFM\cite{xDeepFMLian2018}&0.8052&&&&&&&&0.8400&\\\hline
xDeepFM\cite{xDeepFMLian2018}&0.8052&&&&&&&0.8400\\\hline
% PNN\cite{ProductBasedNNQu2016}&0.7700&&&&&&&0.7661&&\\\hline
FNFM\cite{Field-AwareZhang2019}&0.7470&&&&&&&\\\hline
FiBiNET\cite{FIBINETHuang2019}&0.8021&0.7803&&&&&&\\\hline
PNN\cite{ProductBasedNNQu2016}&0.7700&&&&&&0.7661&\\\hline
% FNN\cite{FNNZhang2016}&&&&&&&&0.7071&&\\\hline
FNN\cite{FNNZhang2016}&&&&&&&0.7071&\\\hline
% AutoInt\cite{AutoIntSong2019}&0.8061&0.7752&&&&&1M(0.8456)&&&\\\hline
AutoInt\cite{AutoIntSong2019}&0.8061&0.7752&&&&0.8456\tnote{e}&&\\\hline
% FAT-DeepFFM\cite{FAT-DeepFFMZhang2019}&0.8104&0.7863&&&&&&&&\\\hline
FAT-DeepFFM\cite{FAT-DeepFFMZhang2019}&\textbf{0.8104}&0.7863&&&&&&\\\hline
FGCNN\cite{FeatureGenLiu2019}&0.8022&0.7883&&&&&&\\\hline
RippleNet\cite{RippleNetWang2018}&&&&&&0.9210\tnote{e}&&0.6780\\\hline
DKN\cite{DKNWang2018}&&&&&&&&0.6570\\\hline
Fi-GNN\cite{FiGNNLi2019}&0.8082&\textbf{0.8120}&&&&&&\\\hline
KNI\cite{End-to-EndQu2019}&&&&0.9238\tnote{c}&&\textbf{0.9704}\tnote{d},\space\textbf{0.9449}\tnote{e}&&\\\hline
AutoGroup\cite{AutoGroupLiu2020}&0.8028&0.7915&&&&&0.7859&\\\hline
AutoFIS\cite{AutoFISLiu2020}&0.8009&0.7852&&&&&&\\\hline

%\multirow{5}{*}{\shortstack{Classification \\ Based}\newline }&
% \centering{FNN\cite{FNNZhang2016}}             & \checkmark &  &  &  \\
% %&
% \centering{PNN\cite{ProductBasedNNQu2016}}             & \checkmark &  & \checkmark & \checkmark \\
% %&
% \centering{DIN\cite{DINZhou2018}}            & \checkmark &  & \checkmark &  \\
% %&
% \centering{DIEN\cite{DIENZhou2019}}            & \checkmark & \checkmark &  &  \\
% %&
% \centering{DSIN\cite{DSINFeng2019}}            & \checkmark & \checkmark &  &  \\
% \bottomrule[1pt]
% \hline
\end{tabular}
% \begin{tablenotes}
% \item[a] the.
% \item[b] the.
% \item[c] the.
% \end{tablenotes}
% \begin{tablenotes}[flushleft]\footnotesize
% \renewcommand{\TPTtagStyle}[1]{\makebox[.6em][l]{#1}}
\begin{tabular}{l}
$^{\text{a}}$ The experiment was performed with no sampling approach on the dataset\\
$^{\text{b}}$ Electronics section  $^{\text{c}}$ Book section
$^{\text{d}}$ Section includes 20M rating instances $^{\text{e}}$ Section includes 1M rating instances\\
\end{tabular}
% \item[a] The experiment was performed with no sampling approach on the dataset
% \item[b] Electronics section  $^{\text{c}}$ Book section
% \item[d] Section includes 20M rating instances $^{\text{e}}$ 

% \end{tablenotes}
% \end{threeparttable}
% \end{small}
\end{table*}
% \vspace{-.5cm}
% \end{tabular}
% \end{table*}

% \begin{tabular}{|l|V{3.5cm}|l|}
%   foo & blah      & bar \\
%   foo & blah blah & bar \\
%   foo & blah blah blah blah blah blah
%                   & bar \\
% \end{tabular}

\subsection{Open-source Implementations}
In this subsection, we collect a set of methods introduced in different published papers in one place. Table \ref{tab:sourceLinks} denotes a list of presented methods covered in this paper. It includes the presented method along with an employed methodology and a link to the corresponding GitHub page of their implementation. %As a common approach, several studies come up with a code link along with their published papers. It is considered to give research communities access to the implementation of proposed algorithms. 
\begin{table*}[h!]
\centering

\caption{Summary of classification based methods with open-source implementations}
% \vspace{-.75\baselineskip}
\label{tab:sourceLinks}
% \footnotesize
\tiny

% \centering
\begin{tabular}{|l|l|p{.23\linewidth}|>{\tiny}p{.44\linewidth}|} 
\hline
Category                                      & Model           & Framework                 & Link                                                                        \\ 
\hline
\multirow{9}{1.2cm}{Factorization
  machines
  ~} & %\cline{2-4}
                                               PNN~\cite{ProductBasedNNQu2016}, KFM~\cite{ProductMultiQu2018}            & Tensorflow,libFFM         & \url{https://github.com/Atomu2014/product-nets-distributed}                                  \\ 
\cline{2-4}
%                                               & KFM~\cite{ProductMultiQu2018} & Tensorflow,libFFM         & \url{https://github.com/Atomu2014/product-nets-distributed}                       \\ 
% \cline{2-4}
                                              & NFM~\cite{NFMHe2017}            & libFM,Tensorflow          & \url{https://github.com/hexiangnan/neural\_factorization\_machine}                \\ 
\cline{2-4}
                                              & FiBiNET~\cite{FIBINETHuang2019}         & DeepCTR, DeepCTR-Torch  & \\ 
\cline{2-4}
                                              & AFM~\cite{AttentionalXiao2017}             & Tensorflow                & \url{https://github.com/hexiangnan/attentional\_factorization\_machine}           \\ 
\cline{2-4}
                                              & Robust-FM~\cite{RobustFMPunjabi2018}       & Spark                     & \url{https://www.dropbox.com/sh/ny6puvtopl98339/AACExLZ0waDL\_ibWhfNItJfGa?dl=0}  \\ 
\cline{2-4}
                                              & FLEN~\cite{FLENChen2019}            & Tensorflow                & \url{https://github.com/aimetrics/jarvis}                                         \\ 
\cline{2-4}                                         
                                              & AutoFIS~\cite{AutoFISLiu2020}            & Tensorflow                & \url{https://github.com/zhuchenxv/AutoFIS}                                         \\ 
\cline{2-4}                
&Others\cite{FMRendle2010,FFMJuan2017,DeepFMGuo2017,Field-AwareZhang2019}               & PyTorchFM                   & \url{https://github.com/rixwew/pytorch-fm}                                        \\ 
                                              
\hline
\multirow{19}{1.2cm}{Deep
  Learning
  ~}         & AutoInt~\cite{AutoIntSong2019}         & Tensorflow                & \url{https://github.com/DeepGraphLearning/RecommenderSystems}                     \\ 
\cline{2-4}
                                              & FGCNN~\cite{FeatureGenLiu2019}           & DeepCTR~\cite{DeepCTR},
  Tensorflow     & \url{https://github.com/shenweichen/DeepCTR}                                      \\ 
\cline{2-4}
                                              & CCPM~\cite{ConvolutionalLiu2015}           & DeepCTR-Torch~\cite{DeepCTRTorch}
  PyTorch   & \url{https://github.com/shenweichen/DeepCTR-Torch}                                \\ 
\cline{2-4}
                                              & xDeepFM~\cite{xDeepFMLian2018}         & Tensorflow                & \url{https://github.com/Leavingseason/xDeepFM}                                    \\ 
\cline{2-4}
                                              & DIN~\cite{DINZhou2018}             & Tensorflow-gpu            & \url{https://github.com/zhougr1993/DeepInterestNetwork}                           \\ 
\cline{2-4}
                                              & DSIN~\cite{DSINFeng2019}           & DeepCTR~\cite{DeepCTR},
  Tensorflow-gpu & \url{https://github.com/shenweichen/DSIN}                                         \\ 
\cline{2-4}
                                              & DIEN~\cite{DIENZhou2019}           & Tensorflow                & \url{https://github.com/mouna99/dien}                                             \\ 
\cline{2-4}
                                              & Wide\&Deep~\cite{WideAndDeepCheng2016}        & Tensorflow                & \url{https://github.com/Tensorflow/models/tree/master/official/r1/wide\_deep}     \\ 
\cline{2-4}
%                                               & DSSM\cite{SS}            & Tensorflow                & \url{https://github.com/baharefatemi/DSSM                                        \\ 
% \cline{2-4}
                                              & FNN~\cite{FNNZhang2016}             & Theano                    & \url{https://github.com/wnzhang/deep-ctr}                                         \\ 
\cline{2-4}
                                              & DSTN~\cite{DeepOuyang2019}            & Tensorflow                & \url{https://github.com/oywtece/dstn}                                             \\ 
\cline{2-4}
                                              & W\&D SSM~\cite{StructuredNiu2018}             & Tensorflow                & \url{https://github.com/niuchenglei/ssm-dnn}                                      \\ 
\cline{2-4}
                                              & Meta-embedding~\cite{WarmupPan2019} & Tensorflow                & \url{https://github.com/Feiyang/MetaEmbedding}                                    \\ 
\cline{2-4}
                                              & DeepMCP
~\cite{RepresentationOuyang2019}         & Tensorflow                & \url{https://github.com/oywtece/deepmcp}                                          \\ 
\cline{2-4}
                                              & FAT-DeepFFM~\cite{FAT-DeepFFMZhang2019}     & PyTorch                   & (unofficial) \url{https://github.com/p768lwy3/torecsys}                           \\ 
\cline{2-4}
                                              & MA-DNN
\cite{ClickThroughOuyang2019}              & Tensorflow                & \url{https://github.com/rener1199/deep\_memory}                                   \\ 
\cline{2-4}
                                              & RocketLaunching~\cite{RocketLaunchingZhou2017} & PyTorch                   & \url{https://github.com/zhougr1993/Rocket-Launching}                              \\ 
\cline{2-4}
                                              & GMP~\cite{OnlineChao2019}             & Tensorflow                & \url{https://github.com/graphmp/graphmp}                                          \\ 
\cline{2-4}
                                              & MIMN
~\cite{PracticePi2019}            & Tensorflow                & \url{https://github.com/UIC-Paper/MIMN}                                           \\ \cline{2-4}
                                              & UBR4CTR
~\cite{UBR4CTRQin2020}            & Tensorflow                & \url{https://github.com/qinjr/UBR4CTR}                                           \\ 
\hline
\multirow{5}{1.2cm}{Network
  based}              & KNI~\cite{End-to-EndQu2019}             & Tensorflow                & \url{https://github.com/Atomu2014/KNI}                                           \\ 
\cline{2-4}
                                              & Fi-GNN~\cite{FiGNNLi2019}          & Tensorflow                & \url{https://github.com/CRIPAC-DIG/Fi\_GNNs}              \\ 
\cline{2-4}
                                              & RippleNet~\cite{RippleNetWang2018}       & Tensorflow                & \url{https://github.com/hwwang55/RippleNet}              \\ 
\cline{2-4}
                                              & DKN~\cite{DKNWang2018}             & Tensorflow                & \url{https://github.com/hwwang55/DKN}                                             \\ 
\cline{2-4}
                                              & KGNN-LS\cite{Knowledge-AwareWang2019}        & Tensorflow                & \url{https://github.com/hwwang55/KGNN-LS}                                         \\
\hline
\end{tabular}
\end{table*}
We intend to facilitate it for  research communities who need to follow the same program setting to compare and evaluate different methods. The majority of the implementation are developed by using Python programming language. In addition to official implementations, there are some toolkits like DeepCTR \cite{DeepCTR} and DeepCTR-Torch \cite{DeepCTRTorch} implemented based on Tensorflow and PyTorch platforms. They are not only provided users with third party implementations of contemporary methods in literature, but also they prepared a platform including several software components to build customized models. In this case researchers can examine different methods under the same input and output interface and apply the similar setting to all approaches.   
% % \subsection{Part One}

% % Lorem ipsum dolor sit amet, consectetur adipiscing elit. Morbi
% % malesuada, quam in pulvinar varius, metus nunc fermentum urna, id
% % sollicitudin purus odio sit amet enim. Aliquam ullamcorper eu ipsum
% % vel mollis. Curabitur quis dictum nisl. Phasellus vel semper risus, et
% % lacinia dolor. Integer ultricies commodo sem nec semper.

% % \subsection{Part Two}

% % Etiam commodo feugiat nisl pulvinar pellentesque. Etiam auctor sodales
% % ligula, non varius nibh pulvinar semper. Suspendisse nec lectus non
% % ipsum convallis congue hendrerit vitae sapien. Donec at laoreet
% % eros. Vivamus non purus placerat, scelerisque diam eu, cursus
% % ante. Etiam aliquam tortor auctor efficitur mattis.

% % \section{Online Resources}

% % Nam id fermentum dui. Suspendisse sagittis tortor a nulla mollis, in
% % pulvinar ex pretium. Sed interdum orci quis metus euismod, et sagittis
% % enim maximus. Vestibulum gravida massa ut felis suscipit
% % congue. Quisque mattis elit a risus ultrices commodo venenatis eget
% % dui. Etiam sagittis eleifend elementum.

% % Nam interdum magna at lectus dignissim, ac dignissim lorem
% % rhoncus. Maecenas eu arcu ac neque placerat aliquam. Nunc pulvinar
% % massa et mattis lacinia.
\subsection{Applications} In this subsection, we briefly review applications typically developed in relation to the user click-through rate prediction. The evaluation of the probability that users make interactions more than a click on promoted item or an ad have been studied from different aspect in many research. 
\paragraph{\underline{Revenue per click prediction}}
%A modern digital advertising is working based on user feedback. In this form advertisers pay publisher platforms to place different ads on their webpages. Following user, there are performance dependent payment models like ost-per-click model advertisers will only be charged if the users click on their ads. The cost-per-conversion option reduces the advertiser’s risk even further by allowing them to pay only if the user takes a predefined action on their website (such as purchasing a product or subscribing to an email list). 
%In digital advertising, there are two metrics like cost-per-click and cost-per-conversion to evaluate the the performance of campaign ads and the value should be paid to advertisers. From the advertisers' perspective, In contrast to these metrics which show the amount of money spent, there is another metric called RPC(revenue per click) to show the earning made from advertising campaigns. A well-known challenge in the RPC prediction problem is that, at the bid-unit level, the data are very sparse. From the perspective of users’ behaviors, the sparsity challenge is twofold. First, for a large number of bid units, only a small number of days record non-zero clicks. Second, among all the bid units that are clicked, the majority does not generate any revenue for the advertiser.
This is an application to show earning of advertisers made from ad campaigns. In this application a metric called RPC(revenue per click) is used to compute the advertising interest based on user feedback in the form of clicks or conversions.
%In digital advertising, metrics like click-through rate and conversion rate are considered as key measures to evaluate the the performance of campaign ads and determine the value should be paid by advertisers to publishers. From the advertisers' perspective, in contrast to these metrics which show the amount of money spent, there is another metric called RPC(revenue per click) to show the earning made from advertising campaigns.
Like the click-through rate prediction, the data sparsity is the common issue from advertiser  perspectives to estimate the revenue based on user click responses. It means that not only the proportion of click events over impression are so small in user behavior histories, but also the number of bid units which receive the click response and lead to revenue is also very few. Although this metric is essentially important to analyze the performance of the online advertising, there are a few studies published in literature regarding this metric because of revenue confidentiality policy adopted by many companies. In the following part, we go over an important work in this domain:

%by adopting empirical Bayes [],
%Although RPC is a vital metric in advertiser bidding decisions, the RPC-related literature is limited, partly because of the confidentiality of revenue data. Among available studies,
Authors in \cite{DynamicYuan2018} proposed a model to dynamically determine the data-driven hierarchy defined for the ad group and campaign and advertisers account. Meanwhile, they presented an empirical bayes method to get inferences through the hierarchical structure. In the context of the sponsored search advertising, bid values typically assigned for keywords to calculate the potential advertiser revenue. However, bid units are formed as the atomic units for the combination of a keyword in addition to match type and an ad group. The performance data were collected on the advertisers’ side for experiments contain daily impressions, clicks, conversions and attributed revenue at the bid-unit level. Therefore, the prediction problem was defined as the prediction of the next day’s RPC for given bid unit, given the historical clicks and revenue data. The features are the hierarchical structure information of the bid units. It encompasses corresponding campaigns, ad groups, and keywords, as well as geo-targeting information at the campaign level, which are shared by the bid units under each campaign. Following an empirical approach to get inferences through the hierarchical structure, they proposed an extended empirical bayes method that was capable of dynamically constructing the hierarchy and used the loss concept in decision tree models for optimization.
\paragraph{\underline{Mistaken click prediction}}
In performance-based business models like the pay-per-click, the quality of click event plays an important role for the revenue made for advertisers. Relying on click events, there is a probability that advertisers are charged by Ad Exchange Network using valueless clicks. These type of clicks is generally considered as accidental clicks quite often happen in mobile devices, when users are confronting interstitial ads. They are interrupted by the ads covering the whole screen. They may click on ads and be directed to a website and bounce back without spending a considerable time. Ignoring these type of click may lead to overestimation of click-through rate values for the mobile advertising.  

To this end, authors in \cite{YouTolomei2018} proposed a data-driven method to detect mistaken clicks on ads. Although from advertisers perspective, valuable clicks are those followed by conversions, but it is not always true for all click events. In the context of Yahoo mobile apps, they categorized the clicks using extra information about the time users spending on the advertiser landing page into three sections accidental, short and long. Decomposing a dwell time distribution into three above classes, they proposed a technique to apply a smooth discounting factor to charge less advertisers with regard to accidental clicks. 

\paragraph{\underline{Fraud in Online Advertising}}
Today, the main principle to target users for online advertising is tailoring ads to user interest profiles. It consequently leads to a billing model to charge advertisers and pay publishers based on how many times targeted users interact with ads. In this business model, publishers register themselves in the ad network to host ad placements and advertisers organize ad campaigns for target users. From publisher’s perspective, the revenue from advertising is directly dependent to number of users interacting with ads in web-pages and the cost paid by advertisers for displaying targeted ads. The rate of investment in online advertising is ascending annually. It tempts some people to commit fraudulent activities. According to common performance based business models like the cost-per-click and cost-per-impression for the sponsored search and display advertising, fraudulent form of clicks and views for textual rich media and video ads has gained a lot of attentions~\cite{UnderstandingMarciel2016}. 

In study conducted in \cite{YourMeng2014}, it is demonstrated that organizing user interest profile relying on user visits have a vulnerability can be used for web-based fraud activities like the cross-site request forgery scripting and click-jacking to embed hidden requests not initiated by users. The increase in adopting IoT based solutions let users connected to Internet from different devices. Compromised devices which do not follow basic security measures can be easy target for such exploitations~\cite{DataSafa2020}. The fraud activities could orchestrate an attack against DoubleClick Ad Exchange Network to a manipulate user interest profile which can modify the publisher revenue. The main attribute of this attack is preparing a mechanism to modify user interest profiles without explicit interactions with ad exchange network and further knowledge about external involving factors. It was planned to generate polluted user profiles worked based on the behavioral targeting and re-targeting which led to present biased targeted ads that in turn need the higher bid price made by advertisers and revenue for the publisher. 

To deal with these smart threatening attacks, different studies have been conducted to take the advantage of machine learning based methods to address challenges from different aspects \cite{CrowdTian2015,DeepThejas2019,DetectingOentaryo2014}. In \cite{DeepThejas2019} authors investigated to use an auto-encoder neural network and GAN to regenerate click events. They designed a neural network model to predict a fake click events by adding some extent of noise to input data. In the context of sponsored search advertising, the threat of fraud crowdsourcing activities was discussed in \cite{CrowdTian2015}. In this case, the fraudulent behaviors including a series of search and click ads is distributed among vast number of web-publishers where fraudulent traffic is buried in the majority of normal traffic. They constructed a graph to represent a click history of users. They then applied a clustering algorithm using a dispersity filter to find the coalitions that attacks are concerted against a common set of advertisers. %Constructing a graph to represent a click history of users, % and remove false positive clusters.  

\paragraph{\underline{Others}}
The application of predicting user response rates is not limited to above applications. They are some studies consider the user response from different aspects to predict short-term final desirable purchase activities \cite{TimeGligorijevic2019,OnlineChao2019} or long-term consistent influence like a branding \cite{ModelBaron2014} in common mobile and display advertising.

\subsection{Future Directions} \label{future_sec}
User response prediction is considered as an important task to evaluate the effectiveness of online advertising and recommender systems. It has led to various research studies to address this problem. In this section, we present a number of possible future directions identified in the recent studies. They can be found useful by researchers in the community to develop the next solutions.
\paragraph{Joint optimization models based on business value metrics}
%As we discussed in the previous sections, 
Implicit feedback like CTR value prediction is commonly used as an objective to select candidate ads for end-users %It is be used as indicator to show user preferences in diverse applications like new-feed and video and social-network based recommendations
\cite{DKNWang2018,PracticalHe2014}. An accurate prediction of CTR values can make a direct impact in the biding value of ad in the display advertising~\cite{SimpleChapelle2015}. Although CTR values can potentially show the user tendency with regard to items and commercials in online provider systems, it is not a definite indicator to show the success of business to gain new customers. %, and there is a need to address the effect of all stakeholders in the real environment. 
Accordingly, business oriented conversion rate values, which convey additional information about user intent with regard to products and services in online systems, are proposed. %Very recently, some studies analyze business value metrics in form of multi-objective optimization. 
For example, by comparing different user conversion feedback values initiated by click incident on items such as add-to-cards, add-to-wishlist and order productsa, order rate is found to be the most robust objective rather compared to add-to-cart to provide relevant personalized recommended list made by e-commerce search~\cite{OnApplicationKarmaker2017}. Pareto efficient learning-to-rank algorithm~\cite{ParetoLin2019} is also proposed to calculate solution by aggregating two loss functions for CTR and Gross Merchandise Value (GMV) prediction using proper constraints. Studies such as \cite{ValueAwarePei2019,Jointly2020} encompass a couple of user interactions as the conversion user feedback. These actions are mapped to profit value which are considered as the reward for the system. In the model training following reinforcement learning, a policy to find better user actions is learnt to maximize accumulated profits and accommodate user preferences. Learning multiple factors with regard to optimize user preferences and commercial profits have a potential to shape the new trend in developing recommendation system algorithms.
% \paragraph{Secure training in recommendation systems}
\paragraph{Neural Architecture Search for User Response Prediction}
Deep neural networks are becoming increasingly popular in recommendation and online advertising, %the structure of deep neural network component has been considered as a design hyper-parameter. As
but training process for them are potentially expensive. For many methods, the structure of neural network are manually set via extensive empirical studies. In our review, we have seen few attempts like \cite{FNNZhang2016,ProductBasedNNQu2016} to present fully connected neural network for user response prediction. To model user interactions with other components in the context of display advertising, diamond shape of multi-layer perception network to have larger hidden layer is suggested by comparing different structures. But broadly speaking, this process is not well-studied in the literature and heavily relies on intuitive understanding of developers and knowledge from the application domain like~\cite{AutoFISLiu2020,AutoGroupLiu2020}. Neural architecture search is to automate the architecture selection process to find the best neural network architecture via an optimization procedure~\cite{NeuralZoph2017}. %. The research field has obtained significant attention after introducing new approach using reinforcement learning to tune an RNN network~\cite{NeuralZoph2017}. 
%Recently some work propose to improve the neural network based model for user response prediction. 
Multi-objective evolutionary optimization~\cite{TowardsSong2020} has recently been proposed to select network architecture, by organizing the search space as a direct acyclic graph, and utilizing learning-to-rank to search and filter out a selection of architectures in each iteration. Future research may consider domain knowledge and business metrics to design efficient and effective neural architecture search strategies for user response prediction.  %The evaluation of system is adopted in terms of the performance of the model to predict click-through rate values. 
%Aside from this study, recommender systems have different challenges in terms of scalability and heterogeneity in input data. This topic needs more studies from different angles.

%\paragraph{Retrain models in user response prediction task and recommender systems}
\paragraph{Online Learning User Response Prediction and Recommender Systems}
%Predictive models are often applied to predict user preferences and the next user interactions in online service providers. In the scenario of social sponsored advertising or digital 
%In order to predict user response, predictive models are continuously trained based on the stream of input data. Practically, the 
Online environment is inherently dynamic that the stream of data are changing over time. It may lead a gradual shift in user preferences which can affect the performance of predictive models. So recommendation systems generally need to apply online learning (retraining mechanisms) to update and tackle new user interactions. This is partially related to cross-domain recommendation approaches that a pre-trained recommendation model is applied to different downstream tasks \cite{ParameterYuan2020}. The full training of model is the straight-forward solution which is only helpful when we have a limited amount of data \cite{IterativeLevine2020}. For online scenarios, selection based retraining and fine-tuning are other types of retraining solutions. The former applies a selection method for sampling of older user interaction and new data to create an updated training data while the latter proposes a transferring strategy to train a model using the new user interaction information. Very recently authors in \cite{HowZhang2020} propose to learn transfer component in a cyclic fashion using meta-learning approach for sequential input data. Nonetheless, this topic is one of the important subjects for practical recommendation systems that needs to be well-addressed in future.
% In addition to specifying the {\itshape template style} to be used in
% formatting your work, there are a number of {\itshape template parameters}
% which modify some part of the applied template style. A complete list
% of these parameters can be found in the {\itshape \LaTeX\ User's Guide.}

% Frequently-used parameters, or combinations of parameters, include:
% \begin{itemize}
% \item {\verb|anonymous,review|}: Suitable for a ``double-blind''
%   conference submission. Anonymizes the work and includes line
%   numbers. Use with the \verb|\acmSubmissionID| command to print the
%   submission's unique ID on each page of the work.
% \item{\verb|authorversion|}: Produces a version of the work suitable
%   for posting by the author.
% \item{\verb|screen|}: Produces colored hyperlinks.
% \end{itemize}

% This document uses the following string as the first command in the
% source file:
% \begin{verbatim}
% \documentclass[acmtog]{acmart}
% \end{verbatim}

% \section{Applications}

% Modifying the template --- including but not limited to: adjusting
% margins, typeface sizes, line spacing, paragraph and list definitions,
% and the use of the \verb|\vspace| command to manually adjust the
% vertical spacing between elements of your work --- is not allowed.

% {\bfseries Your document will be returned to you for revision if
%   modifications are discovered.}

\end{document}